\documentclass[prd,aps,nofootinbib,floats,twocolumn]{revtex4}

\usepackage[dvips]{graphicx}
\usepackage{amssymb}
\usepackage{amsmath}

\usepackage{color}

\begin{document}

\title{The spin expansion for binary black hole mergers: \\
new predictions and future directions}

\author{Latham Boyle and Michael Kesden}

\affiliation{Canadian Institute for Theoretical Astrophysics (CITA)}

\date{December 2007}

\begin{abstract}
  In a recent paper \cite{Boyle:2007sz}, we introduced a spin
  expansion that provides a simple yet powerful way to understand
  aspects of binary black hole (BBH) merger.  This approach relies on
  the symmetry properties of initial and final quantities like the
  black hole mass $m$, kick velocity ${\bf k}$, and spin vector ${\bf
    s}$, rather than a detailed understanding of the merger dynamics.
  In this paper, we expand on this proposal, examine how well its
  predictions agree with current simulations, and discuss several
  future directions that would make it an even more valuable tool.
  The spin expansion yields many new predictions, including several
  {\it exact} results that may be useful for testing numerical codes.
  Some of these predictions have already been confirmed, while others
  await future simulations.  We explain how a relatively small number
  of simulations --- 10 equal-mass simulations, and 16 unequal-mass
  simulations --- may be used to calibrate all of the coefficients in
  the spin expansion up to second order at the minimum computational
  cost.  For a more general set of simulations of given covariance, we
  derive the minimum-variance unbiased estimators for the spin
  expansion coefficients.  We discuss how this calibration would be
  interesting and fruitful for general relativity and astrophysics.
  Finally, we sketch the extension to eccentric orbits.
\end{abstract}

\maketitle

\section{Introduction}

Binary black hole (BBH) merger --- in which two spinning black holes
inspiral due to the emission of gravitational radiation and eventually
merge to form a single spinning black hole --- is one of the most
important problems in classical general relativity, and has
significant ramifications for astrophysics, cosmology, and
gravitational-wave observations.  For decades, analytical and
numerical approaches to the BBH merger problem have been frustrated by
conceptual and technical difficulties associated with solving the
non-linear Einstein equations --- especially during the last few
orbits and final plunge, when the ``luminosity'' in gravitational
radiation is highest.  Dramatic progress came in 2005, as new insights
and increased computational resources finally allowed numerical
relativists to simulate the {\it entire} merger --- including the last
few orbits of inspiral, the plunge, the formation of a common event
horizon, and the ringdown of the final Kerr black hole
\cite{Pretorius:2005gq, Campanelli:2005dd, Baker:2005vv}.  Following
this breakthrough, simulations of BBH merger have produced a number of
remarkable results.

The most surprising and interesting results have been obtained just
within the past year, from simulations of merging black holes with
large initial spins \cite{Campanelli:2006fy, Campanelli:2007ew,
  Schnittman:2007sn, Damour:2007cb, Schnittman:2007nb,
  Schnittman:2007ij, Buonanno:2007sv, Rezzolla:2007rd,
  Gonzalez:2007hi, Campanelli:2007cg, Brugmann:2007zj,
  Herrmann:2007ex, Baker:2007gi, Tichy:2007hk, Herrmann:2007ac,
  Koppitz:2007ev, Pollney:2007ss, Rezzolla:2007xa, Campanelli:2006uy,
  Marronetti:2007wz}.  The result that has received the most attention
is that highly-spinning initial black holes can merge to form a final
black hole with an enormous recoil velocity --- as large as
$4,000~{\rm km/s}$ --- relative to the binary's center-of-momentum
frame \cite{Gonzalez:2007hi, Campanelli:2007cg, Brugmann:2007zj}.  The
idea of a supermassive black hole rocketing through its host galaxy at
such a speed has understandably caught the attention of many
astrophysicists!

In this paper, we highlight a second surprising result.  Despite the
complicated and non-linear dynamics of the merger process, the final
state of the merger seems, in some sense, to be an unexpectedly simple
and smooth function of the initial state of the binary.  We would like
to make this statement more quantitative and precise.

In a recent paper \cite{Boyle:2007sz}, we introduced a ``spin
expansion'' formalism for understanding aspects of BBH merger.  We
expand on this proposal in several ways in the present paper.  Here is
a recap of the basic idea.  Even though the merger is a messy
non-linear process, it is useful to regard it as a map from a simple
initial state (two well separated Kerr black holes with mass ratio $q
\equiv M_{b}/M_{a}$ and dimensionless spins ${\bf a}$ and ${\bf b}$)
to a simple final state (a final Kerr black with mass $m$, spin vector
${\bf s}$ and kick velocity ${\bf k}$).  Given any final quantity $f$
({\it e.g.}\ $m$, ${\bf k}$, or ${\bf s}$), we can Taylor expand the
function $f(q,{\bf a},{\bf b})$ around ${\bf a}={\bf b}=0$, and use
symmetry arguments to dramatically reduce the number of independent
terms at each order.  When compared with published simulation results,
this ``spin expansion'' seems to rapidly converge: the leading-order
terms yield a surprisingly good first approximation, the
next-to-leading-order terms give an even better approximation, and so
on.  In the present paper, we present these points in detail, and
explore some of their implications and extensions.

\subsection{Some advantages of the spin expansion}

{\it How does the spin expansion complement other approaches to the
  BBH merger problem?}

{\bf 1.} First, it is {\it simple}.  Previous approaches --- notably
the post-Newtonian approximation and full numerical relativity ---
are highly technical and sophisticated and have taken decades to
develop.  By contrast, we believe that the derivations in this
paper will be accessible, even to physicists with no prior expertise
in this area.

{\bf 2.} Second, it is {\it general}.  In this paper, we focus on
applying the spin expansion to the final black hole's mass, kick, and
spin ($m,{\bf k},{\bf s}$), but we expect that analogous arguments may
be useful for studying {\it any} other final observable with well
defined transformation properties under the simple symmetries $\{R, P,
X\}$ discussed below.  This may also include the multipoles of
the gravitational-wave signal emitted by the binary --- this is
currently a speculation, and remains a topic for future work.

{\bf 3.} Third, it is {\it conceptually distinct}.  Previous
approaches attempt to follow the {\it dynamics} of the merger --- by
solving the Einstein equations numerically, or through some analytical
approximation.  By contrast, our approach is to consider the map
directly from the initial state to the final state, and thereby ``leap
over'' the complicated merger dynamics in between.  To constrain this
map, we rely purely on {\it symmetry} arguments, together with the
{\it assumption} (supported by simulations) that the Taylor expansion
of the map around ${\bf a}={\bf b}=0$ converges rapidly.  Therefore,
our approach clarifies which aspects of the final state are due to the
complicated non-linearities of Einstein's equations, and which aspects
follow from more elementary considerations.

{\bf 4.} Fourth, it is {\it practical} for cosmological and
astrophysical applications.  For example, a cosmological simulation of
galaxy merger may also wish to track the corresponding supermassive
black holes, since their feedback may be important for galactic
structure and evolution.  Of course, it would be hopeless to follow
the BBH dynamics in detail --- the dynamical time near final merger is
too short relative to the other timescales in the problem.  Instead,
one is likely to resort to a simplifying algorithm: {\it e.g.}\ when
the two holes get sufficiently close, they are replaced by a single
hole with appropriate quantities $\{m,{\bf k},{\bf s}\}$.  The fact
that the spin expansion maps the initial state (well before merger)
directly to the final state (well after merger) makes it well suited
for these types of problems.

{\bf 5.} Fifth, it is {\it efficient}.  To fully specify the initial
BBH spin configuration, we must specify 6 numbers --- 3 components
each for the initial spins ${\bf a}$ and ${\bf b}$.  What is the most
economical way to map out this 6-dimensional space with numerical
simulations?  If we crudely put 10 grid points along each direction,
we would need $10^{6}$ simulations --- an impossibly large value,
since each simulation is very computationally expensive.  On the other
hand, we can use the spin expansion to map out the {\it same}
6-dimensional space, at second or third-order accuracy, with only
${\cal O}(10)$ simulations --- a huge computational savings!  This
issue is treated in detail in Sec.~\ref{S:prog}.  In this sense, the
spin expansion acts like a kind of ``data compression,'' describing
the 6-dimensional space of initial spins more succinctly, without {\it
  over}simplifying it.

{\bf 6.} Sixth, it is {\it valid through the entire merger.}  The spin
expansion is based on {\it exact} symmetries of general relativity,
which are equally valid during all stages of BBH merger: inspiral,
plunge, and ringdown.  Hence, it is particularly well suited for
questions about how the final state depends on the initial state in
BBH merger.  By contrast, the post-Newtonian approximation is an
expansion of the Einstein equations in $v/c$, and treats BBHs as pairs
of interacting point particles.  It breaks down during the late stages
of the merger --- the last orbits, plunge, and ringdown, when the
black holes begin to orbit with relativistic velocities and then cease
to be two separate entities.  Since these late stages emit
gravitational waves copiously, and play a crucial role in determining
the properties of the final black hole, the post-Newtonian formalism
is {\it not} well suited for predicting the final state of BBH merger.
(On the other hand, it is excellently suited for describing the binary
and its gravitational-wave emission when the black holes are well
separated and orbiting non-relativistically.)

{\bf 7.} Seventh, it is {\it predictive}.  As we show in detail in
this paper, the spin expansion makes a host of detailed (and
successful) quantitative predictions for the results of simulations.
Many of these predictions are new and distinct from the predictions of
other analytical approaches.  They reveal interesting features of the
simulations that might not have been noticed otherwise.

{\bf 8.} These predictions are {\it derived} --- they are not guesses.
This is to be contrasted with expressions for the final kick velocity
\cite{Campanelli:2007cg, Campanelli:2007ew, Baker:2007gi,
  Herrmann:2007ex} which were {\it inspired} by post-Newtonian
equations, but are ultimately empirical fitting formulae which are
{\it not} derived.  Indeed, if these formulae continued to hold in
general, it would be quite amazing.  They are linear in the initial
spins, and we will highlight several effects which appear to be
quintessentially non-linear in the spins, and hence {\it not} captured
by the post-Newtonian-inspired fitting formulae.

In this subsection, we have made an aggressive case for the merits of
the spin expansion.  We must also stress the obvious point that the
spin expansion merely {\it complements} the other approaches to BBH
merger --- it does not replace them!  It should be clear that
numerical simulations, post-Newtonian techniques, and
post-Newtonian-inspired fitting formulae offer a huge amount of
dynamical information and insights which cannot be obtained from the
spin expansion alone.  Nonetheless, the spin expansion provides
unique understanding, as we hope will become clear in the course
of this paper.

\subsection{Observational motivations}

The process of BBH merger is of great observational interest, since it
is expected to govern the final evolution of both {\it stellar-mass}
BBHs (produced in stellar collapse) and {\it supermassive} BBHs
(produced whenever galaxies merge).

Gravitational waves from merging stellar-mass BBHs are an important
source for the ground-based detector LIGO \cite{LIGO}, while
gravitational waves from merging supermassive BBHs are a primary source
for the space-based detector LISA \cite{LISA}.  These gravitational
waves will provide
unprecedented tests of strong-field general relativity, and open a new
window onto exotic and previously invisible astrophysical phenomena.
For example, after the two initial black holes form a common event
horizon, they are predicted to ``ring down'' (like a bell) to a final
quiescent Kerr black hole.  This ring-down signal is an important
observable for future gravitational wave detectors, and may be thought
of as a superposition of so-called ``quasi-normal modes.''  The
observed spectrum of quasinormal modes depends on the quantities
$\{m,{\bf k},{\bf s}\}$ characterizing the final black hole, so the
ability to predict these quantities may play an important role in
interpreting these observations.

The quantities $\{m, {\bf k}, {\bf s}\}$ characterizing the final
black hole are also interesting astrophysically.  These quantities may
be probed observationally via the x-ray spectrum emitted from the
inner edge of the accretion disk around a black hole.  From the
standpoint of {\it supermassive} BBH merger, they are believed to be
linked to a variety of observables, including: ({\it i}) the quasar
luminosity function \cite{Kauffmann:1999ce,Cattaneo:1999bf}; ({\it ii})
the location of a quasar with respect to its host galaxy
\cite{Loeb:2007wz}; ({\it iii}) the orientation and shape of jets in
active galactic nuclei \cite{DennettThorpe:2001vy,Merritt:2002hc};
({\it iv}) the correlation between black hole mass and velocity
dispersion in the surrounding stellar bulge
\cite{Gebhardt:2000fk,Tremaine:2002js};
({\it v}) the density profile in the centers of galaxies
\cite{Milosavljevic:2001dd,Gualandris:2007nm}.  Supermassive BBH
merger is part of an interconnected web of astrophysical issues ---
the structure of this web is still poorly understood, and many
fascinating questions remain.

Finally, since the present paper is concerned with the merger of {\it
  spinning} black holes, we should note that many --- perhaps even
{\it most} --- astrophysical black holes are indeed expected to have
significant spin.  On theoretical grounds, black holes grown by gas
accretion are expected to have spins near the maximum Kerr limit
\cite{Bardeen:1970, Thorne:1974ve}.  These predictions are supported
by recent observations of active galactic nuclei (AGN) by the X-ray
observatory {\it XMM-Newton}.  Features in the Fe-K$\alpha$ line of
the Seyfert 1.2 galaxy MCG---06-30-15 are best modeled by a black hole
with Kerr parameter $a=0.989_{-0.002}^{+0.009}$ at 90\% confidence
(where $a=1$ would correspond to the maximal Kerr value)
\cite{Brenneman:2006hw}.  Other observations suggesting significant
black hole spin include \cite{Elvis:2001bn, Wang:2006bz}.  While
torques from accreting gas tend to align the orbital and spin angular
momenta with the large-scale gas flow in gas-rich ``wet'' mergers
\cite{Bogdanovic:2007hp}, no such mechanisms exist in gas-poor ``dry''
mergers \cite{Schnittman:2004vq} making studies of generic initial BBH
spin configurations essential for understanding these systems.

\subsection{Outline of this paper}

This paper is organized as follows.  In Sec.~\ref{S:formalism} we
review the formalism of the spin expansion introduced in our {\it
  Letter} \cite{Boyle:2007sz}.  We use this formalism in
Sec.~\ref{S:exact} to identify several particularly symmetric initial
spin configurations for which a subset of the final quantities $f \in
\{ m, s_i, k_i \}$ {\it identically} vanish.  These configurations may
be useful to numerical relativists to help identify systematic errors
in their codes that violate these symmetry constraints.  We compare
the predictions of our spin expansion to simulations of these
configurations and some less symmetric ones in Sec.~\ref{approx}.  Not
only is the spin expansion {\it consistent} with all existing
simulations, but it allows us to discover {\it qualitatively new}
non-linear effects by specifying the spin dependence these effects must
take.  These {\it non-linear} spin effects reveal the inadequacy of
existing ``Kidder'' fitting formulae for kick velocities
\cite{Campanelli:2007cg,Hinder:2007qu} modeled after the purely {\it
  linear} terms in post-Newtonian expressions for the instantaneous
loss of linear momentum.  Though existing simulations verify these
exciting predictions of the spin expansion, they fail to fully
constrain many of the terms at second order and beyond.  In
Sec.~\ref{S:prog}, we propose a {\it new} series of simulations that
will allow us to calibrate the coefficients of all terms to second
order.  We hope that the promise of our approach and its successes
described in this paper will motivate numerical relativists to
undertake these simulations in the near future. Finally, in
Sec.~\ref{S:disc}, we take stock of what has been accomplished in this
paper and what remains to be done.

Supplementary information has been organized into a series of
appendices.  We compile the relevant results of recently published
simulations for convenience in Appendix~\ref{datatables}.  Lengthy
equations relating different spin expansions in Sec.~\ref{approx} are
relegated to Appendix~\ref{grossness} for clarity of presentation.
Appendix~\ref{3rdOrderCal} provides third-order terms in the spin
expansion to extend the second-order calibration procedure described
in Sec.~\ref{S:prog}.  In Appendix~\ref{S:MVE} we show how systematic
uncertainties in simulations affect our estimates of the coefficients
in the spin expansion.  In particular, for a general set of
simulations of given covariance, we derive the minimum-variance
unbiased estimators for the spin expansion coefficients.  Finally, in
Appendix~\ref{S:ecc}, we briefly remark on the generalization of the
spin expansion to initially eccentric orbits.

\section{The spin-expansion formalism}
\label{S:formalism}

For a brisk introduction to the spin expansion, see the first two
pages of \cite{Boyle:2007sz}.  In this section, we introduce the same
formalism at a more leisurely pace, providing more detailed
explanations along the way.\footnote{If you find the presentation in
  this section suspiciously simplistic or Newtonian, we request your
  patience.  Although we regard this simplicity as one of the key
  virtues of the spin expansion, in a forthcoming paper we make
  contact with the full 3+1 formulation of general relativity, and
  also with the post-Newtonian expansion.  In the meantime, the proof
  is in the pudding: we hope that Section \ref{approx} will convince
  you that the spin expansion is powerfully explanatory and predictive
  when compared with existing simulations.}

\subsection{Preliminaries}
\label{S:prelim}

\begin{figure}
  \begin{center} \includegraphics[width=3.1in]{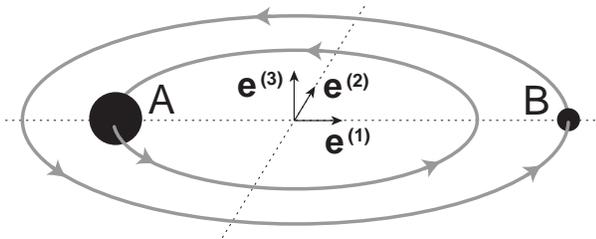} \end{center}
  \caption{A BBH system, including the orthonormal triad
  defined in the text.}  \label{triadfig}
\end{figure}

Imagine two black holes, $A$ and $B$, in a circular
orbit\footnote{Gravitational radiation carries away energy more
  efficiently than angular momentum, and hence circularizes BBH orbits
  \cite{Peters:1963ux}.  Thus, most astrophysically relevant systems
  are expected to circularize long before merger.  With the exception
  of a few papers ({\it e.g.}  \cite{Pretorius:2007jn, Grigsby:2007fu,
    Sperhake:2007gu, Hinder:2007qu, Washik:2008jr}), simulations of
  BBH mergers thus far have focused on circular orbits.  Nevertheless,
  in Appendix \ref{S:ecc}, we will explain the extension of our
  formalism to non-circular (eccentric) orbits.}, as shown in
Fig.~\ref{triadfig}.
Assume that $A$ and $B$ are initially far apart --- far enough that
they may be thought of as two Kerr black holes, characterized by
masses ($M_{a}$, $M_{b}$) and spins (${\bf S}_{a}$, ${\bf S}_{b}$).
Let us work in the center-of-momentum frame of the complete system
(including the gravitational radiation, and the momentum that it
carries).  The orbit gradually shrinks due to gravitational-wave
emission, until $A$ and $B$ eventually merge.  After the merger, the
spacetime quickly settles down to a final Kerr black hole with mass
$M_{f}$, spin ${\bf S}_{f}$, and recoil velocity (or ``kick
velocity'') ${\bf k}$ relative to the center-of-momentum frame.

Next recall that classical general relativity has a trivial
one-parameter rescaling symmetry.  Starting from a given solution, we
can obtain another solution by rescaling every physical quantity $X$
according to its mass dimension $d_{X}^{}$: $X\to\lambda^{d_{X}}X$,
where $\lambda$ is an arbitrary positive number.\footnote{Throughout
this paper, we work in ``geometrical units'' with $G_{N}^{}=c=1$, so
that each physical quantity has units of mass to some power, called
its ``mass dimension.''  For example, angular momentum has mass
dimension $=2$, while distance and time both have mass dimension
$=1$.}  Of course, rescaling the solution in this way is closely
related to rescaling the basic unit of mass. For our purposes, this
freedom is more distracting than interesting: from now on, we will
work exclusively with {\it dimensionless} quantities, which are
unaffected by such a rescaling.  In particular, it is useful to define
the dimensionless initial mass ratio
\begin{equation}
  q\equiv M_{b}/M_{a},
\end{equation}
the dimensionless initial spins
\begin{equation}
  {\bf a}\equiv{\bf S}_{a}/M_{a}^{2}\qquad\qquad
  {\bf b}\equiv{\bf S}_{b}/M_{b}^{2},
\end{equation}
the dimensionless final mass
\begin{equation}
  m\equiv M_{f}/(M_{a}+M_{b})
\end{equation}
and the dimensionless final spin
\begin{equation}
  {\bf s}\equiv{\bf S}_{f}/M_{f}^{2}.
\end{equation}

\subsection{Initial configuration of a black hole binary}
\label{S:initial_config}

To fully specify the initial configuration of this binary system, how
much information do we need to provide?

Since we are only interested in dimensionless quantities, we can fully
specify the initial state of the binary in terms of 8 numbers
$\{\psi,q,a_{i}^{},b_{i}^{}\}$ as follows.  First choose an inspiral
parameter $\psi$, by which we mean a dimensionless quantity that
varies monotonically along the orbit of the binary during the
adiabatic inspiral phase.  For example, $\psi$ could be the
(dimensionless) orbital separation $r/(M_{a}\!+\!M_{b})$, or the
(dimensionless) magnitude of the orbital angular momentum
$L/(M_{a}\!+\!M_{b})^{2}$.  At the ``initial instant'' ({\it i.e.}  at
some particular value of $\psi$), define an orthonormal triad $\{{\bf
e}_{}^{(1)}$, ${\bf e}_{}^{(2)}$, ${\bf e}_{}^{(3)}\}$ as shown in
Fig.~\ref{triadfig}: ${\bf e}_{}^{(3)}$ points along the orbital
angular momentum, ${\bf e}_{}^{(1)}$ points from $A$ to $B$, and ${\bf
e}_{}^{(2)}={\bf e}_{}^{(3)}\!\times{\bf e}_{}^{(1)}$.  This same
triad is conventionally introduced in post-Newtonian studies of
spinning compact binaries (see {\it e.g.} \cite{Kidder:1995zr,
Tagoshi:2000zg, Faye:2006gx, Blanchet:2006gy}).  Now we can specify
the initial state of the binary, at the initial instant $\psi$, by
giving 7 more numbers --- namely the dimensionless mass ratio $q$, and
the initial spin components
\begin{equation}
  \label{ab_components}
  a_{i}^{}\equiv{\bf a}\cdot{\bf e}_{}^{(i)}\qquad\qquad
  b_{i}^{}\equiv{\bf b}\cdot{\bf e}_{}^{(i)}
\end{equation}
relative to the orthonormal triad.  

We can think of the 7 numbers $\{q,a_{i}^{},b_{i}^{}\}$ as
parameterizing the 7-dimensional space of physically-distinct black
hole binaries (in circular orbit).  Each ``point'' in this
7-dimensional space corresponds to an inspiral trajectory.  On each
trajectory, there is an ``initial instant'' labeled by $\psi$, at
which the 7 numbers $\{q,a_{i}^{},b_{i}^{}\}$ are to be specified.

\subsection{Symmetry considerations for binary systems}
\label{S:sym}

Now let us consider how various final (post-merger) quantities can
depend on the initial quantities presented in the previous subsection.

Long after the merger, let $f$ denote a final quantity of interest.
For the purposes of illustration, we will focus in this section on a
particular set of final quantities $f \in \{m,k_{i}^{},s_{i}^{}\}$ ---
namely, the final Kerr black hole's dimensionless mass $m$, and the
components
\begin{equation}
  \label{ks_components}
  k_{i}^{}\equiv{\bf k}\cdot{\bf e}_{}^{(i)}\qquad\qquad
  s_{i}^{}\equiv{\bf s}\cdot{\bf e}_{}^{(i)}
\end{equation}
of its final kick and spin --- relative to the orthonormal triad
defined at the initial time $\psi$ as explained in the previous
subsection.  Our presentation will hopefully be general enough to make
it clear how to apply the same formalism to various other final
quantities of interest, such as the (dimensionless) total radiated
energy $\epsilon_{{\rm rad}} \equiv E_{{\rm rad}}/(M_{a}\!+\!M_{b})$, or
the (dimensionless) total radiated angular momentum ${\bf j}_{{\rm
rad}} \equiv {\bf J}_{{\rm rad}}^{}/(M_{a}\!+\!M_{b})^{2}$.

Any dimensionless final quantity $f$ is a function of the initial
quantities $\{q,a_{i}^{},b_{i}^{}\}$ and the initial instant $\psi$
at which these quantities were specified,
\begin{equation}
  f=f(\psi,q,a_{i}^{},b_{i}^{}).
\end{equation}
The goal of this section is to constrain the function
$f(\psi,q,a_{i}^{},b_{i}^{})$ as much as possible, using symmetry
arguments alone.  We will consider 3 simple transformations of the
binary system: rotation ``$R$,'' parity ``$P$,'' and exchange ``$X$.''
Once we know how the initial and final quantities transform under $R$,
$P$, and $X$, we can constrain the map $f(\psi,q,a_{i}^{},b_{i}^{})$
by requiring it to relate initial quantities to final ones in a way
consistent with their respective transformation laws.

First consider the transformations $R$ and $P$.  $R$ is a global
3-dimensional rotation of the entire binary system (as if it were a
single rigid body), and $P$ is a global parity transformation which
reflects every point in the binary through the origin (the center of
mass).  How do the initial quantities
$\{\psi, q, a_{i}^{}, b_{i}^{}\}$ transform under $R$ and $P$?  The
quantities $\{{\bf e}_{}^{(1)}, {\bf e}_{}^{(2)}\}$ are both {\it
vectors}, while the quantities $\{{\bf e}_{}^{(3)}, {\bf a}, {\bf
b}\}$ are all {\it pseudovectors}.  The dot product of two vectors or
two pseudovectors is a {\it scalar}, which is invariant under both $R$
and $P$.  The dot product of a vector and a pseudovector is a {\it
pseudoscalar}, which is invariant under $R$ and flips sign under $P$.
Thus, the quantities $\{a_{1}^{},a_{2}^{}, b_{1}^{},b_{2}^{}\}$ are
pseudoscalars, while the quantities $\{a_{3}^{},b_{3}^{}\}$ are
scalars.  In summary, the initial spin components transform under $P$
as
\begin{equation}
  \begin{array}{rcl}
    a_{i}^{}&\to&\tilde{a}_{i}^{} \\
    b_{\;\!i}^{}&\to&\tilde{b}_{\;\!i}^{}
  \end{array}
\end{equation}
where we have introduced the convenient notation
\begin{equation}
  \begin{array}{rcl}
    \tilde{a}_{i}&\equiv&\{-a_{1}^{},-a_{2}^{},+a_{3}^{}\} \\
    \tilde{b}_{\;\!i}&\equiv&\{-b_{\;\!1}^{},-b_{\;\!2}^{},+b_{\;\!3}^{}\}.
  \end{array}
\end{equation}
Note that the initial mass ratio $q$ is also a scalar, and the
parameter $\psi$ may be chosen to be a scalar: for example, the
magnitude $r/(M_{a}\!+\!M_{b})$ of the initial separation, or the
magnitude $L/(M_{a}+M_{b})^{2}$ of the initial orbital angular
momentum.

For the rest of this paper, we restrict our attention to final
quantities $f$ that are either scalars (like $\{m, k_{1}^{}, k_{2}^{},
s_{3}^{}\}$) or pseudoscalars (like $\{s_{1}^{}, s_{2}^{},
k_{3}^{}\}$).  In other words, we focus on final quantities $f$ that
are invariant under $R$, and transform under $P$ as:
\begin{equation}
  \label{def_pmP}
  f\to(\pm)_{P}^{}f,
\end{equation}
where $(\pm)_{P}^{}\!=\!+1$ when $f$ is a scalar, and
$(\pm)_{P}^{}\!=\!-1$ when $f$ is a pseudoscalar.  The
function $f(\psi, q, a_{i}^{}, b_{i}^{})$ automatically respects $R$
(since the initial and final quantities are both invariant under $R$).
In order to be consistent with $P$, it must satisfy the constraint
\begin{equation}
  \label{gen_P_constraint}
  f(\psi,q,a_{i}^{},b_{i}^{})=(\pm)_{P}^{}f(\psi,q,\tilde{a}_{i}^{},
  \tilde{b}_{i}^{}).
\end{equation}

Finally consider an ``exchange transformation'' $X$, which leaves the
physical system absolutely unchanged, but simply swaps the labels of
the two black holes, $A\leftrightarrow B$.  How do the initial
quantities $\{\psi,q,a_{i}^{},b_{i}^{}\}$ transform under $X$?  The
parameter $\psi$ is invariant, and the mass ratio transforms as
$q\to1/q$.  The initial spin vectors are swapped, ${\bf
a}\leftrightarrow{\bf b}$, while the triad elements transform as
\begin{equation}
  \{{\bf e}_{}^{(1)},{\bf e}_{}^{(2)},{\bf e}_{}^{(3)}\}\to
  \{-{\bf e}_{}^{(1)},-{\bf e}_{}^{(2)},+{\bf e}_{}^{(3)}\}.
\end{equation}
Therefore the initial spin components transform as
\begin{equation}
  \begin{array}{rcl}
    a_{i}^{}&\to&\tilde{b}_{\;\!i}^{} \\
    b_{\;\!i}^{}&\to&\tilde{a}_{i}^{}.
  \end{array}
\end{equation}
Now suppose that the final quantity $f$ transforms under $X$ as:
\begin{equation}
  \label{def_pmX}
  f\to(\pm)_{X}^{}f,
\end{equation}
where $(\pm)_{X}^{}\!=\!+1$ when $f$ is ``even'' under exchange (like
$\{k_{3}^{}, s_{3}^{}, m\}$), and $(\pm)_{X}^{}\!=\!-1$ when $f$ is
``odd'' under exchange (like $\{k_{1}^{}, k_{2}^{}, s_{1}^{},
s_{2}^{}\}$).  In order to be consistent with $X$, the function
$f(\psi, q, a_{i}^{}, b_{i}^{})$ must satisfy the constraint
\begin{equation}
  \label{gen_X_constraint}
  f(\psi,q,a_{i}^{},b_{i}^{})=(\pm)_{X}^{}
  f(\psi,1/q,\tilde{b}_{i}^{},\tilde{a}_{i}^{}).
\end{equation}
Equivalently, but more conveniently, if $f$ transforms under the
combined transformation $PX$ as:
\begin{equation}
  \label{def_pmPX}
  f\to(\pm)_{PX}^{}f,
\end{equation}
then the function $f(\psi,q,a_{i}^{},b_{i}^{})$ must satisfy the
constraint:
\begin{equation}
  \label{gen_PX_constraint}
  f(\psi,q,a_{i}^{},b_{i}^{})=(\pm)_{PX}^{}
  f(\psi,1/q,b_{i}^{},a_{i}^{}).
\end{equation}
Note that $(\pm)_{PX}^{}$ is just given by the product
\begin{equation}
  \label{pmPX_from_pmP_pmX}
  (\pm)_{PX}^{}=(\pm)_{P}^{}(\pm)_{X}^{}.
\end{equation}

We emphasize that Eqs.~(\ref{gen_P_constraint}) and
(\ref{gen_PX_constraint}) capture the key results of this subsection.
To illustrate these formulae, let us apply them to the final
quantities $f\in\{m,k_{i}^{},s_{i}^{}\}$.

First, in Table \ref{T:transformations}, we collect the corresponding
values of $(\pm)_{P}^{}$, $(\pm)_{X}^{}$, and $(\pm)_{PX}^{}$ for $f
\in \{m, k_{i}^{}, s_{i}^{}\}$.
\begin{table}
  \begin{center}
    \begin{tabular}{|l|c|c|c|c|}
      \hline
      $f$ & $m$, $s_{3}$ & $s_{1}^{}$, $s_{2}^{}$ &
      $k_{1}^{}$, $k_{2}^{}$ & $k_{3}^{}$ \\
      \hline \hline
      $(\pm)_{P}^{}$ & $+$ & $-$ & $+$ & $-$ \\
      \hline
      $(\pm)_{X}^{}$ & $+$ & $-$ & $-$ & $+$ \\
      \hline
      $(\pm)_{PX}^{}$ & $+$ & $+$ & $-$ & $-$ \\
      \hline
    \end{tabular}
  \end{center}
  \caption{Transformation under $P$, $X$, and $PX$,
    for various final quantities $f$.}
  \label{T:transformations}
\end{table}
These values are easy to check.  For example, consider the component
$s_{1}^{}={\bf s}\cdot{\bf e}_{}^{(1)}$.  Under a parity
transformation $P$, the pseudovector quantity ${\bf s}$ (an angular
momentum) is unchanged, while the vector quantity ${\bf e}_{}^{(1)}$
(the direction from $A$ to $B$) flips sign, so their dot product
$s_{1}^{}$ is a pseudoscalar: $(\pm)_{P}^{}=-1$.  Under an exchange
transformation $X$, which merely changes the labels $A\leftrightarrow
B$, the final angular momentum ${\bf s}$ is clearly unchanged, but the
triad element ${\bf e}_{}^{(1)}$ (the direction from $A$ to $B$) flips
sign, so their dot product $s_{1}^{}$ is odd under exchange:
$(\pm)_{X}^{}=-1$.

Now, using Eq.~(\ref{gen_P_constraint}), together with the
$(\pm)_{P}^{}$ row in Table \ref{T:transformations}, we find that
parity $P$ implies that the final quantities $f \in \{m, k_{i}^{},
s_{i}^{}\}$ must obey the following constraints:
\begin{subequations}
  \begin{eqnarray}
    \begin{array}{rcl}
      m(\psi,q,a_{i}^{},b_{i}^{})\!&\!=\!&\!
      +m(\psi,q,\tilde{a}_{i}^{},\tilde{b}_{i}^{}) \\
    \end{array} \\
    \begin{array}{rcl}
      k_{1}^{}(\psi,q,a_{i}^{},b_{i}^{})\!&\!=\!&\!
      +k_{1}^{}(\psi,q,\tilde{a}_{i}^{},\tilde{b}_{i}^{}) \\
      k_{2}^{}(\psi,q,a_{i}^{},b_{i}^{})\!&\!=\!&\!
      +k_{2}^{}(\psi,q,\tilde{a}_{i}^{},\tilde{b}_{i}^{}) \\
      k_{3}^{}(\psi,q,a_{i}^{},b_{i}^{})\!&\!=\!&\!
      -k_{3}^{}(\psi,q,\tilde{a}_{i}^{},\tilde{b}_{i}^{}) \\
    \end{array} \\
    \begin{array}{rcl}
      s_{1}^{}(\psi,q,a_{i}^{},b_{i}^{})\!&\!=\!&\!
      -s_{1}^{}(\psi,q,\tilde{a}_{i}^{},\tilde{b}_{i}^{}) \\
      s_{2}^{}(\psi,q,a_{i}^{},b_{i}^{})\!&\!=\!&\!
      -s_{2}^{}(\psi,q,\tilde{a}_{i}^{},\tilde{b}_{i}^{}) \\
      s_{3}^{}(\psi,q,a_{i}^{},b_{i}^{})\!&\!=\!&\!
      +s_{3}^{}(\psi,q,\tilde{a}_{i}^{},\tilde{b}_{i}^{}) \\
    \end{array}
  \end{eqnarray}
\end{subequations}
Using Eq.~(\ref{gen_PX_constraint}), together with the 
$(\pm)_{PX}^{}$ row in Table \ref{T:transformations}, we find 
that exchange symmetry (or, more correctly, $PX$) implies
that the final quantities $f \in \{m, k_{i}^{}, s_{i}^{}\}$ must obey
the following constraints:
\begin{subequations}
  \begin{eqnarray}
    \begin{array}{rcl}
      m(\psi,q,a_{i}^{},b_{i}^{})\!&\!=\!&\!
      +m(\psi,1/q,b_{i}^{},a_{i}^{}) \\
    \end{array} \\
    \begin{array}{rcl}
      k_{1}^{}(\psi,q,a_{i}^{},b_{i}^{})\!&\!=\!&\!
      -k_{1}^{}(\psi,1/q,b_{i}^{},a_{i}^{}) \\
      k_{2}^{}(\psi,q,a_{i}^{},b_{i}^{})\!&\!=\!&\!
      -k_{2}^{}(\psi,1/q,b_{i}^{},a_{i}^{}) \\
      k_{3}^{}(\psi,q,a_{i}^{},b_{i}^{})\!&\!=\!&\!
      -k_{3}^{}(\psi,1/q,b_{i}^{},a_{i}^{}) \\
    \end{array} \\
    \begin{array}{rcl}
      s_{1}^{}(\psi,q,a_{i}^{},b_{i}^{})\!&\!=\!&\!
      +s_{1}^{}(\psi,1/q,b_{i}^{},a_{i}^{}) \\
      s_{2}^{}(\psi,q,a_{i}^{},b_{i}^{})\!&\!=\!&\!
      +s_{2}^{}(\psi,1/q,b_{i}^{},a_{i}^{}) \\
      s_{3}^{}(\psi,q,a_{i}^{},b_{i}^{})\!&\!=\!&\!
      +s_{3}^{}(\psi,1/q,b_{i}^{},a_{i}^{}) \\
    \end{array}
  \end{eqnarray}
\end{subequations}

\subsection{Series expansions for the final observables}
\label{S:series}

Symmetry considerations impose important constraints on the maps
$f(\psi,q,a_{i}^{},b_{i}^{})$, but to make further progress it is
useful to Taylor expand these maps about ${\bf a}={\bf
b}=0$.\footnote{In performing this Taylor expansion we assume that the
map $f(\psi,q,a_{i}^{},b_{i}^{})$ is analytic in the neighborhood of
${\bf a}={\bf b}=0$.  While recent work by Pretorius and Khurana
\cite{Pretorius:2007jn} suggests that certain finely tuned eccentric
orbits can be exponentially sensitive to initial conditions, stable
circular orbits of non-spinning BBHs should remain circular until the
final plunge and merger.}  In terms of the spin components $\{
a_{i}^{}, b_{i}^{} \}$, this ``spin expansion'' can be written
in the form
\begin{equation}
  \label{f_expansion}
  f=f_{}^{m_{1}^{}m_{2}^{}m_{3}^{}|n_{1}^{}n_{2}^{}n_{3}^{}}(\psi,q)
  a_{1}^{m_{1}}a_{2}^{m_{2}}a_{3}^{m_{3}}
  b_{1}^{n_{1}}b_{2}^{n_{2}}b_{3}^{n_{3}},
\end{equation}
where separate summations over the 6 different indices
$\{m_{1}^{},m_{2}^{},m_{3}^{},n_{1}^{},n_{2}^{},n_{3}^{}\}$ from $0$
to $\infty$ are implied.  Note that the expansion coefficients
$f_{}^{m_{1}m_{2}m_{3}|n_{1}n_{2}n_{3}}$ are now independent of
$a_{i}^{}$ and $b_{i}^{}$, but still depend on $\psi$ and $q$.  Naive
counting suggests that the number of terms in these expansions should
grow rapidly with increasing order in the initial spins (1
zeroth-order term, 6 first-order terms, 21 second-order terms, and so
on).  However, the transformation requirements imposed by $P$ and $X$
significantly reduce the number of terms that actually appear.

Under a parity transformation $P$, the quantities $\{a_{1}^{}, a_{2}^{},
b_{1}^{}, b_{2}^{}\}$ change sign, implying that individual terms in
the expansion are multiplied by $(-1)^{\gamma}$, where we have defined
\begin{equation}
  \label{def_gamma}
  \gamma\equiv m_{1}^{}+m_{2}^{}+n_{1}^{}+n_{2}^{}.
\end{equation}
Eq.~(\ref{gen_P_constraint}) therefore implies the {\it Parity
Constraint}:
\begin{equation} 
  \label{gen_P_constraint_series}
  f_{}^{m_{1}^{}m_{2}^{}m_{3}^{}|n_{1}^{}n_{2}^{}n_{3}^{}}(q)
  \!=\!(\pm)_{P}^{}(-1)^{\gamma}
  f_{}^{m_{1}^{}m_{2}^{}m_{3}^{}|n_{1}^{}n_{2}^{}n_{3}^{}}(q).
\end{equation}
This may be restated as the {\it Parity Rule}:
\begin{itemize}
\item {\it Only terms with even (odd) $\gamma$ can appear in the
spin expansion of a (pseudo)scalar $f$.}
\end{itemize}
Additionally, if we expand both sides of Eq.~(\ref{gen_PX_constraint})
and equate terms with the same dependence on the initial spin
components, we obtain the {\it Exchange Constraint}:
\begin{equation}
  \label{gen_PX_constraint_series}
  f_{}^{m_{1}^{}m_{2}^{}m_{3}^{}|n_{1}^{}n_{2}^{}n_{3}^{}}(\psi,q)
  \!=\!(\pm)_{PX}^{}
  f_{}^{n_{1}^{}n_{2}^{}n_{3}^{}|m_{1}^{}m_{2}^{}m_{3}^{}}(\psi,1/q).\;
\end{equation}

Let us again illustrate these results by applying them to the final
quantities $f \in \{m,k_{i}^{},s_{i}^{}\}$.  We start by writing the 
expansion as
\begin{subequations} \label{expansions}
  \begin{eqnarray}
    \begin{array}{rcl}
      m\!&\!=\!&\!m_{}^{m_{1}^{}m_{2}^{}m_{3}^{}|
        n_{1}^{}n_{2}^{}n_{3}^{}}(q,\psi)
      a_{1}^{m_{1}}a_{2}^{m_{2}}a_{3}^{m_{3}}
      b_{1}^{n_{1}}b_{2}^{n_{2}}b_{3}^{n_{3}}\quad
    \end{array} \\
    \begin{array}{rcl}
      k_{1}^{}\!&\!=\!&\!k_{1}^{\;\;\!m_{1}^{}m_{2}^{}m_{3}^{}|
        n_{1}^{}n_{2}^{}n_{3}^{}}(q,\psi)
      a_{1}^{m_{1}}a_{2}^{m_{2}}a_{3}^{m_{3}}
      b_{1}^{n_{1}}b_{2}^{n_{2}}b_{3}^{n_{3}} \\
      k_{2}^{}\!&\!=\!&\!k_{2}^{\;\;\!m_{1}^{}m_{2}^{}m_{3}^{}|
        n_{1}^{}n_{2}^{}n_{3}^{}}(q,\psi)
      a_{1}^{m_{1}}a_{2}^{m_{2}}a_{3}^{m_{3}}
      b_{1}^{n_{1}}b_{2}^{n_{2}}b_{3}^{n_{3}} \\
      k_{3}^{}\!&\!=\!&\!k_{3}^{\;\;\!m_{1}^{}m_{2}^{}m_{3}^{}|
        n_{1}^{}n_{2}^{}n_{3}^{}}(q,\psi)
      a_{1}^{m_{1}}a_{2}^{m_{2}}a_{3}^{m_{3}}
      b_{1}^{n_{1}}b_{2}^{n_{2}}b_{3}^{n_{3}}\quad
    \end{array} \\
    \begin{array}{rcl}
      s_{1}^{}\!&\!=\!&\!s_{1}^{\;\,m_{1}^{}m_{2}^{}m_{3}^{}|
        n_{1}^{}n_{2}^{}n_{3}^{}}(q,\psi)
      a_{1}^{m_{1}}a_{2}^{m_{2}}a_{3}^{m_{3}}
      b_{1}^{n_{1}}b_{2}^{n_{2}}b_{3}^{n_{3}} \\
      s_{2}^{}\!&\!=\!&\!s_{2}^{\;\,m_{1}^{}m_{2}^{}m_{3}^{}|
        n_{1}^{}n_{2}^{}n_{3}^{}}(q,\psi)
      a_{1}^{m_{1}}a_{2}^{m_{2}}a_{3}^{m_{3}}
      b_{1}^{n_{1}}b_{2}^{n_{2}}b_{3}^{n_{3}} \\
      s_{3}^{}\!&\!=\!&\!s_{3}^{\;\,m_{1}^{}m_{2}^{}m_{3}^{}|
        n_{1}^{}n_{2}^{}n_{3}^{}}(q,\psi)
      a_{1}^{m_{1}}a_{2}^{m_{2}}a_{3}^{m_{3}}
      b_{1}^{n_{1}}b_{2}^{n_{2}}b_{3}^{n_{3}}\quad
    \end{array}
  \end{eqnarray}
\end{subequations} 

The parity constraint, (\ref{gen_P_constraint}) or
(\ref{gen_P_constraint_series}), implies that only terms
with the correct parity (even $\gamma$) can appear in the expansions of
the scalar quantities $\{m,k_{1}^{},k_{2}^{}, s_{3}^{}\}$.  Terms with the
wrong parity (odd $\gamma$) must vanish.  Conversely, only terms
with {\it odd} $\gamma$ have the correct parity to appear in the expansions
of the pseudoscalar quantities $\{s_{1}^{},s_{2}^{},k_{3}^{}\}$; terms with
wrong parity (even $\gamma$) must vanish.

The exchange constraint, (\ref{gen_PX_constraint}) or
(\ref{gen_PX_constraint_series}), implies that the remaining
coefficients must satisfy the following constraints:
\begin{subequations} \label{series_exchange_constraint}
  \begin{eqnarray}
    \begin{array}{rcl}
      m_{}^{m_{1}^{}m_{2}^{}m_{3}^{}|n_{1}^{}n_{2}^{}n_{3}^{}}(q)
      \!=\!+m_{}^{n_{1}^{}n_{2}^{}n_{3}^{}|m_{1}^{}m_{2}^{}m_{3}^{}}
      (1/q)
    \end{array} \\
    \begin{array}{rcl}
      k_{1}^{\;\;\!m_{1}^{}m_{2}^{}m_{3}^{}|n_{1}^{}n_{2}^{}n_{3}^{}}(q)
      \!=\!-k_{1}^{\;\;\!n_{1}^{}n_{2}^{}n_{3}^{}|m_{1}^{}m_{2}^{}m_{3}^{}}
      (1/q) \\
      k_{2}^{\;\;\!m_{1}^{}m_{2}^{}m_{3}^{}|n_{1}^{}n_{2}^{}n_{3}^{}}(q)
      \!=\!-k_{2}^{\;\;\!n_{1}^{}n_{2}^{}n_{3}^{}|m_{1}^{}m_{2}^{}m_{3}^{}}
      (1/q) \\
      k_{3}^{\;\;\!m_{1}^{}m_{2}^{}m_{3}^{}|n_{1}^{}n_{2}^{}n_{3}^{}}(q)
      \!=\!-k_{3}^{\;\;\!n_{1}^{}n_{2}^{}n_{3}^{}|m_{1}^{}m_{2}^{}m_{3}^{}}
      (1/q)
    \end{array} \\
    \begin{array}{rcl}
      s_{1}^{\;\,m_{1}^{}m_{2}^{}m_{3}^{}|n_{1}^{}n_{2}^{}n_{3}^{}}(q)
      \!=\!+s_{1}^{\;\,n_{1}^{}n_{2}^{}n_{3}^{}|m_{1}^{}m_{2}^{}m_{3}^{}}
      (1/q) \\
      s_{2}^{\;\,m_{1}^{}m_{2}^{}m_{3}^{}|n_{1}^{}n_{2}^{}n_{3}^{}}(q)
      \!=\!+s_{2}^{\;\,n_{1}^{}n_{2}^{}n_{3}^{}|m_{1}^{}m_{2}^{}m_{3}^{}}
      (1/q) \\
      s_{3}^{\;\,m_{1}^{}m_{2}^{}m_{3}^{}|n_{1}^{}n_{2}^{}n_{3}^{}}(q)
      \!=\!+s_{3}^{\;\,n_{1}^{}n_{2}^{}n_{3}^{}|m_{1}^{}m_{2}^{}m_{3}^{}}
      (1/q)
    \end{array}
  \end{eqnarray}
\end{subequations}
where, for brevity, we have not displayed the $\psi$ dependence on
each side.

Note that the exchange constraint (\ref{gen_PX_constraint_series})
imposes duality relations between coefficients at $q$ and $1/q$.
Without loss of generality, we can focus on the region of initial
parameter space with $0\leq q\leq 1$, since spin expansions in the
region with $1<q<\infty$ may simply be obtained via the duality
relations (\ref{gen_PX_constraint_series}) or
(\ref{series_exchange_constraint}).

In the equal-mass case, $q\!=\!1\!=\!1/q$, so the duality relations
directly relate previously independent coefficients.  In particular,
these relations require the kick-velocity coefficients
$k_{i}^{m_{1}^{}m_{2}^{}m_{3}^{}|n_{1}^{}n_{2}^{}n_{3}^{}}$ with
$\{m_{1},m_{2},m_{3}\}=\{n_{1},n_{2},n_{3}\}$ to vanish for $q=1$
since the components $k_{i}^{}$ have $(\pm)_{PX}^{}\!=\!-1$. This
result is consistent with the famous kick formula of Fitchett
\cite{Fitchett:1983}, who was one of the first to calculate the
gravitational-wave kick resulting from the merger of non-spinning BBHs
in the Newtonian approximation.  In his honor, we would like to
name the non-spinning kick coefficients $k_{i}^{000|000}$ ``the
coeFitchetts.''

\section{Exact Results and Special Configurations} 
\label{S:exact}

In this section, we will highlight several {\it exact} results which
follow from the symmetry considerations developed in the previous
section.  Since they are exact, these predictions may be useful for
testing numerical codes which compute BBH mergers numerically.

To derive and summarize the results, it is helpful to think about the
operations $P$ and $X$ as elements of a discrete group $G$ of
operators acting on binary systems ${\cal S}$.  This approach
conveniently generalizes to include other discrete symmetries like
charge conjugation $C$.\footnote{Charge conjugation $C$ is an exact
symmetry of classical general relativity and electromagnetism.  Under
$C$, the charge $Q$ of a black hole changes sign,
$Q\to-Q$.}${}^{,}$\footnote{Though black hole charge is expected to be
negligible in astrophysical contexts (and certainly much smaller than
the Kerr-Newman bound), it could be important in microphysical
contexts --- for example if TeV-scale quantum gravity leads to
black-hole production in high-energy accelerators, such as the Large
Hadron Collider (LHC), soon to begin operation at CERN
\cite{Dimopoulos:2001hw}.}  The three operators $P$, $X$, and $C$ all
square to unity and commute:
\begin{equation}
  \begin{array}{c}
    $P$^{2}=$X$^{2}=$C$^{2}=1, \\
    $[P,X]=[P,C]=[X,C]=0$,
  \end{array}
\end{equation}
so they generate the Abelian group
$G=\mathbb{Z}_{2}\times\mathbb{Z}_{2}\times\mathbb{Z}_{2}$, with eight
elements:\footnote{Time reversal $T$ is another exact symmetry of
classical general relativity and electromagnetism.  One might hope
that we could use $T$ to impose further constraints on binary black
hole merger, but unfortunately we cannot.  Black hole merger is
inherently a dissipative process; we are interested in black holes
that emit gravitational waves and inspiral --- not those which absorb
gravitational waves and outspiral!}
\begin{equation}
G \equiv \{1,P,X,PX,C,PC,XC,PXC\}.
\end{equation}
The non-trivial elements of this group transform the initial and final
quantities in a BBH merger as summarized in Table~\ref{T:ops}.
\begin{table}
\begin{center}
\begin{tabular}[t]{|c||c|c|c|c|c|c|c|}
\hline
& $P$ & $X$ & $PX$ & $C$ & $PC$ & $XC$ & $PXC$ \\ 
\hline \hline
$q$ & $q$ & $1/q$ & $1/q$  & $q$ & $q$ & $1/q$ & $1/q$ \\ 
\hline
$a_{1}^{}$ & $-a_{1}^{}$ & $-b_{1}^{}$ &
$+b_{1}^{}$ & $+a_{1}^{}$ & $-a_{1}^{}$ &
$-b_{1}^{}$ & $+b_{1}^{}$ \\
\hline
$a_{2}^{}$ & $-a_{2}^{}$ & $-b_{2}^{}$ &
$+b_{2}^{}$ & $+a_{2}^{}$ & $-a_{2}^{}$ &
$-b_{2}^{}$ & $+b_{2}^{}$ \\ 
\hline
$a_{3}^{}$ & $+a_{3}^{}$ & $+b_{3}^{}$ & $+b_{3}^{}$ & 
$+a_{3}^{}$ & $+a_{3}^{}$ & $+b_{3}^{}$ & $+b_{3}^{}$ \\ 
\hline
$b_{1}^{}$ & $-b_{1}^{}$ & $-a_{1}^{}$ & $+a_{1}^{}$ & 
$+b_{1}^{}$ & $-b_{1}^{}$ & $-a_{1}^{}$ & $+a_{1}^{}$ \\ 
\hline
$b_{2}^{}$ & $-b_{2}^{}$ & $-a_{2}^{}$ & $+a_{2}^{}$ & 
$+b_{2}^{}$ & $-b_{2}^{}$ & $-a_{2}^{}$ & $+a_{2}^{}$ \\ 
\hline
$b_{3}^{}$ & $+b_{3}^{}$ & $+a_{3}^{}$ & $+a_{3}^{}$ & 
$+b_{3}^{}$ & $+b_{3}^{}$ & $+a_{3}^{}$ & $+a_{3}^{}$ \\ 
\hline
$Q_{a}^{}$ & $+Q_{a}^{}$ & $+Q_{b}^{}$ & $+Q_{b}^{}$ & 
$-Q_{a}^{}$ & $-Q_{a}^{}$ & $-Q_{b}^{}$ & $-Q_{b}^{}$ \\ 
\hline
$Q_{b}^{}$ & $+Q_{b}^{}$ & $+Q_{a}^{}$ & $+Q_{a}^{}$ & 
$-Q_{b}^{}$ & $-Q_{b}^{}$ & $-Q_{a}^{}$ & $-Q_{a}^{}$ \\ 
\hline \hline
$m$ & $+m$ & $+m$ & $+m$ & $+m$ & $+m$ & $+m$ & $+m$ \\
\hline
$k_{1}^{}$ & $+k_{1}^{}$ & $-k_{1}^{}$ & $-k_{1}^{}$ & 
$+k_{1}^{}$ & $+k_{1}^{}$ & $-k_{1}^{}$ & $-k_{1}^{}$ \\ 
\hline
$k_{2}^{}$ & $+k_{2}^{}$ & $-k_{2}^{}$ & $-k_{2}^{}$ & 
$+k_{2}^{}$ & $+k_{2}^{}$ & $-k_{2}^{}$ & $-k_{2}^{}$ \\ 
\hline
$k_{3}^{}$ & $-k_{3}^{}$ & $+k_{3}^{}$ & $-k_{3}^{}$ & 
$+k_{3}^{}$ & $-k_{3}^{}$ & $+k_{3}^{}$ & $-k_{3}^{}$ \\ 
\hline
$s_{1}^{}$ & $-s_{1}^{}$ & $-s_{1}^{}$ & $+s_{1}^{}$ & 
$+s_{1}^{}$ & $-s_{1}^{}$ & $-s_{1}^{}$ & $+s_{1}^{}$ \\ 
\hline
$s_{2}^{}$ & $-s_{2}^{}$ & $-s_{2}^{}$ & $+s_{2}^{}$ & 
$+s_{2}^{}$ & $-s_{2}^{}$ & $-s_{2}^{}$ & $+s_{2}^{}$ \\ 
\hline
$s_{3}^{}$ & $+s_{3}^{}$ & $+s_{3}^{}$ & $+s_{3}^{}$ & 
$+s_{3}^{}$ & $+s_{3}^{}$ & $+s_{3}^{}$ & $+s_{3}^{}$ \\ 
\hline
$Q_{f}^{}$ & $+Q_{f}^{}$ & $+Q_{f}^{}$ & $+Q_{f}^{}$ & 
$-Q_{f}^{}$ & $-Q_{f}^{}$ & $-Q_{f}^{}$ & $-Q_{f}^{}$ \\ 
\hline
\end{tabular}
\caption{Transformations of the initial and final observables
  listed in the first column under the group of operations $G$ formed
  from parity $P$, exchange $X$, and charge conjugation $C$ and listed
  in the first row.}
\label{T:ops}
\end{center}
\end{table}

Each column in Table \ref{T:ops} (beyond the first) is identified with
an operator $g\in G$, and establishes a relationship between two {\it
physically distinct} BBH systems ${\cal S}$ and ${\cal S}'$ related by
${\cal S}' = g{\cal S}$, ${\cal S} = g{\cal S}'$.  For example, the
second column ($P$) relates {\it any} system ${\cal S}$ with initial
spin components $\{ a_{i}^{}, b_{i}^{} \}$ to a second system ${\cal
S}'$ with initial spins
\begin{equation}
  \begin{array}{rcl}
    \{a_{1}', a_{2}', a_{3}'\}\!&\!=\!&\!
    \{-a_{1}^{},-a_{2}^{},+a_{3}^{}\} \\
    \{ b_{\;\!1}', b_{\;\!2}', b_{\;\!3}'\}\!&\!=\!&\!
    \{-b_{\;\!1}^{},-b_{\;\!2}^{},+b_{\;\!3}^{}\} \, .
  \end{array}
\end{equation}
The final quantities for the system ${\cal S}'$ will be given in terms
of those for ${\cal S}$ by
\begin{equation} \label{SStF}
  \begin{array}{rcl}
    \{k_{1}',k_{2}',k_{3}'\}\!&\!=\!&\!
    \{+k_{1}^{},+k_{2}^{},-k_{3}^{}\} \\
    \{s_{1\;\!}',s_{2\;\!}',s_{3}'\}\!&\!=\!&\!
    \{-s_{1\;\!}^{},-s_{2\;\!}^{},+s_{3}^{}\} \\
    \{m',Q_{f}'\}\!&\!=\!&\!\{m,Q_{f}^{}\} \, .
  \end{array}
\end{equation}
Some of the 7 predicted relationships represented by the 7 columns of
Table~\ref{T:ops} may be useful to numerical relativists for
identifying errors in their numerical codes that fail to respect the
given symmetries.  For example, $P$ demands that simulations of the
systems ${\cal S}$ and ${\cal S}'$ must yield final quantities that
are related by Eq.~(\ref{SStF}).  Any deviations from this
relationship would reveal systematic errors in the simulations that
need to be corrected.

\begin{table}
\begin{center}
\begin{tabular}[t]{|c|p{3.5cm}|p{3.5cm}|}
\hline
& input & output \\
\hline\hline
$P$ & $a_{1}^{}\!=\!a_{2}^{}\!=\!b_{1}^{}\!=\!b_{2}^{}\!=\!0$ & 
$s_{1}^{}\!=\!s_{2}^{}\!=\!k_{3}^{}\!=\!0$ \\
\hline
$X$ & $(a_{1}^{},a_{2}^{})\!=\!-(b_{1}^{},b_{2}^{})$,$\qquad\;$
$a_{3}^{}\!=\!b_{3}^{}$, $q\!=\!1$, $Q_{a}\!=\!Q_{b}$ &
$k_{1}^{}\!=\!k_{2}^{}\!=\!s_{1}^{}\!=\!s_{2}^{}\!=\!0$ \\
\hline
$PX$ & $(a_{1}^{},a_{2}^{},a_{3}^{})\!=\!(b_{1}^{},b_{2}^{},b_{3}^{})$,
$q=1$, $Q_{a}^{}\!=\!Q_{b}^{}$ & 
$k_{1}^{}\!=\!k_{2}^{}\!=\!k_{3}^{}\!=\!0$ \\
\hline
$C$ & $Q_{a}^{}\!=\!Q_{b}^{}\!=\!0$ & $Q_{f}^{}\!=\!0$ \\
\hline
$PC$ & $a_{1}^{}\!\!=\!\!a_{2}^{}\!\!=\!\!b_{1}^{}\!\!=\!\!b_{2}^{}
\!\!=\!\!0$,$\qquad$ $Q_{a}^{}\!=\!Q_{b}^{}\!=\!0$ 
& $s_{1}^{}\!=\!s_{2}^{}\!=\!k_{3}^{}\!=\!Q_{f}^{}\!=\!0$ \\
\hline
$XC$ & $(a_{1}^{},a_{2}^{})\!=\!-(b_{1}^{},b_{2}^{})$,$\qquad\;$
$a_{3}^{}\!=\!b_{3}^{}$, $q\!=\!1$, $Q_{a}^{}\!=\!-Q_{b}^{}$ & 
$k_{1}^{}\!=\!k_{2}^{}\!=\!s_{1}^{}\!=\!s_{2}^{}\!=\!Q_{f}^{}
\!=\!0$ \\
\hline
$PXC$ &
$(a_{1}^{},a_{2}^{},a_{3}^{})\!=\!(b_{1}^{},b_{2}^{},b_{3}^{})$,
$q\!=\!1$, $Q_{a}^{}=-Q_{b}^{}$ & $k_{1}^{}\!=\!k_{2}^{}\!=\!
k_{3}^{}\!=\!Q_{f}^{}\!=\!0$ \\
\hline
\end{tabular}
\caption{Special initial configurations (``input''), and the 
corresponding predictions for the final state (``output'').}
\label{T:special_configs}
\end{center}
\end{table}
Each non-trivial element $g\in G$ not only establishes a dual system
${\cal S}' = g{\cal S}$ for {\it all} BBH systems ${\cal S}$, but also
defines a ``special'' configuration ${\cal S}_{g}$ that is its {\it
own} dual, {\it i.e.}  $g {\cal S}_{g} = {\cal S}_{g}$.  Since the
initial state is invariant under $g$, the final state must also be
invariant, so we can conclude that any final quantity $f$ that is {\it
not} invariant under $g$ must vanish.  For example, consider the
second column ($P$) in Table \ref{T:ops}.  We see that, if the initial
configuration satisfies $a_{1}^{} =a_{2}^{} =b_{1}^{} =b_{2}^{} =0$,
then a parity transformation $P$ leaves this initial configuration
invariant.  Therefore, we can conclude that the corresponding final
state must satisfy $s_{1}^{}=s_{2}^{}=k_{3}^{}=0$, since these 3
quantities are {\it not} invariant under $P$.  Similarly, from each of
the 7 columns in Table \ref{T:ops}, we can read off a ``special''
initial configuration, and the corresponding predictions for the final
state of the system.  These 7 special configurations, and their
consequences, are summarized in Table
\ref{T:special_configs}.

\section{Testing our Expansions with Existing Simulations}
\label{approx}

Although the previous section's exact results are interesting, the
real power of the spin expansion is in the much larger set of {\it
approximate} predictions it makes for generic spin configurations.

Currently, published simulations of BBH mergers can be subdivided into
5 different classes of initial spin configurations.  For each of these
classes, we compare in detail the predictions of the spin expansion
with existing numerical results.  As we shall see, the spin expansion
makes new and successful predictions in each case, and provides a
simple and systematic way of {\it deriving} features of existing
simulations that were previously opaque.

Since rigorous systematic errors are not yet available for many of the
simulated data sets analyzed in this section, our analysis of these
simulations will by necessity be correspondingly non-rigorous.  In
particular, we will use the following rather heuristic procedure.  For
each data set, we compute the $\chi^{2}$ per degree of freedom
($\chi^{2}$/d.o.f.).  We do {\it not} attribute meaning to the overall
value of the $\chi^{2}$/d.o.f.  (since, in many cases, we have had to
guess error bars for the simulated data).  However, if including a new
term predicted by the spin expansion leads to a large {\it fractional}
reduction in the $\chi^{2}$/d.o.f., we interpret this reduction as
admittedly non-rigorous evidence for this new term.  Note that an
overall rescaling of the error bars will {\it not} lead to a
fractional reduction in the $\chi^{2}$/d.o.f., so the conclusions in
this section should be largely insensitive to our estimates for the
error bars.

In any case, statistical rigor is {\it not} the point of this section.
Our goal is to illustrate our formalism through a few worked examples,
to demonstrate its power to explain currently available simulations,
and to make several predictions for future simulations.  It will
usually be clear (by eye) that our leading-order and
next-to-leading-order formulae provide a good explanation of the basic
qualitative features seen in the simulated data thus far.

We have summarized our use of simulations in
Appendix~\ref{datatables}, where we also explain our estimates for the
corresponding error bars -- which are often just crude guesses.  The
crudeness of these guesses sometimes leads to impossibly small
$\chi^2$/d.o.f.\ for our fits.  This is primarily because genuine
errors in the simulations are {\it systematic}, whereas published
works (and our own analysis) treat these errors as {\it statistical}.
Systematic errors that preserve the symmetries of the configuration
will be well fit by our expansions, regardless of their size, albeit
with erroneous numerical values for the best-fit coefficients.

In this section, we will often encounter 3-component quantities
$(x_{1}^{},x_{2}^{},x_{3}^{})$ where we wish to treat the ``1'' and
``2'' components together, without the ``3'' component.  Thus, it is
convenient to introduce the notation
\begin{equation}
  {\bf x}_{\perp}^{}=(x_{1}^{},x_{2}^{})
\end{equation}
as these components are {\it perpendicular} to the orbital angular
momentum vector.  So, for example, ${\bf
a}_{\perp}^{}=(a_{1}^{},a_{2}^{})$, ${\bf k}_{\perp}^{}=(k_{1}^{},
k_{2}^{})$, and ${\bf k}_{\perp}^{001|000}=(k_{1}^{001|000},
k_{2}^{001|000})$.  From a notational standpoint, ${\bf x}_{\perp}^{}$
acts like a ``2-vector'' in the sense that
\begin{equation}
  \big|{\bf x}_{\perp}^{}\big|\equiv\!
  \sqrt{x_{1}^{2}\!+\!x_{2}^{2}}\, ,\qquad
  {\bf x}_{\perp}^{}\!\!\cdot{\bf y}_{\perp}^{}
  \!=\!x_{1}^{}y_{1}^{}\!\!+\!x_{2}^{}y_{2}^{}.
\end{equation}

\subsection{Case \#1: Non-precessing spins, \newline
($q=1$ and ${\bf a}\propto{\bf b}\propto{\bf e}_{}^{(3)}$)}
\label{Case_1}

First consider the case in which the black holes have equal mass
($q=1$), and both spins are aligned (or anti-aligned) with the orbital
angular momentum:
\begin{equation} \label{Case1Spins}
  \left(\begin{array}{c} a_{1}^{} \\ a_{2}^{} \\ a_{3}^{}
    \end{array}\right)\!=\!
  \left(\begin{array}{c} 0 \\ 0 \\ a_{3}^{} 
    \end{array}\right),\qquad\quad
  \left(\begin{array}{c} b_{1}^{} \\ b_{2}^{} \\ b_{3}^{}
    \end{array}\right)\!=\!
  \left(\begin{array}{c} 0 \\ 0 \\ b_{3}^{}
    \end{array}\right).
\end{equation}
This corresponds to the special configuration identified with the
operator $P$ in the previous section, for which $k_{3}^{}$ and ${\bf
s}_{\perp}^{}$ exactly vanish.  What about {\it non}-vanishing
observables like ${\bf k}_{\perp}^{}$, $s_{3}^{}$, and $m$?  

The zeroth-order terms in the $s_{3}^{}$ and $m$ expansions are
non-vanishing, and physically correspond to the spin and mass of the
final black hole produced in the merger of two non-spinning holes
$A$ and $B$.  However, exchange $X$ requires the zeroth order term
in the ${\bf k}_{\perp}^{}$ expansion (the ``coeFitchett''
${\bf k}_{\!\perp}^{000|000}$) to vanish, as discussed in
section~\ref{S:series}.  To capture the leading-order (LO),
next-to-leading-order (NLO), and next-to-next-to-leading-order
(NNLO) behavior, we therefore expand $s_{3}^{}$ and $m$ to second
order in the initial spins and ${\bf k}_{\perp}^{}$ to third order.

Using the parity and exchange constraints,
(\ref{gen_P_constraint_series}) and
(\ref{gen_PX_constraint_series}), to equate or eliminate
coefficients, our expansions (\ref{expansions}) for this configuration
become:
\begin{subequations} \label{case1_exp}
  \begin{eqnarray}
    \label{k_perp_case1}
    {\bf k}_{\!\perp}^{}\!\!&\!=\!&\!
    {\bf k}_{\!\perp}^{001|000}(a_{3}^{}\!-\!b_{3}^{})\!+\!
    {\bf k}_{\!\perp}^{002|000}(a_{3}^{2}\!-\!b_{3}^{2}) \nonumber\\
    \!&\!+\!&\!{\bf k}_{\!\perp}^{003|000}(a_{3}^{3}\!-\!b_{3}^{3})
    \!+{\bf k}_{\!\perp}^{002|001}(a_{3}^{2}b_{3}^{}\!-\!b_{3}^{2}
    a_{3}^{}),\qquad \\
    \label{s_3_case1}
    s_{3}^{}\!&\!=\!&\!s_{3}^{\;\,000|000}
    \!+\!s_{3}^{\;\,001|000}(a_{3}^{}\!+\!b_{3}^{}) \nonumber \\
    \!&\!+\!&\!s_{3}^{\;\,002|000}(a_{3}^{2}\!+\!b_{3}^{2})
    \!+\!s_{3}^{\;\,001|001}a_{3}b_{3}, \\
    \label{m_case1}
    m\!&\!=\!&\!m_{}^{000|000}\!+\!
    m_{}^{001|000}(a_{3}\!+\!b_{3}) \nonumber \\
    \!&\!+\!&\!m_{}^{002|000}(a_{3}^{2}\!+\!b_{3}^{2})
    \!+\!m_{}^{001|001}a_{3}^{}b_{3}^{}.
  \end{eqnarray}
\end{subequations}
Unfortunately for us, currently published simulations only report
the magnitude $|{\bf k}_{\perp}^{}|=\sqrt{k_{1}^{2}+k_{2}^{2}}$, 
not the individual components $k_{1}^{}$ and $k_{2}^{}$.  Taking the
magnitude of the expansion (\ref{k_perp_case1}) for ${\bf
k}_{\perp}^{}$ and Taylor expanding to third order in the initial
spins yields
\begin{eqnarray}
  \label{Case1kickmag}
  \big|{\bf k}_{\!\perp}^{}\big|\!&\!=\!&\!
  \big|{\bf k}_{\!\perp}^{001|000}\big|\;
  \big|a_{3}^{}\!-\!b_{3}^{}\big| 
  \nonumber\\
  \!&\!\times\!&\!\Big[1\!+\!A(a_{3}^{}\!+\!b_{3}^{})
  \!+\!B(a_{3}^{2}\!+\!b_{3}^{2})\!+\!Ca_{3}^{}b_{3}^{}\Big].
\end{eqnarray}
For convenience, we have introduced new coefficients $A$, $B$,
and $C$, which may be expressed in terms of the original expansion
coefficients ${\bf k}_{\!\perp}^{m_{1}m_{2}m_{3}|n_{1}n_{2}n_{3}}$.
These expressions are given in Appendix~\ref{Case1App}.
  
It is interesting to note that, although $|{\bf k}_{\!\perp}^{}|$ is
{\it even} under both $P$ and $X$ (just like
$s_{3}^{}$ and $m$), its expansion (\ref{Case1kickmag}) is different
from the $s_{3}$ and $m$ expansions (\ref{s_3_case1},
\ref{m_case1}).  This is a reflection of the fact that, although
$|{\bf k}_{\!\perp}^{}|$ has the same transformation properties as
  $s_{3}^{}$ and $m$, it is a composite quantity constructed from more
``fundamental'' quantities ($k_{1}^{}$ and $k_{2}^{}$) with
different transformation properties.  Looking more closely, we
notice that the expression inside square brackets in
Eq.~(\ref{Case1kickmag}) bears a close formal resemblance to the
expansions (\ref{s_3_case1}, \ref{m_case1}): there is a constant
term at LO, a term proportional to $(a_{3}^{}+b_{3}^{})$ at NLO, and
two terms proportional to $(a_{3}^{2}+b_{3}^{2})$ and
$a_{3}^{}b_{3}^{}$ at NNLO.  The compositeness of $|{\bf k}_{\!\perp}^{}|$
manifests itself through the overall factor of
$|a_{3}^{}-b_{3}^{}|$ in front of the square brackets in
(\ref{Case1kickmag}).

Eqs.~(\ref{Case1kickmag}), (\ref{s_3_case1}), and (\ref{m_case1}) make
detailed quantititive predictions for the final kicks, spins, and
masses --- let's see how they stack up against actual simulations!

Many groups have published results from the simulations of binary
mergers with initial spins aligned or anti-aligned with the angular
momentum direction ${\bf e}_{}^{(3)}$ \cite{Baker:2005vv,
  Herrmann:2007ac, Pollney:2007ss, Rezzolla:2007xa, Marronetti:2007wz,
  Berti:2007nw}.  While all groups begin their simulations with the
binary on a quasi-circular orbit, they make different choices for the
initial dimensionless orbital separation $r/(M_{a}^{}+M_{b}^{})$.  As
explained in Section~\ref{S:formalism}, our expansion coefficients
(\ref{expansions}) are defined at a fixed value of the inspiral
parameter $\psi$, which in this case is the dimensionless orbital
separation $r/(M_{a}^{}+M_{b}^{})$.  Connecting simulations performed
at different values of $\psi$ will in general require careful use of
post-Newtonian approximations as discussed in a forthcoming paper.
However, for the special case considered here (where ${\bf
  a}\propto{\bf b}\propto{\bf e}_{}^{(3)}$), the initial spins will
not precess and their projection onto the orthonormal triad $\{{\bf
  e}_{}^{(1)}$, ${\bf e}_{}^{(2)}$, ${\bf e}_{}^{(3)}\}$ will not vary
with orbital phase.  As such, it is possible in principle to jointly
fit all the simulations even though they do not all correspond to the
same initial separation.  In practice, there are systematic
differences between numerical codes \footnote{For example, Pollney
  {\it et al.}  \cite{Pollney:2007ss} discusses the necessity of
  properly choosing an integration constant corresponding to the
  linear momentum acquired before the simulations begins.  Rezzolla
  {\it et al.}  \cite{Rezzolla:2007xa} suggests that this integration
  constant may be responsible for the discrepancy between their
  results and those of \cite{Herrmann:2007ac}.} which make a combined
fit to all existing simulations unreliable.  In
Appendix~\ref{Case1TabApp}, we describe our choice of simulations to
test Eqs.~(\ref{Case1kickmag}, \ref{s_3_case1}, \ref{m_case1}).

\begin{figure}
  \begin{center}
    \includegraphics[width=3.5in]{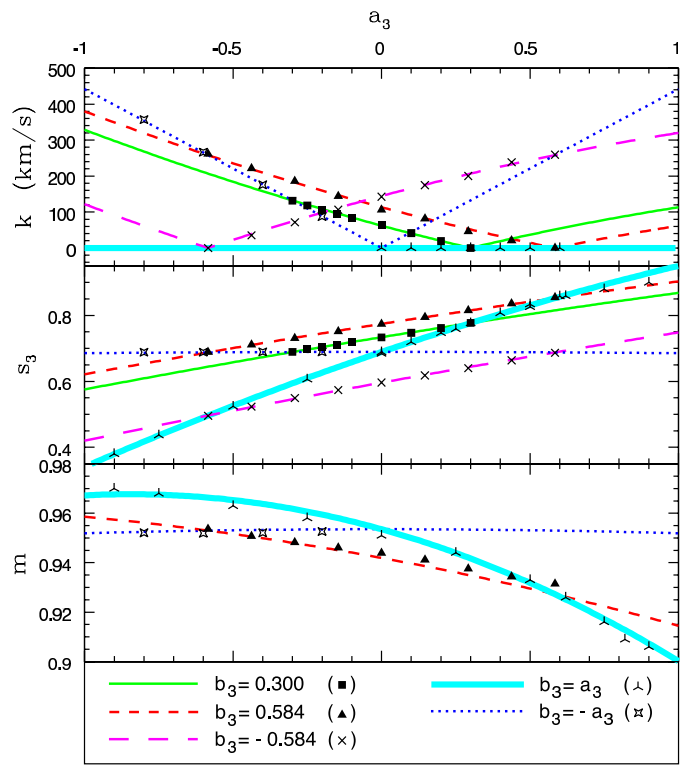}
  \end{center}
  \caption{Case \#1 best-fit curves.  The top, center, and bottom
    panels show, respectively, the kick velocity $|{\bf
      k}_{\perp}^{}|$, dimensionless spin $s_{3}^{}$, and
    dimensionless mass $m$ of the final black hole.  The data points
    appearing in each panel are described in
    Appendix~\ref{Case1TabApp}.  The curves in the top, center, and
    bottom panels, correspond, respectively, to
    Eqs.~(\ref{Case1kickmag}), (\ref{s_3_case1}), and (\ref{m_case1}),
    with the coefficients given in Table~\ref{zspin_results}.  Each
    curve has a fixed value of $b_{3}$ given by the legend at the
    bottom of the figure, with $a_{3}$ varying along the abscissa.
    For presentation purposes, we have switched $a_3$ and $b_3$ for
    points on the magenta (long-dashed) curve; this exchange does not
    affect the values of $|{\bf k}_{\perp}|$ and $s_3$.}
  \label{Case1fig}
\end{figure}

First consider the final kick $|{\bf k}_{\perp}^{}|$.  At leading
order, Eq.~(\ref{Case1kickmag}) predicts that $|{\bf k}_{\perp}^{}|$
should be proportional to $|a_{3}\!-\!b_{3}|$; this approximately
linear behavior has been noticed in simulations in several previous
papers \cite{Herrmann:2007ac, Pollney:2007ss, Baker:2007gi}.  At
next-to-leading order, Eq.~(\ref{Case1kickmag}) predicts that $|{\bf
  k}_{\perp}^{}|$ should receive a small additive correction
proportional to $|a_{3}-b_{3}| (a_{3}+b_{3})$.  This prediction is
well supported by the simulations in the following sense.  When we fit
the 28 simulations in \cite{Rezzolla:2007xa} with non-zero values for
$|{\bf k}_{\perp}^{}|$ to the leading-order term in
Eq.~(\ref{Case1kickmag}), there is only one fitting parameter (namely
the magnitude $|{\bf k}_{\perp}^{001|000}|$), and the fit is rather
poor $\chi^{2}/{\rm d.o.f.}=38.2/(28-1)\approx 1.4$.  Then, when we
include the next-to-leading-order term, there is one more fitting
parameter (namely $A$), and the fit improves dramatically:
$\chi^{2}/{\rm d.o.f.}=5.14/(28-2)\approx0.2$.  We interpret this
significant drop in the $\chi^{2}/{\rm d.o.f.}$ as strong evidence for
the second-order term in Eq.~(\ref{Case1kickmag}).\footnote{Pollney
  {\it et al.}  \cite{Pollney:2007ss} also noted a significant
  deviation from linearity in the kick magnitudes for initially
  (anti-)aligned spins, although their guess for the correction term
  differs from that derived via our formalism.  Rezzolla {\it et al.}
  \cite{Rezzolla:2007xa} arrived at the {\it same} second-order
  fitting formula as we did using similar considerations; our
  parameters $|{\bf k}_{\perp}^{001|000}|$ and $A$ correspond to
  $|c_1|$ and $c_2/c_1$ in their notation.}  This NLO fit, with only
two fitting parameters ($|{\bf k}_{\!\perp}^{}|$ and $A$), is
displayed in the top panel of Fig.~\ref{Case1fig}.

Is there also evidence for the next-to-next-to-leading-order
(NNLO) terms in Eq.~(\ref{Case1kickmag})?  When we include these
final two terms, there are two more fitting parameters ($B$ and
$C$), and the fit again improves significantly: $\chi^{2}/{\rm
  d.o.f.}=2.2/(28-4)\approx0.09$.  Our best-fit values for the 4
fitting parameters in Eq.~(\ref{Case1kickmag}) are shown in
Table~\ref{zspin_results}.
\begin{table}
  \begin{center}
    \begin{tabular}{|c|l|}
      \hline
      Data set & Best-fit coefficients \\
      \hline
      $|{\bf k}_{\perp}^{}|$ & $|{\bf k}_{\perp}^{001|000}|
      \approx 221~{\rm km/s}$ \\
      (28 simulations) & $A\approx-0.205$ \\
      & $B\approx-0.091$ \\
      & $C\approx-0.201$ \\
      \hline
      $s_{3}^{}$ & $s_{3}^{000|000}\approx0.6893$ \\
      (48 simulations) & $s_{3}^{001|000}\approx0.1524$ \\
      & $s_{3}^{001|001}\approx-0.0195$ \\
      & $s_{3}^{002|000}\approx-0.0121$ \\
      \hline
      $m$ & $m_{}^{000|000}\approx0.9530$ \\
      (24 simulations) & $m_{}^{001|000}\approx-0.0167$ \\
      & $m_{}^{001|001}\approx-0.0083$ \\
      & $m_{}^{002|000}\approx-0.0052$ \\
      \hline
    \end{tabular}
  \end{center}
  \caption{Case \#1 best-fit parameters, from fitting
    Eqs.~(\ref{Case1kickmag}, \ref{s_3_case1}, \ref{m_case1}) to the
    final kicks, spins, and masses of the simulations described in
    Appendix~\ref{Case1TabApp}.  These fits are displayed in
    Fig.~\ref{Case1fig}.}
  \label{zspin_results}
\end{table}

Next consider the final spin $s_{3}^{}$ and the final mass $m$.  It is
convenient to treat these two quantities in parallel, since their
expansion (\ref{s_3_case1}) and (\ref{m_case1}) are formally identical
to each other.  Eqs.~(\ref{s_3_case1}) and (\ref{m_case1}) predict
that at zeroth order the final spin $s_{3}^{}$ and final mass $m$
should be equal, respectively, to the final spin $s_{3}^{000|000}$ and
final mass $m_{}^{000|000}$ from the merger of two {\it non-spinning}
black holes.  At first order, there should be a linear correction
proportional to ($a_{3}^{}\!+\!b_{3}^{}$), and at second order there
should be two corrections: one proportional to $a_{3}^{}b_{3}^{}$, and
the other proportional to
($a_{3}^{2}\!+\!b_{3}^{2}$).\footnote{Rezzolla {\it et al.}
  \cite{Rezzolla:2007xa} argued that BBHs with equal and opposite
  spins $(a_3 = -b_3)$ should behave as if they were non-spinning, so
  that only a single term $p_2 (a_{3}^{}\!+\!b_{3}^{})^2$ should
  appear at second order.  According to this conjecture our
  coefficients should be related as $s_{3}^{002|000} = 1/2 \,
  s_{3}^{001|001} = p_2$.}

These predictions are again supported by the simulations.
First consider $s_{3}^{}$.  By itself, the leading-order term
$s_{3}^{000|000}$ gives a poor fit [$\chi^{2}/{\rm
  d.o.f.}=3,661/(48-1)=77.9$] to the 48 simulations of
\cite{Rezzolla:2007xa} and \cite{Marronetti:2007wz}.  When we include
the NLO term, $s_{3}^{001|000}(a_{3}^{}+b_{3}^{})$, the fit improves
dramatically [$\chi^{2}/{\rm d.o.f.}=16.8/(48-2)=0.366$].  Finally,
when we add the NNLO terms, $s_{3}^{002|000}(a_{3}^{2}+b_{3}^{2})$ and
$s_{3}^{001|001}a_{3}^{}b_{3}^{}$, the fit is even better
[$\chi^{2}/{\rm d.o.f.}=0.0386$].  The fit to $m$ using
Eq.~(\ref{m_case1}) is closely analogous: the zeroth-order fit is
again lousy ($\chi^{2}/{\rm d.o.f.}=436/(24-1)=19.0$); the first-order
fit is much improved ($\chi^{2}/{\rm d.o.f.}=46.8/(24-2)=2.13$); and
the second-order fit is better still ($\chi^{2}/{\rm
  d.o.f.}=5.30/(24-4)=0.265$).  Thus, in both the $s_{3}$ and $m$ data
sets, we have clear evidence for both next-to-leading-order (NLO) and
next-to-next-to-leading-order corrections.\footnote{The final spins
  and masses are equally well fit by a single second-order term
  proportional to $(a_{3}^{}\!+\!b_{3}^{})^2$ as suggested by Rezzolla
  {\it et al.}  \cite{Rezzolla:2007xa}.  Though post-Newtonian
  approximations may suggest that the spin-spin term proportional to
  $a_{3}^{} b_{3}^{}$ comes in at higher order, we restrict ourselves
  to what can be asserted purely on the basis of our formalism.} Our
best-fit parameters are shown in Table~\ref{zspin_results}, and the
corresponding fits to $s_{3}^{}$ and $m$ are plotted in
Fig.~\ref{Case1fig}, in the middle and bottom panels, respectively.

Though we have used numerical simulations to calibrate the values of
the coefficients in our spin expansions, physical intuition provides
some insight into these values.  Test particles with orbital angular
mometum aligned with the spin of a Kerr black hole have innermost
stable circular orbits (ISCOs) with smaller radii, and therefore emit
more gravitational radiation before merger.  Similar behavior has been
observed for comparable-mass black holes in numerical simulations
\cite{Campanelli:2006uy}.  This suggests, as we have indeed found by
comparing to simulations, that the coefficient $m_{}^{\;\!001|000}$
should be negative: more orbits implies more energy carried away in
gravitational waves and a smaller final mass.  The impact on the final
spin is more ambiguous; aligned BBHs should convey more spin angular
momentum to the final black hole, but less orbital angular mometum
because of their smaller ISCOs.  Since $s_{3}^{\;\!001|000}$ is
positive we conclude that the former effect dominates over the latter
though its comparatively small value evidences significant
cancellation.

\subsection{Case \#2: ``Superkick'' configuration \newline
  ($q=1$, ${\bf a}=-{\bf b}$, $a_{3}=b_{3}=0$)} \label{Case_2}

We next consider the case in which the black holes have equal mass
($q=1$) and equal and opposite spins lying in the orbital plane:
\begin{equation} \label{Case2Spins}
  \left(\!\begin{array}{c} a_{1}^{} \\ a_{2}^{} \\ a_{3}^{} 
    \end{array}\!\right)\!=\!
  \left(\!\!\begin{array}{c} +a\cos\phi \\ 
      +a\;\!\sin \phi \\ 0 \end{array}\!\!\right),\qquad
  \left(\!\begin{array}{c} b_{1}^{} \\ b_{2}^{} \\ b_{3}^{}
    \end{array}\!\right)\!=\!
  \left(\!\!\begin{array}{c} -a\cos\phi \\ 
      -a\;\!\sin \phi \\ 0 \end{array}\!\!\right).
\end{equation}
As shown in Section~\ref{S:exact} (Table \ref{T:special_configs}),
this is an example of the special configuration associated with the
exchange operator $X$, for which ${\bf k}_{\perp}^{}$ and ${\bf
  s}_{\perp}^{}$ exactly vanish.  What about the non-vanishing
quantities $k_{3}^{}$, $s_{3}^{}$, and $m$?

The ``superkick'' initial spin configuration (\ref{Case2Spins}) is
parameterized by only two numbers ($a$ and $\phi$), instead of six
($a_{i}^{}$ and $b_{i}^{}$).  As a result, when we substitute
Eq.~(\ref{Case2Spins}) into the original spin expansions of
Eq.~(\ref{expansions}), many of the terms become degenerate with one
another.  It is convenient to eliminate this degeneracy by collecting
all terms with the same dependence on $a$ and $\phi$.  Then the spin
expansion takes the form
\begin{subequations}
  \label{both_forms}  
  \begin{equation}
    \label{1stform}
    f\!=\!\sum_{i=0}^{\infty}\sum_{j=0}^{i}
    \big[{}^{s\!}f_{}^{(i,j)}a_{}^{i}\sin(j\phi)\!+\!
    {}^{c\!}f_{}^{(i,j)}a_{}^{i}\cos(j\phi)\big],
  \end{equation}
  where $f$ represents one of the non-vanishing final observables
  ({\it e.g.}\ $k_{3}^{}$, $s_{3}^{}$, or $m$).  The coefficients
  ${}^{s\!}f_{}^{(i,j)}$ and ${}^{c\!}f_{}^{(i,j)}$ are finite
  linear combinations of the original spin-expansion coefficients
  $f_{}^{m_{1}m_{2}m_{3}|n_{1}n_{2}n_{3}}$.  These linear
  combinations can be derived explicitly by equating the two
  expansions (\ref{expansions}) and (\ref{1stform}) term by term;
  the first few combinations are provided in Appendix
  \ref{Case2App}.
  
  For situations in which $\phi$ varies while $a$ remains fixed, we
  should go one step beyond Eq.~(\ref{1stform}) by collecting all terms
  with the same $\phi$ dependence.  Then Eq.~(\ref{1stform}) is
  rewritten in the form
  \begin{equation}
    \label{2ndform}
    f\!=\!\sum_{j=0}^{\infty}\big[{}^{s\!}f_{}^{(j)}
    \sin(j\phi)\!+\!{}^{c\!}f_{}^{(j)}\cos(j\phi)\big]\, ,
  \end{equation}
\end{subequations}
where we have defined the coefficients
\begin{equation}
  \label{angle_coeff}
  ^{s\!}f_{}^{(j)}\equiv\sum_{i=j}^{\infty}{}^{s\!}f_{}^{(i,j)}a_{}^{i}
  \qquad\quad
  ^{c\!}f_{}^{(j)}\equiv\sum_{i=j}^{\infty}{}^{c\!}f_{}^{(i,j)}a_{}^{i} \, .
\end{equation}
We stress that Eq.~(\ref{1stform}) is the {\it same} series as
Eq.~(\ref{2ndform}).  The first form (\ref{1stform}) is appropriate
for situations where $a$ and $\phi$ both vary, whereas the second form
(\ref{2ndform}) is appropriate for situations where $\phi$ varies but
$a$ does not.

Symmetry considerations further restrict the terms that can appear in
the expansions (\ref{both_forms}).  Applying parity $P$ to the initial
configuration (\ref{Case2Spins}) is equivalent to the transformation
$\phi\to\phi+\pi$ which sends
$\{\sin(j\phi),\cos(j\phi)\}\to\{(-1)^{j}\sin(j\phi),
(-1)^{j}\cos(j\phi)\}$.  It follows that
\begin{itemize}
\item {\it In the superkick configuration (\ref{Case2Spins}), only
    terms with even (odd) $i$ and $j$ can appear in the expansions
    (\ref{both_forms}) for a scalar (pseudoscalar) quantity $f$.}
\end{itemize}
Since the superkick initial spin configuration (\ref{Case2Spins}) is
invariant under $X$, an exchange transformation does not yield any
further constraints.

These simple considerations lead to detailed quantitative predictions.
To illustrate this point, let us start by displaying some of the
leading terms in the expansions (\ref{1stform}) for the non-vanishing
observables $k_{3}^{}$, $s_{3}^{}$, and $m$:
\begin{subequations} \label{k3_final}
  \begin{eqnarray}
    \label{k3_superkick}
    k_{3}^{}\!&\!\!=\!\!&\!
    [{}^{s}k_{3}^{(1,1)}a_{}^{1}\!+\!{}^{s}k_{3}^{(3,1)}a_{}^{3}
    \!+\!{\cal O}(a_{}^{5})]\sin\;\!(\phi) \nonumber\\
    \!&\!\!+\!\!&\!
    [{}^{c}k_{3}^{(1,1)}a_{}^{1}\!+\!{}^{c}k_{3}^{(3,1)}a_{}^{3}
    \!+\!{\cal O}(a_{}^{5})]\cos(\phi) \nonumber\\
    \!&\!\!+\!\!&\!
    [{}^{s}k_{3}^{(3,3)}a_{}^{3}\!+\!{}^{s}k_{3}^{(5,3)}a_{}^{5}
    \!+\!{\cal O}(a_{}^{7})]\sin\;\!(3\phi) \nonumber\\
    \!&\!\!+\!\!&\!
    [{}^{c}k_{3}^{(3,3)}a_{}^{3}\!+\!{}^{c}k_{3}^{(5,3)}a_{}^{5}
    \!+\!{\cal O}(a_{}^{7})]\cos(3\phi) \nonumber\\
    \!&\!\!+\!\!&\!\ldots \\
    \label{s3_superkick}
    s_{3}^{}\!&\!\!=\!\!&\!
    [{}^{c}s_{3}^{\,(0,0)}a_{}^{0}+{}^{c}s_{3}^{\,(2,0)}a_{}^{2}
    +{\cal O}(a_{}^{4})]\cos(0\phi) \nonumber\\
    \!&\!\!+\!\!&\!
    [{}^{s}s_{3}^{\,(2,2)}a_{}^{2}+{}^{s}s_{3}^{\,(4,2)}a_{}^{4}
    +{\cal O}(a_{}^{6})]\sin\;\!(2\phi) \nonumber\\
    \!&\!\!+\!\!&\!
    [{}^{c}s_{3}^{\,(2,2)}a_{}^{2}+{}^{c}s_{3}^{\,(4,2)}a_{}^{4}
    +{\cal O}(a_{}^{6})]\cos(2\phi) \nonumber\\
    \!&\!\!+\!\!&\!\ldots \\
    \label{m_superkick}
    m\!&\!\!=\!\!&\![{}^{c}m_{}^{(0,0)}a_{}^{0}\!+\!
    {}^{c}m_{}^{(2,0)}a_{}^{2}\!+\!{\cal O}(a^{4})]
    \cos(0\phi) \nonumber\\
    \!&\!\!+\!\!&\![{}^{s}m_{}^{(2,2)}a_{}^{2}\!+\!
    {}^{s}m_{}^{(4,2)}a_{}^{4}\!+\!{\cal O}(a^{6})]
    \sin\;\!(2\phi) \nonumber\\
    \!&\!\!+\!\!&\![{}^{c}m_{}^{(2,2)}a_{}^{2}\!+\!
    {}^{c}m_{}^{(4,2)}a_{}^{4}\!+\!{\cal O}(a^{6})]
    \cos(2\phi) \nonumber\\
    \!&\!\!+\!\!&\!\ldots
  \end{eqnarray}
\end{subequations}
In these equations, we have explicitly displayed the $\cos(0\phi)=1$
factors to emphasize the similarity between the $j=0$ and $j\ne0$
terms.

First consider the predictions of Eq.~(\ref{k3_superkick}) for the
kick velocity $k_{3}^{}$.  The leading-order (${\cal O}(a_{}^{1})$)
terms predict that this kick should vary as $\sin(\phi+{\rm phase})$
at a fixed value of $a$, and that the {\it amplitude} of this sinusoid
should scale {\it linearly} with $a$.  The next-to-leading-order
(${\cal O}(a_{}^{3})$) terms predict that this leading $\sin(\phi+{\rm
  phase})$ behavior should receive a small additive correction of the
form $\sin(3\phi+{\rm phase})$ with amplitude proportional to
$a_{}^{3}$.  Next consider the predictions of Eq.~(\ref{s3_superkick})
for the final spin $s_3$.  The leading-order (${\cal O}(a_{}^{0})$)
term predicts that the final spin is a constant, independent of both
$a$ and $\phi$: $s_{3}^{}\propto a_{}^{0}\cos(0\phi)$.  The
next-to-leading-order (${\cal O}(a_{}^{2})$) terms predict that the
leading $\cos(0\phi)$ behavior should receive a small additive
correction of the form $\sin(2\phi+{\rm phase})$ with amplitude
proportional to $a_{}^{2}$.  Finally, since Eq.~(\ref{m_superkick}) is
formally identical to Eq.~(\ref{s3_superkick}), $m$ should behave in
the same way as $s_{3}^{}$.

\begin{figure}
  \begin{center}
    \includegraphics[width=3.5in]{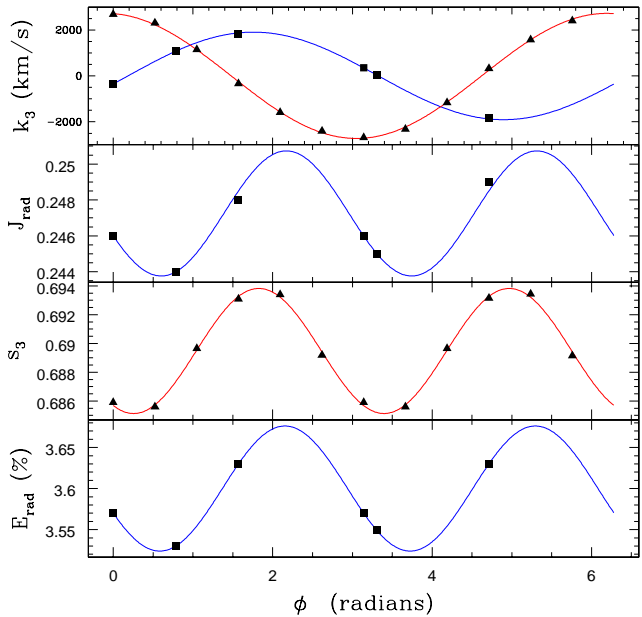}
  \end{center}
  \caption{Case \#2 best-fit curves.  Panels from top to bottom show the kick
    velocity $k_3$, the radiated angular momentum $J_{\rm rad}$, the
    final spin $s_3$, and the percentage of radiated energy $E_{\rm
      rad}$, all plotted against the angle $\phi$ between the initial
    spin ${\bf a}$ and ${\bf e}^{(1)}$.  The blue curves show fits to
    the square data points taken from \cite{Campanelli:2007cg}, while
    the red curves show fits to the triangle data points of
    \cite{Brugmann:2007zj}.  The curves for $k_3$ only show the
    first-order terms in Eq.~(\ref{k_3_case2}), while those for
    $J_{\rm rad}$, $s_3$, and $E_{\rm rad}$ include all terms up to
    second-order.  The $\chi^2/{\rm d.o.f.}$ and best-fit coefficients
    are listed in Table~\ref{super_results}.}
  \label{Case2fig}
\end{figure}

We test these predictions using the published simulations of
\cite{Campanelli:2007cg,Brugmann:2007zj}, with relevant data and
errors described in Appendix~\ref{Case2TabApp}.  Although both papers
estimate the final kicks $k_{3}^{}$, the simulations in
\cite{Campanelli:2007cg} start at a different orbital separation
$r/(M_{a}+M_{b})$ from those in \cite{Brugmann:2007zj}.  In
Sec.~\ref{Case_1}, we were able to jointly fit simulations with
different initial separations because the relevant spin-expansion
coefficients were insensitive to the initial separation.
Unfortunately, in the present configuration (\ref{Case2Spins}), the
relevant expansion coefficients {\it are} sensitive to the initial
separation.  As we discuss in a future paper, it should be possible to
connect the expansion coefficients at different initial separations
using post-Newtonian techniques.  For the time being, though, we must
perform separate fits for the simulation sets $\{ A_{i} \}$ of
\cite{Campanelli:2007cg} and $\{ B_{i} \}$ of \cite{Brugmann:2007zj}.
As the initial spin magnitude $a$ is fixed within each data set, we
should use the spin expansion in the form (\ref{2ndform}).  Thus, for
the non-vanishing observables $k_{3}^{}$, $s_{3}^{}$, and $m$, we have
the expressions:
\begin{subequations} \label{case2_exp}
  \begin{eqnarray}
    \label{k_3_case2}
    k_{3}^{}\!&\!\!=\!\!&\!{}^{s}k_{3}^{(1)}\sin(\phi)\!+\!
    {}^{c}k_{3}^{(1)}\cos(\phi) \nonumber \\
    \!&\!\!+\!\!&\!{}^{s}k_{3}^{(3)}\sin(3\phi)\!+\!{}^{c}k_{3}^{(3)} 
    \cos(3\phi)+{\cal O}(a_{}^{5}), \\
    \label{s_3_case2}
    s_{3}^{}\!&\!\!=\!\!&\!{}^{c}s_{3}^{\;(0)}+{}^{s}s_{3}^{\;(2)}
    \sin(2\phi)\!+{}^{c}s_{3}^{\;(2)}\cos(2\phi)\!+\!{\cal O}(a_{}^{4}), \\
    \label{m_case2}
    m\!&\!\!=\!\!&\!{}^{c}m^{(0)}\!+\!{}^{s}m^{(2)}\sin(2\phi)\!+\!
    {}^{c}m^{(2)}\cos(2\phi)\!+\!{\cal O}(a_{}^{4})\qquad
  \end{eqnarray}
\end{subequations}
Instead of reporting $s_{3}^{}$ and $m$, Campanelli {\it et
  al.} \cite{Campanelli:2007cg} report the percentage of the initial
energy carried away by gravitational radiation, $\% E_{{\rm rad}}$,
and the dimensionless radiated angular momentum $J_{{\rm rad}}/M^{2}$.
Since both of these quantities are scalars, with $P=+1$ and $X=+1$,
their expansions are exactly analogous to the expansions for $m$ and
$s_{3}^{}$ in Eqs.~(\ref{s_3_case2}) and (\ref{m_case2}).  The
best-fit coefficients from our leading-order and next-to-leading-order
fits to these quantities are listed in Table~\ref{super_results}, and
the leading-order fits themselves are displayed in
Fig.~\ref{Case2fig}.
\begin{table}
  \begin{small}
    \begin{center}
      \begin{tabular}{|c||c|c|c|c|}
        \hline
        Data Set & $A$ & $A$ & $B$ & $B$ \\
        \hline \hline
        $^{c}k_{3}^{1}$ & $-350$ & $-323$ & $3753$ & $2714$ \\
        \hline
        $^{s}k_{3}^{1}$ & $1876$ & $1837$ & $-343$ & $-246$ \\
        \hline
        $^{c}k_{3}^{3}$ & --- & $-29.4$ & --- & $-20.8$ \\
        \hline
        $^{s}k_{3}^{3}$ & --- & $2.73$ & --- & $83.5$ \\
        \hline \hline
        $^{c}(J_{\rm rad}/M^2)^{0}$ & $0.2471$ & $0.2471$ & --- & --- \\
        \hline
        $^{c}(J_{\rm rad}/M^2)^{2}$ & --- & $-0.0012$ & --- & --- \\
        \hline
        $^{s}(J_{\rm rad}/M^2)^{2}$ & --- & $-0.0027$ & --- & --- \\
        \hline \hline
        $^{c}s_{3}^{0}$ & --- & --- & $0.6895$ & $0.6895$ \\
        \hline
        $^{c}s_{3}^{2}$ & --- & --- & --- & $-0.0038$ \\
        \hline
        $^{s}s_{3}^{2}$ & --- & --- & --- & $-0.0021$ \\
        \hline \hline
        $^{c}(\%E_{\rm rad})^{0}$ & $3.587$ & $3.600$ & --- & --- \\
        \hline
        $^{c}(\%E_{\rm rad})^{2}$ & --- & $-0.0301$ & --- & --- \\
        \hline
        $^{s}(\%E_{\rm rad})^{2}$ & --- & $-0.0695$ &--- & --- \\
        \hline
      \end{tabular}
    \end{center}
  \end{small}
  \caption{Fits for Case \#2.  The first column lists the coefficients
    being determined.  Kick velocities are in units of km/s while the
    remaining coefficients are dimensionless.  The second column 
    provides the best-fit values for these coefficients when the 
    lowest-order terms are fit to data set $A$, the simulations of 
    \cite{Campanelli:2007cg}.  Expansions for the radiated angular 
    momentum $J_{\rm rad}/M^2$ and energy $\%E_{\rm rad}$ have 
    zeroth-order terms while first order is lowest for $k_3$. The 
    third column lists best-fit values for next-to-lowest order fits; 
    this is second order for $J_{\rm rad}/M^2$ and $\%E_{\rm rad}$ and 
    third order for $k_3$.  The fourth and fifth columns list the 
    corresponsing values of coefficients for data set $B$, the 
    simulations of \cite{Brugmann:2007zj}.  This data set provides 
    the spin $s_3$ of the final black hole instead of 
    $J_{\rm rad}/M^2$ and $\%E_{\rm rad}$.  The expansion for $s_3$ 
    has zeroth and second-order terms.}
  \label{super_results}
\end{table}
\begin{figure}
  \begin{center}
    \includegraphics[width=3.5in]{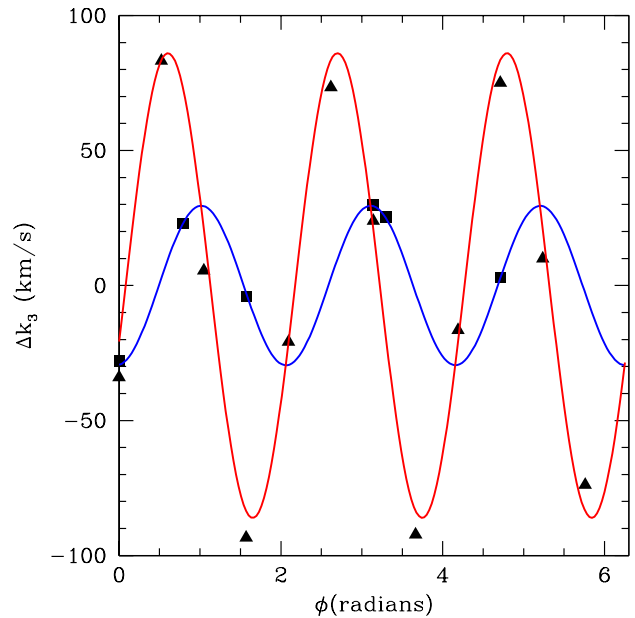}
  \end{center}
  \caption{The residuals $\Delta k_3$ after the first-order terms of
    Eq.~(\ref{k_3_case2}) are subtracted from the simulated final
    kicks.  As in the top panel of Fig.~\ref{Case2fig}, the blue
    curves show fits to the square data points taken from
    \cite{Campanelli:2007cg}, while the red curves show fits to the
    triangle data points of \cite{Brugmann:2007zj}.  These curves
    consist of the third-order terms of Eq.~(\ref{k_3_case2}) with the
    coefficients listed in Table~\ref{super_results}.}
  \label{Residfig}
\end{figure}

First focus on the kick velocity $k_{3}^{}$.  The top panel in
Fig.~\ref{Case2fig} clearly reveals the $\sin(\phi+{\rm phase})$
behavior predicted (at leading order) by the spin expansion, and
previously noted in Eq.~(1) of \cite{Campanelli:2007cg} and Eq.~(6) of
\cite{Brugmann:2007zj}.  Although the spin expansion cannot predict
the phases of the sine waves in this panel, it predicts that (at
leading order) their amplitudes should be proportional to the spin
magnitude $a$.  This prediction is also supported by the simulations:
the ratio of the best-fit values of $[(^{s}k_{3}^{(1)})^2 +
(^{c}k_{3}^{(1)})^2]^{1/2}$ for the simulations $\{ A_i \}$ of
\cite{Campanelli:2007cg} and $\{ B_i \}$ of \cite{Brugmann:2007zj} is
$0.684$, not far off from the ratio of their spins $0.515/0.723 =
0.712$.

The {\it next-to-leading order} predictions of the spin expansion for
$k_{3}^{}$ also appear to be confirmed.  To see this, we subtract the
leading-order $\sin(\phi+{\rm phase})$ prediction from the simulated
$k_{3}^{}$ data, and plot the residuals in Fig.~\ref{Residfig}.  As
predicted, these residuals oscillate as ${\rm sin}(3\phi+{\rm
  phase})$.  Furthermore, the amplitude of this residual oscillation
scales as $a_{}^{3}$, as predicted.  The ratio of the
best-fit values of $[(^{c}k_{3}^{(3)})^2 + (^{s}k_{3}^{(3)})^2]^{1/2}$
for the simulations $\{ A_i \}$ and $\{ B_i \}$ is $0.343$, which is
close to the spin ratio cubed, $(0.515/0.723)^3 = 0.361$.

Thus, we have seen that the spin expansion succeeds in reproducing
previously observed aspects of the $k_{3}^{}$ data, and also in
predicting previously unrecognized features.

The remaining three panels of Fig.~\ref{Case2fig} show observables
whose $\phi$-dependence agrees closely with what we expect for scalars
like $s_3$ and $m$ in Eqs.~(\ref{s_3_case2}) and (\ref{m_case2}).  At
zeroth-order they are independent of $\phi$, while at next-to-leading
order the predicted $\sin(2\phi+{\rm phase})$ contribution
appears.\footnote{Br$\ddot{{\rm u}}$gmann {\it et al.}
  \cite{Brugmann:2007zj} noted a $\sin(2\phi+{\rm phase})$ dependence
  in $s_3$, but concluded that it might be a non-physical systematic
  error, since they could not find a consistent $\sin(\phi+{\rm
    phase})$ contribution (see their Fig.~13, and corresponding
  discussion).  Our formalism explains why the $\sin(\phi+{\rm
    phase})$ term is forbidden, and suggests that the oscillation
  which they observe is probably a real physical effect, not a
  systematic error.}  Unfortunately for us, \cite{Campanelli:2007cg}
and \cite{Brugmann:2007zj} chose to publish different observables so
we cannot test whether the amplitude of these double-frequency terms
really scales like $a^2$ as predicted.  Hopefully, future simulations
performed at different values of $a$ will address this question.
  
\subsection{Case \#3: Herrmann {\it et al} ``B-series'' \newline
($q=1$, ${\bf a}=-{\bf b}$, $a_{2}^{}=b_{2}^{}=0$)}
\label{Case_3}

We next consider the ``B-series'' simulations of Herrmann {\it et al.}
\cite{Herrmann:2007ex}.  This configuration consists of equal-mass
black holes with equal-magnitude, oppositely directed spins lying in
the $({\bf e}^{(1)},{\bf e}^{(3)})$ plane:
\begin{equation} \label{Case3Spins}
  \left(\!\begin{array}{c} a_{1}^{} \\ a_{2}^{} \\ a_{3}^{}
    \end{array}\!\right)\!=\!
  \left(\!\!\begin{array}{c} +a\;\!\sin \phi \\ 0 \\
      +a\cos\phi \end{array}\right),\qquad
  \left(\!\begin{array}{c} b_{1}^{} \\ b_{2}^{} \\ b_{3}^{}
    \end{array}\!\right)\!=\!
  \left(\!\!\begin{array}{c} -a\;\!\sin \phi \\ 0 \\
      -a\cos\phi \end{array}\!\!\right).
\end{equation}
Though this is obviously a highly symmetric configuration, it is not
one of the special configurations derived in Section~\ref{S:exact} and
therefore has non-vanishing values for all of the final observables
$f\in\{m, {\bf k}_{\!\perp}, k_{3}^{}, {\bf s}_{\!\perp}^{},
s_{3}^{}\}$.  As in Case \#2, the initial spin configuration
(\ref{Case3Spins}) is parameterized by two numbers ($a$ and $\phi$)
instead of six $\{ a_i, b_i \}$, so that the original spin expansions
(\ref{expansions}) become highly degenerate.  Therefore, just as in
Case \#2, we should remove these degeneracies by collecting terms with
the same dependence on $a$ and $\phi$, and rewriting the expansion in
the form (\ref{1stform}), or equivalently (\ref{2ndform},
\ref{angle_coeff}).  For convenience, Appendix \ref{Case3App} provides
explicit expressions for the ``new'' coefficients
$\{{}^{c\!}f^{(i,j)}, {}^{s\!}f^{(i,j)}\}$ in terms of the original
spin-expansion coefficients $f^{m_1 m_2 m_3|n_1 n_2 n_3}$, up to third
order in $a$.
  
Since the B-series configuration (\ref{Case3Spins}) has different
symmetries from the superkick configuration (\ref{Case2Spins}), there
are different rules governing which terms appear in the expansions
(\ref{both_forms}) for each observable.  We can discover these rules
as follows.  Applying a parity transformation $P$ to the initial
configuration (\ref{Case3Spins}) is equivalent to sending
$\phi\to-\phi$, and hence $\{\sin(j\phi),\cos(j\phi)\}
\to\{-\sin(j\phi),\cos(j\phi)\}$.  From this we infer that
\begin{itemize}
\item {\it In the B-series configuration (\ref{Case3Spins}), only
    cosine (sine) terms can appear in the expansions
    (\ref{both_forms}) for a scalar (pseudoscalar) quantity $f$.}
\end{itemize}
Applying a combined parity and exchange transformation $PX$ to the
initial configuration (\ref{Case3Spins}) is equivalent to the
transformation $\phi\to\phi+\pi$, and therefore
$\{\sin(j\phi),\cos(j\phi)\}\!\to\!\{(-1)^{j}\sin(j\phi),
(-1)^{j}\cos(j\phi)\}$.  We thus infer that
\begin{itemize}
\item {\it In the B-series configuration (\ref{Case3Spins}), only
    terms with even (odd) $i$ and $j$ can appear in the expansions
    (\ref{both_forms}) for a quantity $f$ that is even (odd) under
    $PX$.}
\end{itemize}

From these simple considerations, a host of interesting predictions
follow.\footnote{Herrmann {\it et al.}  \cite{Herrmann:2007ex}
  proposed a general kick formula [their Eq.~(5)] inspired by the
  Kidder formula for the emission of linear momentum
  \cite{Kidder:1995zr}.  This formula is linear in the initial spins
  and agrees with ours to this order.  A key difference is that they
  claim their formula is only valid when the initial spin components
  are determined at {\it entrance}, when the binary reaches the
  ``last'' orbit or plunge.  We claim that symmetry and the inclusion
  of higher-order terms allows our formula to be a valid approximation
  at {\it any} initial separation.}  To illustrate this point,
consider the expansions for $f\in\{k_{i}^{}, s_{i}^{}, m\}$:
\begin{subequations}
  \label{Case3_expansions}
  \begin{eqnarray}
    {\bf k}_{\!\perp}^{}\!&\!\!=\!\!&\![
    {}^{c}{\bf k}_{\!\perp}^{\,(1,1)}a_{}^{1}+
    {}^{c}{\bf k}_{\!\perp}^{\,(3,1)}a_{}^{3}\!+\!{\cal O}(a_{}^{5})]
    \cos(\phi) \nonumber\\
    \!&\!\!+\!\!&\![
    {}^{c}{\bf k}_{\!\perp}^{\,(3,3)}a_{}^{3}+
    {}^{c}{\bf k}_{\!\perp}^{\,(5,3)}a_{}^{5}\!+\!{\cal O}(a_{}^{7})]
    \cos(3\phi) \nonumber\\
    \!&\!\!+\!\!&\!\ldots \\
    k_{3}^{}\!&\!\!=\!\!&\![
    {}^{s}k_{3}^{\,(1,1)}a_{}^{1}+{}^{s}k_{3}^{\,(3,1)}a_{}^{3}
    \!+\!{\cal O}(a_{}^{5})]\sin(\phi) \nonumber\\
    \!&\!\!+\!\!&\![
    {}^{s}k_{3}^{\,(3,3)}a_{}^{3}+{}^{s}k_{3}^{\,(5,3)}a_{}^{5}
    \!+\!{\cal O}(a_{}^{7})]\sin(3\phi) \nonumber\\
    \!&\!\!+\!\!&\!\ldots \\
    {\bf s}_{\!\perp}^{}\!&\!\!=\!\!&\![
    {}^{s}{\bf s}_{\!\perp}^{\;(2,2)}a_{}^{2}+
    {}^{s}{\bf s}_{\!\perp}^{\;(4,2)}a_{}^{4}\!+\!
    {\cal O}(a_{}^{6})]\sin(2\phi) \nonumber\\
    \!&\!\!+\!\!&\![
    {}^{s}{\bf s}_{\!\perp}^{\;(4,4)}a_{}^{4}+
    {}^{s}{\bf s}_{\!\perp}^{\;(6,4)}a_{}^{6}\!+\!
    {\cal O}(a_{}^{8})]\sin(4\phi) \nonumber\\
    \!&\!\!+\!\!&\!\ldots \\
    s_{3}^{}\!&\!\!=\!\!&\![
    {}^{c}s_{3}^{\;(0,0)}a_{}^{0}+{}^{c}s_{3}^{\;(2,0)}a_{}^{2}
    \!+\!{\cal O}(a_{}^{4})]\cos(0\phi) \nonumber\\
    \!&\!\!+\!\!&\![
    {}^{c}s_{3}^{\;(2,2)}a_{}^{2}+{}^{c}s_{3}^{\;(4,2)}a_{}^{4}
    \!+\!{\cal O}(a_{}^{6})]\cos(2\phi) \nonumber\\
    \!&\!\!+\!\!&\!\ldots \\
    m\!&\!\!=\!\!&\![
    {}^{c}m_{}^{(0,0)}a_{}^{0}\!+\!{}^{c}m_{}^{(2,0)}a_{}^{2}
    \!+\!{\cal O}(a_{}^{4})]\cos(0\phi) \nonumber \\
    \!&\!\!+\!\!&\![
    {}^{c}m_{}^{(2,2)}a_{}^{2}\!+\!{}^{c}m_{}^{(4,2)}a_{}^{4}
    \!+\!{\cal O}(a_{}^{6})]\cos(2\phi) \nonumber \\
    \!&\!\!+\!\!&\!\ldots 
  \end{eqnarray}
\end{subequations}
Each of these equations makes specific predictions (at leading order
in $a$, at next-to-leading order, and so on) for how the final-state
quantities should vary as a function of $a$ and $\phi$.  We hope that
in the future many of these predictions will be tested in detail.

Currently, there are 7 simulations to which we can compare our
predictions: the 6 ``B-Series'' simulations in \cite{Herrmann:2007ex},
plus one additional ($\phi=0$) simulation from an earlier work by the
same group \cite{Herrmann:2007ac}.  We summarize the relevant
simulations in Appendix~\ref{Case3TabApp}.  Although the authors
report results for the final kick, as well as the final radiated
energy and angular momentum from each simulation, the radiated energy
and angular momentum values are not accurate enough to constrain the
spin expansion beyond the trivial zeroth-order (constant) term.
Therefore, we only consider the final kick.  Since the initial spin
magnitude is fixed ($a=0.6$ in all 7 simulations) and only $\phi$
varies, we should rewrite the expansions (\ref{Case3_expansions}) in
the form (\ref{2ndform}).  In particular, the expansions for the final
kick become
\begin{subequations}
  \begin{eqnarray}
    \label{Case3_kperp}
    {\bf k}_{\!\perp}^{}\!&\!\!=\!\!&\!{}^{c}{\bf k}_{\!\perp}^{(1)}
    \cos(\phi)\!+\!{}^{c}{\bf k}_{\!\perp}^{(3)}\cos(3\phi)
    \!+\!{\cal O}(a_{}^{5}) \\
    \label{Case3_k3}
    k_{3}^{}&\!\!=\!\!&\!{}^{s}k_{3}^{(1)\;\!}\sin\;\!(\phi)\!+\!
    {}^{s}k_{3}^{(3)}\sin\;\!(3\phi)\!+\!{\cal O}(a_{}^{5})
  \end{eqnarray}
\end{subequations}
As in Case \#1, only the magnitudes of the final kicks $|{\bf k}|$
have been published.\footnote{Herrmann {\it et al.}
  \cite{Herrmann:2007ex} show the individual components of ${\bf k}$
  in their Fig.~10, but without sufficient numerical precision to
  allow us to constrain higher-order terms.}  Combining
Eqs.~(\ref{Case3_kperp}) and (\ref{Case3_k3}), we find that
$|{\bf k}|^{2}={\bf k}_{\!\perp}^{2}+k_{3}^{2}$ has the expansion:
\begin{equation} \label{Case3_ksqr}
  |{\bf k}|^{2}=A_{0}+A_{2}\cos(2\phi)+A_{4}\cos(4\phi)
  +{\cal O}(a_{}^{6}), 
\end{equation}
where the amplitudes $A_i$ are given by
\begin{subequations} \label{case3_amps}
  \begin{eqnarray}
    \label{case3_amp1}
    A_{0}\!&\!\!\equiv\!\!&\!\frac{1}{2}\Big[|{}^{c}{\bf k}_{\!\perp}^{(1)}|^2 
    \!+\!({}^{s\!}k_{3}^{(1)})^{2}\Big] \\
    \label{case3_amp2}
    A_{2}\!&\!\!\equiv\!\!&\!\frac{1}{2}\Big[|{}^{c}{\bf k}_{\!\perp}^{(1)}|^2
    \!-\!({}^{s\!}k_{3}^{(1)})^{2}\Big]\!+\!{}^{c}{\bf k}_{\!\perp}^{(1)}
    \!\!\cdot\!{}^{c}{\bf k}_{\!\perp}^{(3)}\!\!+\!{}^{s\!}k_{3}^{(1)}
    {}^{s\!}k_{3}^{(3)}\qquad \\
    \label{case3_amp3}
    A_{4}\!&\!\!\equiv\!\!&\!{}^{c}{\bf k}_{\!\perp}^{(1)}\!\!\cdot\! 
    {}^{c}{\bf k}_{\!\perp}^{(3)}\!\!-\!{}^{s\!}k_{3}^{(1)}{}^{s\!}k_{3}^{(3)}.
  \end{eqnarray}
\end{subequations}
Note that the expansion (\ref{Case3_ksqr}) only contains terms of the
form ${\rm cos}(2n\phi)$, just as we would expect for a quantity with
$P=+1$ and $PX=+1$, like $m$ or $s_{3}^{}$.  However, since $|{\bf
  k}|_{}^{2}$ is a composite quantity constructed from the more
``fundamental'' quantities ${\bf k}_{\perp}^{}$ and $k_{3}^{}$, the
$a_{}^{0}$ term is missing from the coefficient $A_{0}$, and instead
the leading order term in $A_{0}$ is proportional to $a_{}^{2}$.  This
is a crucial physical difference between the expansion for $|{\bf
  k}|^{2}$ and the expansions for $m$ and $s_{3}^{}$ which both
contain a non-vanishing $a_{}^{0}$ term.

\begin{figure}
  \begin{center}
    \includegraphics[width=3.5in]{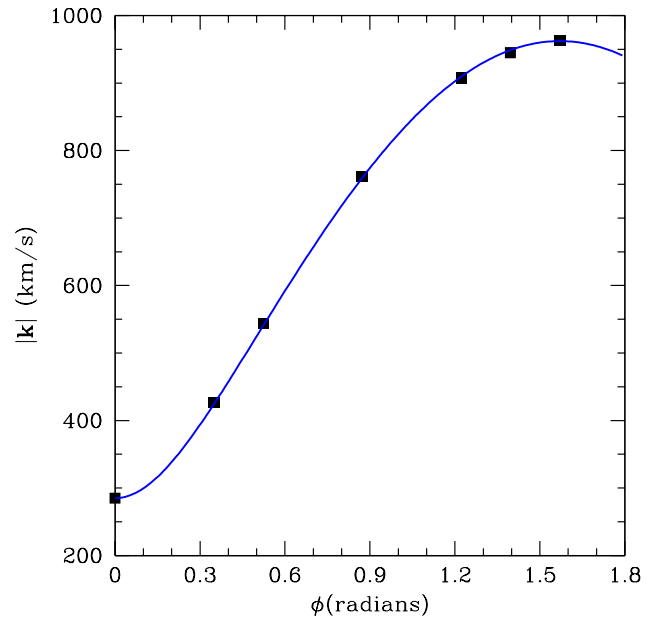}
  \end{center}
  \caption{The kick magnitudes $|{\bf k}|$ as a function of polar
    angle $\phi$ for the Case \#3 configuration given by
    Eq.~(\ref{Case3Spins}).  The square points correspond to the
    simulations listed in Table~\ref{Case3_table}, while the blue
    curve shows the second-order fit to Eq.~(\ref{Case3_ksqr}) with
    amplitudes listed in Table~\ref{Case3_results}.}
  \label{Case3fig}
\end{figure}

When we fit the 7 simulations to the second-order and fourth-order
forms of Eq.~(\ref{Case3_ksqr}), the $\chi^{2}/{\rm d.o.f.}$ is
$8.36\times10^{-4}$ and $3.30\times10^{-4}$, respectively.  The
corresponding best-fit values for the amplitudes $A_{i}$ are listed in
Table~\ref{Case3_results}, and the second-order fit itself is
displayed in Fig.~\ref{Case3fig}.\footnote{Herrmann {\it et al.}
  \cite{Herrmann:2007ex} found that the angle $\phi$ (their
  $\hat{\theta}$) remains unchanged during the final few orbits of
  inspiral, allowing them to apply their kick formula using the {\it
    initial} spin components rather than those at entrance.  Their
  quantities $\{ V_{\rm max}^x, V_{\rm max}^y,V_{\rm max}^z \}$
  correspond directly to our $\{ ^{c}k_{1}^1, ^{c}k_{2}^1, ^{s}k_{3}^1
  \}$, leading to values $A_0 = 5.02 \times 10^5$ (km/s)$^2$, $A_0 =
  -4.20 \times 10^5$ (km/s)$^2$ consistent with our results.}  We see
that the $A_{0}$ and $A_{2}$ terms already provide an excellent fit to
the simulations of \cite{Herrmann:2007ex}, and there is only marginal
evidence for the $A_{4}$ term.  In order to definitively detect the
presence of the $A_{4}$ term, additional simulations would be needed.
\begin{table}
  \begin{small}
    \begin{center}
      \begin{tabular}{|c|c|c|c|}
        \hline
        Order & $A_0$ & $A_2$ & $A_4$ \\
        \hline \hline
        up to $a^2$ & $5.04 \times 10^5$ & $-4.23 \times 10^5$ & --- \\ 
        \hline
        up to $a^4$ & $5.05 \times 10^5$ & $-4.20 \times 10^5$ &
        $-3.92 \times 10^3$ \\ 
        \hline
      \end{tabular}
    \end{center}
  \end{small}
  \caption{Fits of the kick magnitudes $|{\bf k}|^{\rm num}$ listed in
    Table~\ref{Case3_table} to the fitting formula of .
    The first column lists the $a$-dependence of the
      highest-order term appearing in the fit; remaining columns
    provide the best-fit values for the amplitudes $A_i$ in units of
    (km/s)$^2$.  The first row describes a fit to the two
      second-order terms in Eq.(\ref{Case3_ksqr}), while the second
      row provides best-fit values to all three terms that appear to
    fourth order in $a$.}
  \label{Case3_results}
\end{table}

\subsection{Case \#4: Herrmann {\it et al} ``S-series''}
\label{Case_4}

In this section we consider the ``S-Series'' simulations of Herrmann
{\it et al} \cite{Herrmann:2007ex}, in which the black holes have
equal mass ($q=1$), and spins initially given by
\begin{equation}
  \left(\!\begin{array}{c} a_{1}^{} \\ a_{2}^{} \\ a_{3}^{}
    \end{array}\!\right)\!=\!
  \left(\begin{array}{c}
      -a \\ 0 \\ 0 \end{array}\right),\qquad\quad
  \left(\!\begin{array}{c} b_{1}^{} \\ b_{2}^{} \\ b_{3}^{}
    \end{array}\!\right)\!=\!
  \left(\begin{array}{c} a\;\!\sin\phi \\ 0 \\ a\cos\phi 
    \end{array}\right).
\end{equation}
This is the least symmetric spin configuration that we have considered
so far.  Since the configuration is parameterized by two numbers ($a$
and $\phi$) instead of six ($a_{i}^{}$ and $b_{i}^{}$), we should
again use the expansions (\ref{both_forms}).  In particular, since all
22 simulations in this series have the same spin magnitude $a = 0.6$,
we should write the expansion in the form (\ref{2ndform}).  Due to the
lack of symmetry, most of the expansion coefficients
$\{{}^{c\!}f_{}^{(j)},{}^{s\!}f_{}^{(j)}\}$ are non-vanishing.  In
Appendix~\ref{Case4App}, we provide relations between these ``new''
expansion coefficients and the original spin expansion coefficients
$f_{}^{m_{1}m_{2}m_{3}|n_{1}n_{2}n_{3}}$.

\begin{figure}
  \begin{center}
    \includegraphics[width=3.5in]{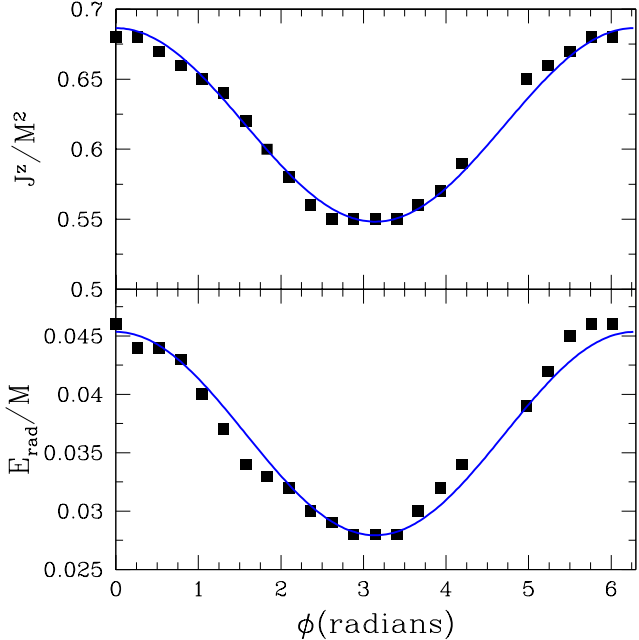}
  \end{center}
  \caption{The $z-$compnent of the final spin $J_{\rm final}^z$ and
    radiated energy $E_{\rm rad}$ as functions of the polar angle
    $\phi$ between the spin ${\bf b}$ of black hole $B$ and the
    orbital angular momentum in the configuration of Case \#4.  The
    square points correspond to the simulations listed in
    Table~\ref{Case4_table}, while the blue curves show the linear fit
    to Eq.~(\ref{2ndform}) with the first-order coefficients listed in
    Table~\ref{case4_JEres}.}
  \label{Case4JE}
\end{figure}

We first consider the final spin $J_{\rm final}^z$ in the ${\bf
  e}^{(3)}$ direction, and the total energy $E_{\rm rad}$ emitted in
gravitational radiation during the inspiral.  Both of these quantities
have expansions of the same form as that of the final mass $m$, so it
is convenient to analyze them in parallel.  As indicated by the fits
listed in Table~\ref{case4_JEres}, there is little evidence for terms
beyond linear order in $a$ in either expansion.  These linear-order
fits are shown in Fig.~\ref{Case4JE}.
\begin{table}
  \begin{small}
    \begin{center}
      \begin{tabular}{|c||c|c|c|c|}
        \hline
        Order & 0th & 1st & 2nd & 3rd \\
        \hline \hline
        $\chi^2/{\rm d.o.f.}$ & $0.321$ & $3.96\!\times\!10^{-3}$ &
        $1.29\!\times\!10^{-3}$ & $9.95\!\times\!10^{-4}$ \\
        \hline
        $^{c}(J_{\rm final}^z/M^2)^0$ & $0.610$ & $0.617$ & $0.618$ & 
        $0.618$ \\
        \hline
        $^{c}(J_{\rm final}^z/M^2)^1$ & --- & $6.91\!\times\!10^{-2}$ &
        $6.84\!\times\!10^{-2}$ & $6.85\!\times\!10^{-2}$ \\
        \hline
        $^{s}(J_{\rm final}^z/M^2)^1$ & --- & --- & 
        $-5.95\!\times\!10^{-3}$ & $-5.78\!\times\!10^{-3}$ \\
        \hline
        $^{c}(J_{\rm final}^z/M^2)^2$ & --- & --- & 
        $-5.15\!\times\!10^{-3}$ & $-5.26\!\times\!10^{-3}$ \\
        \hline
        $^{s}(J_{\rm final}^z/M^2)^2$ & --- & --- & --- & 
        $7.31\!\times\!10^{-5}$ \\
        \hline
        $^{c}(J_{\rm final}^z/M^2)^3$ & --- & --- & --- &
        $-2.34\!\times\!10^{-3}$ \\ 
        \hline \hline
        $\chi^2/{\rm d.o.f.}$ & $1.476$ & $0.0508$ & $0.0174$ & 
        $0.0161$ \\ 
        \hline
        $^{c}(E_{\rm rad}/M)^0$ & $3.45\!\times\!10^{-2}$ & 
        $3.66\!\times\!10^{-2}$ & $3.68\!\times\!10^{-2}$ & 
        $3.68\!\times\!10^{-2}$ \\ 
        \hline
        $^{c}(E_{\rm rad}/M)^1$ & --- & $8.71\!\times\!10^{-3}$ &
        $8.82\!\times\!10^{-3}$ & $8.82\!\times\!10^{-3}$ \\ 
        \hline
        $^{s}(E_{\rm rad}/M)^1$ & --- & --- & $-1.36\!\times\!10^{-3}$ &
        $-1.41\!\times\!10^{-3}$ \\ 
        \hline
        $^{c}(E_{\rm rad}/M)^2$ & --- & --- & $2.57\!\times\!10^{-4}$ &
        $3.25\!\times\!10^{-4}$ \\ 
        \hline
        $^{s}(E_{\rm rad}/M)^2$ & --- & --- & --- & 
        $-1.37\!\times\!10^{-4}$ \\ 
        \hline
        $^{c}(E_{\rm rad}/M)^3$ & --- & --- & --- & 
        $3.43\!\times\!10^{-4}$ \\ 
        \hline
      \end{tabular}
    \end{center}
  \end{small}
  \caption{Fits for the $z$-component of the final black hole spin
    $J_{\rm final}^z/M^2$ and radiated energy $E_{\rm rad}/M$in Case \#4.
    The first column gives the observable being fitted.  The second
    column listes the $\chi^2$/d.o.f. for the best fit.  Remaining
    columns provide the best-fit values for the listed parameters.}
  \label{case4_JEres}
\end{table}
These same linear terms, of the form $X^{000|001} b_3$, appeared in
our analysis of (anti-)aligned configurations (Case \#1).  They can be
understood physically by the same arguments presented in the final
paragraph of that subsection.  Though estimates of $E_{\rm rad}$ are
not available for Case \#1, we can attempt a quantitative comparison
between the final spins $J_{\rm final}^z/M^2$ here and values of $s_3$
provided for that configuration.  Equating these expansions
term-by-term at zeroth and first order in $a$, we find
\begin{subequations}
  \label{Case1&4comp}
  \begin{eqnarray}
    ^{c}(J_{\rm final}^z/M^2)^{0} &\simeq& s_{3}^{000|000} 
    (m^{000|000})^2 \, , \\
    ^{c}(J_{\rm final}^z/M^2)^{1} &\simeq& \big[ s_{3}^{001|000}
      (m^{000|000})^2 \\ \nonumber
      && + 2s_{3}^{000|000} m^{000|000} m^{001|000} \big] a \, .
  \end{eqnarray}
\end{subequations}
Inserting the appropriate values from Tables~\ref{zspin_results} and
\ref{case4_JEres} into the right and left-hand sides of
Eq.~(\ref{Case1&4comp}) respectively yields close agreement between
$^{c}(J_{\rm final}^z/M^2)^{0} = 0.617$ and $s_{3}^{000|000}
(m^{000|000})^2 = 0.626$ and between $^{c}(J_{\rm final}^z/M^2)^{1} =
6.91 \times 10^{-2}$ and $[s_{3}^{001|000} (m^{000|000})^2 +
2s_{3}^{000|000} m^{000|000} m^{001|000}] a = 6.96 \times 10^{-2}$.
This agreement is well within the systematic errors attributed to each
series of simulations, and suggests that our approach of decoupling
spin-dependent effects term-by-term shows promise.  Further
simulations will be necessary to determine how well this promise is
fulfilled.

\begin{figure}
  \begin{center}
    \includegraphics[width=3.5in]{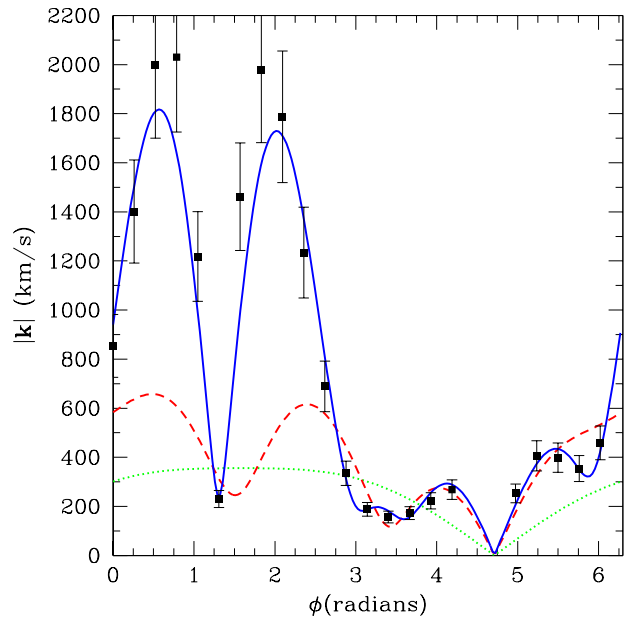}
  \end{center}
  \caption{The magnitude $|{\bf k}|$ of the recoil velocity in km/s as a
    function of the polar angle $\phi$ between the spin ${\bf b}$ of
    black hole $B$ and the orbital angular momentum in the
    configuration of Case \#4.  The square points correspond to the
    simulations listed in Table~\ref{Case4_table}, while the green
    (dotted), red (dashed), and blue (solid) curves show the best fits
    of Eq.~(\ref{case4_kmag}) to these simulations using terms derived
    from expanding Eqs.~(\ref{case4_kperp}) and (\ref{case4_k3}) to
    first, second, and, third order in $a$ respectively.  The error
    bars correspond to $1\sigma$ errors of 15\% in the kick magnitudes
    $|{\bf k}|$ reported by Herrmann {\it et al}
    \cite{Herrmann:2007ex} as the accuracy of their simulations.}
  \label{Case4K}
\end{figure}

We now turn our attention to the black-hole kicks.  As in the case of
the ``B-series'' considered in the previous subsection, our analysis
is hampered by only having access to the magnitude of the kicks rather
than their individual components.  In order to most clearly illuminate
the degeneracies that remain between terms in our general expansion
(\ref{expansions}), we must unfortunately resort to yet another new
expansion for this configuration.  We expand the individual components
and squared magnitude as
\begin{subequations} \label{Case4kickfit}
  \begin{eqnarray} \label{case4_kperp}
    {\bf k}_{\perp}^{}\!\!&\!\!=\!\!&\!\!\sum_{m=0}^{\infty}
    {\rm cos}_{}^{m}(\theta)[{}^{c}{\bf k}_{\perp}^{(m)}\!+\!
    {}^{s}{\bf k}_{\perp}^{(m)}{\rm sin}(\theta)] \, , \\
    \label{case4_k3}
    k_{3}^{}\!\!&\!\!=\!\!&\!\!\sum_{m=0}^{\infty}{\rm cos}_{}^{m}
    (\theta)[{}^{c\!}k_{3}^{(m)}\!+\!{}^{s\!}k_{3}^{(m)}{\rm sin}(\theta)] 
    \, , \\
    \label{case4_kmag}
    |{\bf k}|^2\!\!&\!\!=\!\!&\!\!\sum_{m=0}^{\infty}{\rm cos}_{}^{m}
    (\theta)[{}^{c\!}K_{}^{(m)}\!+\!{}^{s\!}K_{}^{(m)}{\rm sin}(\theta)] 
    \, .
  \end{eqnarray}
\end{subequations}
As previously, we relate the coefficients in this new expansion to
those in the general expansion in Appendix~\ref{Case4App}.  This new
expansion is useful because to third order, ${}^{c\!}k_{3}^{(0)} =
{}^{s\!}k_{3}^{(0)}$ implying that ${}^{c\!}K_{}^{(0)} =
{}^{s\!}K_{}^{(0)}$ and ${}^{c\!}K_{}^{(1)} = {}^{s\!}K_{}^{(1)}$.
There are therefore only 2 independent terms in the expansion of
$|{\bf k}|^2$ when ${\bf k}_{\perp}$ and $k_{3}$ are expanded to first
order in $a$.  The number of independent terms in the expansion of
$|{\bf k}|^2$ increases to 6 when ${\bf k}_{\perp}$ and $k_{3}$ are
expanded to second order in $a$, and increases again to 10 when the
components are expanded to third order.  The independent coefficients
at each order and their best-fit numerical values are listed in
Table~\ref{Case4_Kres}, while the fits themselves are displayed in
Fig.~\ref{Case4K}.  The error bars in this case are proportional to
the kick magnitudes $|{\bf k}|$ themselves unlike in Cases \#1 and
\#2.  This implies that our best-fit spin expansions will naturally
agree more closely with the smaller values of $|{\bf k}|$.
\begin{table}
  \begin{small}
    \begin{center}
      \begin{tabular}{|c|c|c|c|}
        \hline
        & 1st order & 2nd order & 3rd order \\
        \hline
        $\chi^2/{\rm d.o.f.}$ & $19.5$ & $12.8$ & $1.50$ \\
        \hline
        ${}^{c\!}K_{}^{(0)}$ & $6.33 \times 10^4$ & $3.19 \times 10^4$
        & $4.87 \times 10^5$ \\
        \hline
        ${}^{c\!}K_{}^{(1)}$ & --- & $-6.16 \times 10^4$ & $-3.03 
        \times 10^6$ \\
        \hline
        ${}^{c\!}K_{}^{(2)}$ & $2.75 \times 10^4$ & $7.08 \times 10^5$
        & $2.92 \times 10^6$ \\
        \hline
        ${}^{s\!}K_{}^{(2)}$ & --- & $2.81 \times 10^5$ & $2.53 
        \times 10^6$ \\
        \hline
        ${}^{c\!}K_{}^{(3)}$ & --- & $1.99 \times 10^5$ & $9.99 
        \times 10^6$ \\
        \hline
        ${}^{s\!}K_{}^{(3)}$ & --- & --- & $8.27 \times 10^6$ \\
        \hline
        ${}^{c\!}K_{}^{(4)}$ & --- & $-5.36 \times 10^5$ & $-2.95 
        \times 10^6$ \\
        \hline
        ${}^{s\!}K_{}^{(4)}$ & --- & --- & $-8.70 \times 10^5$ \\
        \hline
        ${}^{c\!}K_{}^{(5)}$ & --- & --- & $-6.53 \times 10^6$ \\
        \hline
        ${}^{c\!}K_{}^{(6)}$ & --- & --- & $-3.05 \times 10^6$ \\
        \hline
      \end{tabular}
    \end{center}
  \end{small}
  \caption{Fits of the squared kick magnitudes $|{\bf k}|^2$ listed in
    Table~\ref{Case4_table} to the fitting formula of
    Eq.(\ref{Case4kickfit}).  The first column shows the coefficients
    being fitted, while the second, third, and fourth columns list the
    numerical values for these coefficients when fits are performed to
    first, second, and third order in $a$ for the individual components
    of the kicks.  The magnitudes of the coefficients are given in
    units of (km/s)$^2$.}
  \label{Case4_Kres}
\end{table}

The angular dependence of the numerically determined kicks in this
configuration is highly non-trivial, and seems to challenge the linear
``Kidder'' kick formulae contemplated previously in the literature
\cite{Campanelli:2007cg,Herrmann:2007ex}.  These formulae can be
reconciled with the numerical results by recognizing that they are
linear not in the {\it initial} spins, but in the spins evaluated at
{\it merger}.  Both $\vec{V}_{\rm recoil}$ in Eq.~(1) of
\cite{Campanelli:2007cg} and ${\bf V}$ in Eq.~(5) of
\cite{Herrmann:2007ex} depend on the angles
\begin{equation} \label{KidAng}
  \Theta_{f}^{(i)} \equiv \cos^{-1} ({\bf \Sigma} \cdot {\bf m}^{(i)})
\end{equation}
between
\begin{equation} \label{Sigma}
  {\bf \Sigma} \equiv M \left( \frac{{\bf b}}{M_b} - \frac{{\bf a}}{M_a}
  \right)
\end{equation}
and the components ${\bf m}^{(i)}$ of an orthonormal triad defined at
merger.  Our triad $ \{ {\bf e}^{(i)} \}$ was defined at the beginning
of the simulation.  For configurations like Case \#4 that lack high
degrees of symmetry, the initial spins appearing in ${\bf \Sigma}$
will precess during the inspiral and the orientation of the triad $\{
{\bf m}^{(i)} \}$ will vary with respect to a fixed frame.  This
implies that the angles $\Theta_{f}^{(i)}$ appearing in the Kidder
formulae implicitly depend on the initial spins, introducing
non-linearity into the formulae.  We attempt to capture this non-linear
spin dependence explicitly in our spin expansions, allowing us to
construct fitting formulae that only depend on genuine {\it initial}
conditions.

The price we pay for making this spin dependence explicit is more
complicated nonlinear fitting formulae, as well as a more cumbersome
explicit procedure for relating expansions calibrated at different
initial stages of the inspiral.  This procedure for relating different
expansions will be provided in the second paper of this series.
An example of how our nonlinear fitting formulae
might arise from the linear Kidder formulae can be seen in the
configurations of Case \#4.  According to Eq.~(5) of
\cite{Herrmann:2007ex},
\begin{equation} \label{Kid3}
  k_3 \propto {\bf \Sigma} \cdot (K_n {\bf m}^{(1)} + K_k {\bf m}^{(2)}) \, .
\end{equation}
Post-Newtonian expansions \cite{Kidder:1995zr} reveal that to linear
order the triad $\{ {\bf m}^{(i)}({\bf a}, {\bf b}) \}$ for equal-mass
spinning BBHs is related to that in the non-spinning case by
\begin{equation} \label{spinlag}
  \left(\begin{array}{c} {\bf m}^{(1)}({\bf a}, {\bf b}) \\
    {\bf m}^{(2)}({\bf a}, {\bf b}) \\ {\bf m}^{(3)}({\bf a}, {\bf b})
  \end{array}\right)=
  \left(\begin{array}{ccc}
    \cos \Psi & \sin \Psi & 0 \\ -\sin \Psi & \cos \Psi & 0 \\
    0 & 0 & 1 \end{array}\right)
  \left(\begin{array}{c} {\bf m}^{(1)}(0,0) \\ {\bf m}^{(2)}(0,0)
    \\ {\bf m}^{(3)}(0,0)
  \end{array}\right)
\end{equation}
where $\Psi$ is linear in $a_3 + b_3$.  If we Taylor expand $\Psi$ in
Eq.~(\ref{spinlag}) then insert both Eqs.~(\ref{Sigma}) and
(\ref{spinlag}) into the Kidder formula (\ref{Kid3}), terms
prooprtional to $a_1 (a_3 + b_3)^n$ emerge.  We thus see one way in
which a formula linear in the spins at {\it merger} can become
nonlinear in the {\it initial} spins.

This indeed may be part of the story explaining the numerical results
in Case \#4.  The most surprising feature of the ``S-Series'' kicks is
the small kick velocity of 230 km/s at $\phi = 75^{\circ}$.  This spin
orientation is quite close to the ``superkick'' configuration ($\phi =
90^{\circ}$) at which one would naively expect the kicks to be
maximized.  While the {\it amplitude} of the Kidder formula for $k_3$
is indeed maximized at $\phi = 90^{\circ}$, Eq.~(\ref{Kid3}) predicts
that the superkick should have a sinusoidal dependence on the
azimuthal angle at merger as seen in Case \#2.  This angle will depend
on the total phase accumulated between the beginning of the simulation
and merger.  The numerical results for the S-Series listed in the
final column of Table~\ref{Case4_table} indicate that the duration of
the inspiral varies as $\cos \phi$ (linear in $a_3 + b_3$).  The
orbital frequency of equal-mass non-spinning BBHs at merger is about
$\omega \simeq 0.15 M^{-1}$ \cite{Buonanno:2006ui}, suggesting a final
orbital period $\tau = 2\pi/\omega \simeq 42 M$.  This estimate of the
period is quite comparable to the difference in merger times $177.3 M
- 138.6 M = 38.7 M$ between the successive peaks in $|{\bf k}|$ at
$\phi = 45^{\circ}$ and $\phi = 105^{\circ}$.  If this effect is
indeed responsible for the observed kicks in Case \#4, it helps to
explain which third order terms are more significant than in previous
cases.  Further simulations are necessary to determine if this
explanation is correct, and how best to account for it in our
formalism.

\subsection{Case \#5: Generically oriented initial spins}
\label{Case_5}

Finally, we consider the set of 8 equal-mass simulations published in
Tichy and Marronetti \cite{Tichy:2007hk}.  Each simulation has a
different initial spin configuration, and most of these configurations
have no particular symmetry.  The initial configurations are {\it not}
chosen according to a pattern, but are instead intended to be
``generic.''  In contrast to previous subsections, there is no natural
way to parameterize them in terms of one or two numbers.  This means
that, again in contrast to previous subsections, we cannot use
symmetries and degeneracies to significantly reduce the ``effective''
number of coefficients in the spin expansion.  We must therefore
truncate our spin expansions at an order for which the number of
independent terms is fewer than the number of simulated data points if
we hope to non-trivially test our formalism.

\begin{figure}
  \begin{center}
    \includegraphics[width=3.5in]{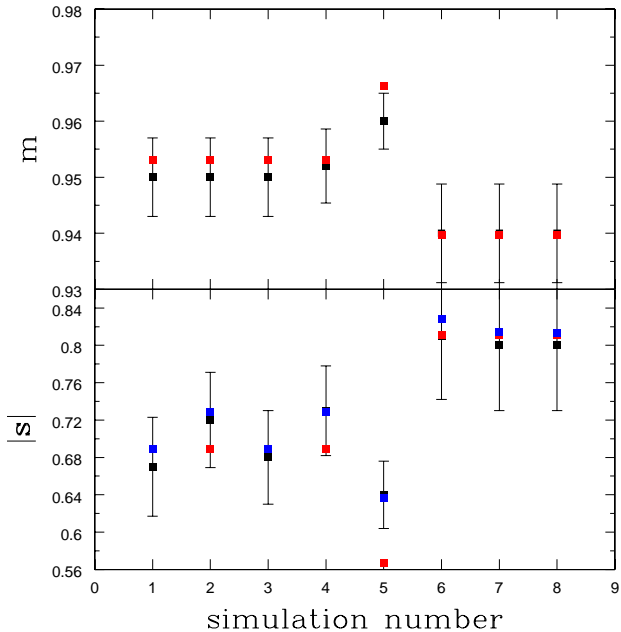}
  \end{center}
  \caption{Final masses $m$ and spin magnitudes $|{\bf s}|$ for the 8
    simulations $A_i$ of Case \#5 listed by simulation number $i$.
    The black points show the numerically determined values and
    $1\sigma$ error bars listed in Appendix~\ref{Case5TabApp}.  The
    red points in the top and bottom panels show the predictions of
    Eqs.~(\ref{Case5m}) and (\ref{Case5approx}) respectively.  These
    use the best-fit values of coefficients obtained in Case \#1 and
    fit {\it no} additional parameters.  The blue points in the bottom
    panel show the best-fit of Eq.~(\ref{Case5smag}) to $|{\bf s}|$,
    with coefficients listed in Table~\ref{Case5_results}.}
  \label{Case5res}
\end{figure}

For each of the 8 simulations in \cite{Tichy:2007hk}, the authors
quote the final kick {\it magnitude} $|{\bf k}|$, the final spin {\it
  magnitude} $|{\bf s}|$, and the final mass $m$.  Is this enough
information to test the spin expansion formalism?  The answer is
``no'' for $|{\bf k}|$, and ``yes'' for $|{\bf s}|$ and $m$.  In the
previous subsections, we have seen that in the general case we should
go to second- or even third-order in the spin expansion to achieve a
good fit for $|{\bf k}|$; but already at second-order, the independent
coefficients in the general spin expansion for $|{\bf k}|$ outnumber
the 8 values of $|{\bf k}|$ provided by \cite{Tichy:2007hk}.  Thus, we
cannot use the $|{\bf k}|$ data to perform a non-trivial test.  The
situation for $m$ (and even for $|{\bf s}|$) is better because, as we
have seen in previous subsections, the linear terms in the spin
expansion seem sufficient to provide a good fit to the data (except in
special symmetric cases where these linear terms vanish, so that the
second-order terms become important).  Let us test whether this simple
behavior continues to hold for the generic initial spin configurations
of \cite{Tichy:2007hk}.  At linear order, the mass and spin magnitude
are given by:
\begin{subequations}
  \label{Case_5_1st_order}
  \begin{eqnarray} \label{Case5m}
    m&\!=\!&\!m^{000|000}\!+\!m^{001|000}(a_3\!+\!b_3) \, , \\
    \label{Case5sperp}
    {\bf s}_{\!\perp}\!&\!=\!&\!{\bf s}_{\!\perp}^{100|000}
    (a_1\!+\!b_1)\!+\!{\bf s}_{\!\perp}^{010|000}(a_2\!+\!b_2) \, , \\
    \label{Case5s3}
    s_{3}^{}&\!=\!&\!s_{3}^{\;\,000|000}\!+\!s_{3}^{\;\,001|000}
    (a_3\!+\!b_3) \, , \\
    \label{Case5smag}
    |{\bf s}|&\!=\!&\!\sqrt{|{\bf s}_{\perp}|^2\!+\!s_{3}^2} \, .
  \end{eqnarray}
\end{subequations}
The four coefficients $\{m_{}^{000|000}, m_{}^{001|000},
  s_{3}^{000|000}, s_{3}^{001|000}\}$ were already accurately
  determined by the large set of simulations in Case \#1.  We
therefore fix these coefficients to the values listed in
Table~\ref{zspin_results}.  Eq.~(\ref{Case5m}) thus predicts the final
mass $m$ {\it with zero free parameters}.  If we Taylor expand
the square root in Eq.~(\ref{Case5smag}) for $|{\bf s}|$, keeping
terms up to linear order in the initial spins, we obtain
\begin{equation} \label{Case5approx}
  |{\bf s}| \approx s_{3}^{000|000} + s_{3}^{001|000} (a_3 + b_3)
\end{equation}
which again has {\it no free parameters}.  Alternatively, if we
  do {\it not} Taylor expand the square root in Eq.~(\ref{Case5smag}),
  then the expression for $|{\bf s}|$ has three free parameters: the
magnitudes of the coefficients ${\bf s}_{\perp}^{100|000}$ and
${\bf s}_{\perp}^{010|000}$ and the angle $\Theta$ between them.  In
Table~\ref{Case5_results}, we list the $\chi^2/{\rm d.o.f.}$ and the
best-fit values of any free parameters from fitting to
Eqs.~(\ref{Case5m}), (\ref{Case5approx}), and (\ref{Case5smag}).  The
fits themselves are shown in Fig.~\ref{Case5res}.
\begin{table}
  \begin{center}
    \begin{tabular}{|c|l|}
      \hline
      Quantity & Results \\
      \hline
      $m$ & $\chi^2/{\rm d.o.f.} = 0.275$ \\
      Eq.~(\ref{Case5m}) & 0 free parameters \\
      \hline
      $|{\bf s}|$ & $\chi^2/{\rm d.o.f.} = 0.671$ \\
      Eq.~(\ref{Case5approx}) & 0 free parameters \\
      \hline
      $|{\bf s}|$ & $\chi^2/{\rm d.o.f.} = 0.0707$ \\
      Eq.~(\ref{Case5smag}) & 3 free parameters: \\
      & $|{\bf s}_{\perp}^{100|000}| \approx 0.1871$ \\
      & $|{\bf s}_{\perp}^{010|000}| \approx 0.2086$ \\
      & $\cos \Theta \approx -0.974$ \\
      \hline
    \end{tabular}
  \end{center}
  \caption{Fits to the final masses $m$ and spin magnitudes $|{\bf s}|$ of the
    simulations listed in Appendix~\ref{Case5TabApp}.  The first row lists the
    $\chi^2/{\rm d.o.f.}$ of fits of Eq.~(\ref{Case5m}); we use the values of
    the coefficients determined in Case \#1 leaving {\it no} free parameters.
    The second row shows fits of Eq.~(\ref{Case5approx}) to $|{\bf s}|$; using
    the appropriate coefficients determined from Case \#1 again leaves no free
    parameters.  The third row fits the full formula of Eq.~(\ref{Case5smag})
    to $|{\bf s}|$; we list the $\chi^2/{\rm d.o.f.}$ and best-fit values 
    of the 3 new parameters.}
  \label{Case5_results}
\end{table}

The goodness of these fits speaks for itself.  The most remarkable
result is that Eqs.~(\ref{Case5m}) and (\ref{Case5approx}) --- which
have no free parameters and only depend on $a_3 + b_3$ --- do an
excellent job (red points in Fig.~\ref{Case5res}).  For all their
importance in generating the ``superkicks,'' the components of the
initial spins in the orbital plane $\{a_{1}^{}, a_{2}^{}, b_{1}^{},
b_{2}^{}\}$ seem to have little effect on the final masses and spins.
Only one simulation (\#5) is fit poorly by the red points; not
surprisingly it is the configuration with the {\it largest} spin
projection in the orbital plane.  The red points {\it overpredict} $m$
by failing to account for the energy carried away by the gravitational
radiation sourced by the planar spins, and {\it underpredict} $|{\bf
  s}|$ by neglecting the contribution of ${\bf s}_{\perp}$.  These
results are provocative, but remain provisional until verified by
further simulations.

\section{Calibrating the spin expansion with new simulations}
\label{S:prog}

In this section, we suggest a relatively small set of simulations
($10$ equal-mass simulations, and $16$ unequal-mass simulations) with
initial spin configurations specially chosen to allow {\it all} of the
spin-expansion coefficients up to second order to be determined
uniquely.  Once the spin expansion is calibrated in this way, it
becomes fully predictive --- {\it i.e.}\ it predicts the simulated
observables (to second-order accuracy), for {\it any} initial spin
configuration.

This section is organized into four subsections.
Subsection~\ref{S:degen} explains why currently available simulations
are {\it not} sufficient to calibrate the spin expansion.
Subsections~\ref{S:10sims} and \ref{S:16sims} suggest an explicit
choice of 10 equal-mass and 16 unequal-mass simulations with initial
spin configurations suitable for this calibration, and provide
corresponding formulae for the spin expansion coefficients in terms of
the results of these simulations.  Subsection~\ref{S:comments}
collects several additional comments for readers who are interested in
pursuing or extending the program suggested here.  The {\it
importance} of performing these simulations is discussed later in
Sec.~\ref{S:disc}.

In Appendix~\ref{S:MVE}, we generalize the calibration procedure to
take account of the inevitable systematic errors arising from the
simulations themselves and from truncating the spin expansion at
finite order.  In particular, given $N$ simulations of known
covariance, we derive the minimum-variance unbiased estimators for the
spin expansion coefficients, as well as the corresponding covariance
matrix for these estimators.

\subsection{Breaking degeneracies}
\label{S:degen}

In the previous section, we examined in detail the predictions of the
spin expansion for the 5 different configurations (Case \#1 through
Case \#5) that have been simulated to date.  Although at first
glance it might seem that the total number of currently available
simulations is more than sufficient to calibrate all of the
expansion coefficients up to second order, in practice we had to
perform new fits for each of the 5 configurations.  These new fits
were required for three distinct reasons.  Firstly, many published
works have only provided the final kick and spin {\it magnitudes}
$|{\bf k}|$ and $|{\bf s}|$, not the individual components $k_{i}^{}$
and $s_{i}^{}$.  Combining the spin expansions for individual
components to predict final magnitudes introduces degeneracies between
terms.

Secondly, many of the available simulation sets study the {\it same}
highly symmetric configurations --- like spins aligned with the
orbital angular momentum (Case \#1), or the superkick configuration
(Case \#2).  These particular configurations have justifiably garnered
much of the attention, both because of their astrophysical interest
and because they cleanly illustrate several key features of spinning
BBH merger.  Nevertheless, since the initial spin components in
these configurations are either purely aligned with the orbital
angular momentum (Case \#1) or purely perpendicular (Case \#2), they
fail to constrain the coupling between aligned and perpendicular spin
components allowed by symmetry beyond linear order.  We found that
such terms were required to attain a good fit to the kicks of Case
\#4, suggesting that further simulations are indeed necessary to
determine the coefficients of these terms.

Thirdly, different groups performed their simulations at {\it
different} values of the initial dimensionless orbital separation
$r/(M_{a}^{}+ M_{b}^{})$.  Recall however from Sec.~\ref{S:formalism}
that our spin-expansion coefficients are defined at a {\it fixed}
value of $r/(M_{a}^{}+M_{b})$ --- or, more generally, at a fixed value
of the dimensionless inspiral parameter $\psi$.  While post-Newtonian
techniques may be able to relate the spin-expansion coefficients at
different values of $\psi$, doing so will add an additional layer of
complication and potential systematic error to the calibration
process.  We treat this issue in a forthcoming paper.
  
As {\it existing} simulations are insufficient to fully constrain the
spin expansion, in the following subsections we explicitly provide the
initial spin configurations for a small number of {\it new}
simulations with which we will be able to achieve the desired
calibration.  Since our spin-expansion coefficients remain functions
of the mass ratio $q$, a new set of simulations is required in
principle at {\it each} value of $q$.  In practice, we hope that the
$q$-dependence of our coefficients is sufficiently smooth that we can
interpolate between coefficients calibrated at a small set of mass
ratios.

\subsection{Equal-mass ($q=1$) BBH mergers}
\label{S:10sims}

Final quantities for equal-mass ($q=1$) BBH mergers are characterized
by their eigenvalues $\pm1$ under parity $P$ and exchange $X$.  In
this subsection, we will use the 4 variables $\{w,x,y,z\}$ to denote
generic final quantities with the 4 possible combinations of these
eigenvalues as shown in Table~\ref{T:wxyz}.
\begin{table}
  \begin{center} \begin{tabular}{|c||c|c|c|c|} 
  \hline 
  & $w$ & $x$ & $y$ & $z$ \\ 
  \hline\hline 
  $P$ & $+1$ & $+1$ & $-1$ & $-1$ \\ 
  \hline 
  $X$ & $+1$ & $-1$ & $+1$ & $-1$ \\ 
  \hline 
  {\it e.g.}\ & $m$, $s_{3}^{}$ & $k_{1}^{}$, $k_{2}^{}$ &
  $s_{1}^{}$, $s_{2}^{}$ & $\;\;\;k_{3}^{}\;\;\;$ \\
  \hline
  \end{tabular} \end{center}
  \caption{Eigenvalues under $P$ and $X$ for the generic final
    quantities $w$, $x$, $y$, and $z$, in subsection \ref{S:10sims},
    along with examples of physical quantities with these 
    transformation properties.}
  \label{T:wxyz}
\end{table}

To second order in the spin expansion, $w$, $x$, $y$, and $z$ are
given by:
\begin{subequations}
  \label{wxyz_2nd_order}
  \begin{eqnarray}
    \label{w_2nd_order}
    w\!&\!\!=\!\!&\!w_{}^{000|000} \nonumber\\
    \!&\!\!+\!\!&\!w_{}^{001|000}(a_{3}^{}\!+\!b_{3}^{})
    \!+\!w_{}^{002|000}(a_{3}^{2}\!+\!b_{3}^{2}) \nonumber\\
    \!&\!\!+\!\!&\!w_{}^{200|000}(a_{1\!}^{2}\!+\!b_{1}^{2})
    \!+\!w_{}^{020|000}(a_{2\!}^{2}\!+\!b_{2}^{2}) \nonumber\\
    \!&\!\!+\!\!&\!w_{}^{110|000}(a_{1\!}^{}a_{2}^{}
    \!+\!b_{1\!}^{}b_{2}^{})\!+\!w_{}^{100|010}(a_{1\!}^{}b_{2\!}^{}
    \!+\!b_{1\!}^{}a_{2}^{}) \nonumber\\
    \!&\!\!+\!\!&\!w_{}^{100|100}a_{1\!}^{}b_{1\!}^{}
    \!+\!w_{}^{010|010}a_{2}^{}b_{2\!}^{}
    \!+\!w_{}^{001|001}a_{3}^{}b_{3}^{}\quad \\
    \label{x_2nd_order}
    x\!&\!\!=\!\!&\!x_{}^{\;\!001|000}(a_{3}^{}\!-\!b_{3}^{})
    \!+\!x_{}^{\;\!002|000}(a_{3}^{2}\!-\!b_{3}^{2}) \nonumber\\
    \!&\!\!+\!\!&\!x_{}^{\;\!200|000}(a_{1\!}^{2}\!-\!b_{1}^{2})
    \!+\!x_{}^{\;\!020|000}(a_{2\!}^{2}\!-\!b_{2}^{2}) \nonumber\\
    \!&\!\!+\!\!&\!x_{}^{\;\!110|000}(a_{1\!}^{}a_{2\!}^{}
    \!-\!b_{1\!}^{}b_{2}^{})\!+\!x_{}^{\;\!100|010}
    (a_{1\!}^{}b_{2\!}^{}\!-\!b_{1\!}^{}a_{2}^{}) \\
    \label{y_2nd_order}
    y\!&\!\!=\!\!&\!y_{}^{100|000}(a_{1\!}^{}\!+\!b_{1}^{})
    \!+\!y_{}^{010|000}(a_{2}^{}\!+\!b_{2}^{}) \nonumber\\
    \!&\!\!+\!\!&\!
    y_{}^{101|000}(a_{1\!}^{}a_{3}^{}\!+\!b_{\;\!1\!}^{}b_{\;\!3}^{})\!+\!
    y_{}^{011|000}(a_{2}^{}a_{3}^{}\!+\!b_{\;\!2}^{}b_{\;\!3}^{}) \nonumber\\
    \!&\!\!+\!\!&\!
    y_{}^{100|001}(a_{1\!}^{}b_{\;\!3}^{}\!+\!b_{\;\!1\!}^{}a_{3}^{})\!+\!
    y_{}^{010|001}(a_{2}^{}b_{\;\!3}^{}\!+\!b_{\;\!2}^{}a_{3}^{})\quad \\
    \label{z_2nd_order}
    z\!&\!\!=\!\!&\!z_{}^{100|000}(a_{1\!}^{}\!-\!b_{1}^{})
    \!+\!z_{}^{010|000}(a_{2}^{}\!-\!b_{2}^{}) \nonumber\\
    \!&\!\!+\!\!&\!
    z_{}^{101|000}(a_{1\!}^{}a_{3}^{}\!-\!b_{\;\!1\!}^{}b_{\;\!3}^{})\!+\!
    z_{}^{011|000}(a_{2}^{}a_{3}^{}\!-\!b_{\;\!2}^{}b_{\;\!3}^{}) \nonumber\\
    \!&\!\!+\!\!&\!
    z_{}^{100|001}(a_{1\!}^{}b_{\;\!3}^{}\!-\!b_{\;\!1\!}^{}a_{3}^{})\!+\!
    z_{}^{010|001}(a_{2}^{}b_{\;\!3}^{}\!-\!b_{\;\!2}^{}a_{3}^{})\quad
  \end{eqnarray}
\end{subequations}
Note that the expansions for $x$, $y$, and $z$ each contain 6
coefficients, while the $w$ expansion has 10 coefficients.

In Table~\ref{T:10sims}, we suggest a set of 10 simulations designed
to determine all the coefficients of Eq.~(\ref{wxyz_2nd_order}) at the
minimum computational cost.  We name these 10 configurations $\{{\rm
  eq}0,\ldots,{\rm eq}9\}$.  The $r$th simulation ($r=0,\ldots,9$) has
initial spin components $\{a_{i}^{(r)}, b_{i}^{(r)}\}$, and
leads to final-state values $w=w_{}^{(r)}$, $x=x_{}^{(r)}$,
$y=y_{}^{(r)}$, and $z=z_{}^{(r)}$.
\begin{table}
  \begin{center} 
    \begin{tabular}{|l|l|} 
      \hline 
      \multicolumn{2}{|c|}{{\rm eq}0: none} \\ 
      \hline 
      {\rm eq}1: $a_{1}^{}$ & {\rm eq}2: $a_{2}^{}$ \\ 
      \hline 
      {\rm eq}3: $a_{1}^{}$, $a_{3}^{}$ & 
      {\rm eq}4: $b_{\;\!1}^{}$, $a_{3}^{}$ \\ 
      \hline 
      {\rm eq}5: $a_{1}^{}$, $a_{2}^{}$, $a_{3}^{}$ & 
      {\rm eq}6: $a_{1}^{}$, $b_{\;\!2}^{}$, $a_{3}^{}$ \\ 
      \hline 
      \multicolumn{2}{|c|}{{\rm eq}7: $a_{1}^{}$, $b_{\;\!1}^{}$} \\ 
      \hline 
      \multicolumn{2}{|c|}{{\rm eq}8: $a_{2}^{}$, $b_{\;\!2}^{}$} \\ 
      \hline 
      \multicolumn{2}{|c|}{{\rm eq}9: $a_{3}^{}$, $b_{\;\!3}^{}$} \\ 
      \hline
      \multicolumn{2}{|c|}{{\rm N.B.} $a_{3}^{(3)}\neq a_{3}^{(4)}$} \\
      \hline 
    \end{tabular}
  \end{center} 
  \caption{A suggested set of 10 simulations (denoted eq0 through eq9)
    which can simultaneously determine all of the coefficients 
    up to second order in the general spin expansions, 
    Eqs.~(\ref{wxyz_2nd_order}), for equal-mass ($q=1$) BBH merger.
    For each simulation, we list the non-vanishing 
    initial spin components.  These spin components may all be chosen 
    independently apart from the requirement that the values of $a_3$ 
    differ between simulations eq3 and eq4 ($a_{3}^{(3)}\neq 
    a_{3}^{(4)}$).  While many different sets of configurations satisfy
    this requirement, it is particularly convenient to choose the set 
    given by Eq.~(\ref{relations_10sims}).}
  \label{T:10sims}
\end{table}

In each of the proposed simulations, only those initial spin
components listed in Table~\ref{T:10sims} are non-zero.  Although all
10 of the simulations $\{{\rm eq}0,\ldots,{\rm eq}9\}$ are necessary
to determine the $w$ expansion coefficients, only the 6 simulations
$\{{\rm eq}1,\ldots,{\rm eq}6\}$ are required to determine the $x$,
$y$, and $z$ coefficients.  Thus, if one judges that a 10-simulation
set is too expensive for one's computational budget, the 6 simulations
$\{{\rm eq}1,\ldots,{\rm eq}6\}$ provide a less ambitious but still
useful alternative (for example, they fully calibrate the expansion
for the kick components $k_{i}^{}$).

The simulations in Table \ref{T:10sims} lift all degeneracies up to
second order in the initial spins.  {\it Any} choice for the
values of the non-vanishing initial spin components $\{
a_{i}^{(r)}, b_{i}^{(r)}\}$ in this table will yield a unique
solution for all of the coefficients, provided $a_{3}^{(3)}\neq
a_{3}^{(4)}$.  We can use this additional freedom to make the final
expressions (\ref{wxyz_inversion}) for the expansion coefficients in
terms of the simulated observables $\{ w_{}^{(r)}, x_{}^{(r)},
y_{}^{(r)}, z_{}^{(r)} \}$ as algebraically simple as possible.
In particular, if we choose the initial non-vanishing spin
components in Table \ref{T:10sims} to satisfy
\begin{equation}
  \label{relations_10sims}
  \begin{array}{rcl}
    \alpha_{1}^{}\!&\equiv&\!+a_{1}^{(1)}\!=\!+a_{1}^{(3)}
    \!=\!+b_{1}^{(4)}\!=\!+a_{1}^{(5)}\!=\!+a_{1}^{(6)} \, , \\
    \alpha_{2}^{}\!&\equiv&\!+a_{2}^{(2)}\!=\!+a_{2}^{(5)}
    \!=\!+b_{2}^{(6)} \, , \\
    \alpha_{3}^{}\!&\equiv&\!+a_{3}^{(3)}\!=\!-a_{3}^{(4)}
    \!=\!+a_{3}^{(5)}\!=\!+a_{3}^{(6)} \, ,
  \end{array}
\end{equation}
then Eqs.~(\ref{wxyz_2nd_order}) can be inverted to yield the simple
result:
\begin{subequations}
  \label{wxyz_inversion}
  \begin{eqnarray}
    \label{w_inversion}
    \!\!&\!\!\!\!&\!\!\begin{array}{rcl}
      w_{}^{000|000}\!&\!\!=\!\!&\!w_{}^{(0)} \\
      w_{}^{200|000}\!&\!\!=\!\!&\!\bar{w}_{}^{(1)}/\alpha_{1}^{2} \\
      w_{}^{020|000}\!&\!\!=\!\!&\!\bar{w}_{}^{(2)}/\alpha_{2}^{2} \\
      w_{}^{110|000}\!&\!\!=\!\!&\![\bar{w}_{}^{(5)}\!-\!
      \bar{w}_{}^{(3)}\!-\!\bar{w}_{}^{(2)}]/\alpha_{1}^{}\alpha_{2}^{} \\
      w_{}^{100|010}\!&\!\!=\!\!&\![\bar{w}_{}^{(6)}\!-\!
      \bar{w}_{}^{(3)}\!-\!\bar{w}_{}^{(2)}]/\alpha_{1}^{}\alpha_{2}^{} \\
      w_{}^{002|000}\!&\!\!=\!\!&\![\bar{w}_{}^{(4)}\!+\!
      \bar{w}_{}^{(3)}\!-\!2\bar{w}_{}^{(1)}]/2\alpha_{3}^{2} \\
      w_{}^{001|000}\!&\!\!=\!\!&\![\bar{w}_{}^{(3)}\!-\!
      \bar{w}_{}^{(4)}]/2\alpha_{3}^{} \\
      w_{}^{100|100}\!&\!\!=\!\!&\![\bar{w}_{}^{(7)}\!-\!
      2\bar{w}_{}^{(1)}]/\alpha_{1}^{2} \\
      w_{}^{010|010}\!&\!\!=\!\!&\![\bar{w}_{}^{(8)}\!-\!
      2\bar{w}_{}^{(2)}]/\alpha_{2}^{2} \\
      w_{}^{001|001}\!&\!\!=\!\!&\![\bar{w}_{}^{(9)}\!-\!
      2\bar{w}_{}^{(3)}\!+\!2\bar{w}_{}^{(1)}]/\alpha_{3}^{2}
    \end{array} \\ 
    \label{x_inversion}
    \!\!&\!\!\!\!&\!\!\begin{array}{rcl}
      x_{}^{200|000}\!&\!\!=\!\!&\!x_{}^{(1)}/\alpha_{1}^{2} \\
      x_{}^{020|000}\!&\!\!=\!\!&\!x_{}^{(2)}/\alpha_{2}^{2} \\
      x_{}^{110|000}\!&\!\!=\!\!&\![x_{}^{(5)}\!-\!x_{}^{(3)}\!-\!
      x_{}^{(2)}]/\alpha_{1}^{}\alpha_{2}^{} \\
      x_{}^{100|010}\!&\!\!=\!\!&\![x_{}^{(6)}\!-\!x_{}^{(3)}\!+\!
      x_{}^{(2)}]/\alpha_{1}^{}\alpha_{2}^{} \\
      x_{}^{002|000}\!&\!\!=\!\!&\![x_{}^{(3)}\!+\!x_{}^{(4)}]
      /2\alpha_{3}^{2} \\
      x_{}^{001|001}\!&\!\!=\!\!&\![x_{}^{(3)}\!-\!x_{}^{(4)}\!-\!
      2x_{}^{(1)}]/2\alpha_{3}^{}
    \end{array}
  \end{eqnarray}
  \begin{eqnarray}
    \label{y_inversion}
    \!\!&\!\!\!\!&\!\!\begin{array}{rcl}
      y_{}^{100|000}\!&\!\!=\!\!&\!y_{}^{(1)}/\alpha_{1}^{} \\
      y_{}^{010|000}\!&\!\!=\!\!&\!y_{}^{(2)}/\alpha_{2}^{} \\
      y_{}^{101|000}\!&\!\!=\!\!&\!+[y_{}^{(3)}\!-\!y_{}^{(1)}]
      /\alpha_{1}^{}\alpha_{3}^{} \\
      y_{}^{100|001}\!&\!\!=\!\!&\!-[y_{}^{(4)}\!-\!y_{}^{(1)}]
      /\alpha_{1}^{}\alpha_{3}^{} \\
      y_{}^{011|000}\!&\!\!=\!\!&\!+[y_{}^{(5)}\!-\!y_{}^{(3)}
      \!-\!y_{}^{(2)}]/\alpha_{2}^{}\alpha_{3}^{} \\
      y_{}^{010|001}\!&\!\!=\!\!&\!+[y_{}^{(6)}\!-\!y_{}^{(3)}
      \!-\!y_{}^{(2)}]/\alpha_{2}^{}\alpha_{3}^{}
    \end{array} \\
    \label{z_inversion}
    \!\!&\!\!\!\!&\!\!\begin{array}{rcl}
      z_{}^{100|000}\!&\!\!=\!\!&\!z_{}^{(1)}/\alpha_{1}^{} \\
      z_{}^{010|000}\!&\!\!=\!\!&\!z_{}^{(2)}/\alpha_{2}^{} \\
      z_{}^{101|000}\!&\!\!=\!\!&\!+[z_{}^{(3)}\!-\!z_{}^{(1)}]
      /\alpha_{1}^{}\alpha_{3}^{} \\
      z_{}^{100|001}\!&\!\!=\!\!&\!+[z_{}^{(4)}\!+\!z_{}^{(1)}]
      /\alpha_{1}^{}\alpha_{3}^{} \\
      z_{}^{011|000}\!&\!\!=\!\!&\!+[z_{}^{(5)}\!-\!z_{}^{(3)}
      \!-\!z_{}^{(2)}]/\alpha_{2}^{}\alpha_{3}^{} \\
      z_{}^{010|001}\!&\!\!=\!\!&\!-[z_{}^{(6)}\!-\!z_{}^{(3)}
      \!+\!z_{}^{(2)}]/\alpha_{2}^{}\alpha_{3}^{}
    \end{array}
  \end{eqnarray}
\end{subequations}
where we have defined $\bar{w}_{}^{(r)}\equiv w_{}^{(r)} -
w_{}^{000|000}$.  

Once Eqs.~(\ref{wxyz_2nd_order}) have been calibrated with the 10
simulations $\{{\rm eq}0,\ldots,{\rm eq}9\}$, they can predict the
results of {\it any} additional equal-mass simulations to second-order
accuracy.

\subsection{Unequal-mass ($q \neq 1$) BBH mergers}
\label{S:16sims}

Calibrating the spin-expansion coefficients for unequal-mass
($q\neq1$) BBH mergers proceeds similarly to the equal-mass case
discussed in the previous subsection.

Without loss of generality, we can choose $0<q<1$, since exchange
symmetry $X$ relates coefficients for mass ratios $q$ and $1/q$.
However, at a fixed value of $q$ ($\neq\!1$), exchange $X$ no longer
restricts the terms appearing in the expansions.  So, for the purposes
of this section, there are only two types of final quantities: scalars
and pseudoscalars.  Scalars (like $\{m, k_{1}^{}, k_{2}^{},
s_{3}^{}\}$) will be represented by the variable $u$, while
pseudoscalars (like $\{s_{1}^{},s_{2}^{},k_{3}^{}\}$) will be
represented by $v$.

To second order in the spin expansion, $u$ and $v$ are given by
\begin{subequations}
  \label{uv_2nd_order}
  \begin{eqnarray}
    u\!&\!\!=\!\!&\!u_{}^{000|000} \nonumber\\
    \!&\!\!+\!\!&\!u_{}^{200|000}a_{1}^{2}\!+\!
    u_{}^{000|200}b_{1}^{2}\!+\!u_{}^{020|000}a_{2}^{2}\!+\!
    u_{}^{000|020}b_{2}^{2} \nonumber\\
    \!&\!\!+\!\!&\!u_{}^{001|000}a_{3}^{}\!+\!u_{}^{000|001}b_{3}^{}
    \!+\!u_{}^{002|000}a_{3}^{2}\!+\!u_{}^{000|002}b_{3}^{2} \nonumber\\
    \!&\!\!+\!\!&\!u_{}^{110|000}a_{1\!}^{}a_{2}^{}
    \!\!+\!u_{}^{000|110}b_{1\!}^{}b_{2}^{}
    \!\!+\!u_{}^{100|010}a_{1\!}^{}b_{2}^{}
    \!\!+\!u_{}^{010|100}b_{1\!}^{}a_{2}^{} \nonumber\\
    \label{u_2nd_order}
    \!&\!\!+\!\!&\!u_{}^{100|100}a_{1\!}^{}b_{1}^{}\!\!+\!u_{}^{010|010}
    a_{2}^{}b_{2}^{}\!\!+\!u_{}^{001|001}a_{3}^{}b_{3}^{}, \\ 
    v\!&\!\!=\!\!&\!v_{}^{100|000}a_{1}^{}\!+\!v_{}^{000|100}b_{1}^{}
    \!+\!v_{}^{010|000}a_{2}^{}\!+\!v_{}^{000|010}b_{2}^{} \nonumber\\
    \!&\!\!+\!\!&\!v_{}^{101|000}a_{1\!}^{}a_{3}^{}
    \!\!+\!v_{}^{000|101}b_{1\!}^{}b_{3}^{}
    \!\!+\!v_{}^{011|000}a_{2}^{}a_{3}^{}  
    \!\!+\!v_{}^{000|011}b_{2}^{}b_{3}^{} \nonumber\\
    \label{v_2nd_order}
    \!&\!\!+\!\!&\!v_{}^{100|001}a_{1\!}^{}b_{3}^{}
    \!\!+\!v_{}^{001|100}b_{1\!}^{}a_{3}^{}
    \!\!+\!v_{}^{010|001}a_{2}^{}b_{3}^{}
    \!\!+\!v_{}^{001|010}b_{2}^{}a_{3}^{}. \nonumber\\
  \end{eqnarray}
\end{subequations}
Note that the expansion for $u$ contains 16 coefficients, while the
expansion for $v$ contains 12 coefficients.

\begin{table}
  \begin{center}
    \begin{tabular}{|l|l|} 
      \hline
      \multicolumn{2}{|c|} {{\rm uneq}0: none} \\ 
      \hline 
      {\rm uneq}1: $\;\;a_{1}^{}$ & {\rm uneq}2: $\;\;a_{2}^{}$ \\ 
      \hline 
      {\rm uneq}3: $\;\;b_{\!\;1}^{}$ & {\rm uneq}4: $\;\;b_{\;\!2}^{}$ \\ 
      \hline 
      {\rm uneq}5: $\;\;a_{1}^{}$, $a_{3}^{}$ & 
      {\rm uneq}6: $\;\;a_{2}^{}$, $a_{3}^{}$ \\ 
      \hline 
      {\rm uneq}7: $\;\;b_{\;\!1}^{}$, $b_{\;\!3}^{}$ & 
      {\rm uneq}8: $\;\;b_{\;\!2}^{}$, $b_{\;\!3}^{}$ \\ 
      \hline 
      {\rm uneq}9: $\;\;a_{1}^{}$, $b_{\;\!1}^{}$, $a_{3}^{}$ & 
      {\rm uneq}10: $a_{2}^{}$, $b_{\;\!2}^{}$, $a_{3}^{}$ \\ 
      \hline 
      {\rm uneq}11: $a_{1}^{}$, $b_{\;\!2}^{}$, $b_{\;\!3}^{}$ & 
      {\rm uneq}12: $b_{\;\!1}^{}$, $a_{2}^{}$, $b_{\;\!3}^{}$ \\ 
      \hline 
      \multicolumn{2}{|c|}{{\rm uneq}13: $a_{1}^{}$, $a_{2}^{}$} \\ 
      \hline
      \multicolumn{2}{|c|}{{\rm uneq}14: $b_{\;\!1}^{}$, $b_{\;\!2}^{}$} \\ 
      \hline 
      \multicolumn{2}{|c|}{{\rm uneq}15: $a_{3}^{}$, $b_{\;\!3}^{}$} \\ 
      \hline 
      \multicolumn{2}{|c|}{{\rm N.B.} $a_{3}^{(5)}\neq a_{3}^{(6)}$, 
      $b_{3}^{(7)}\neq b_{3}^{(8)}$} \\
      \hline
    \end{tabular}
  \end{center}
  
  \caption{A suggested set of 16 unequal-mass ($q\neq1$) simulations
  that constitute a minimum set necessary to calibrate the
  spin-expansion coefficients in Eq.~(\ref{uv_2nd_order}).  
  For each simulation, we list the non-vanishing initial spin 
  components.  These spin components may all be chosen 
  independently, apart from the constraints $a_{3}^{(5)}\neq 
  a_{3}^{(6)}$ and $b_{3}^{(7)}\neq b_{3}^{(8)}$, but it is 
  particularly convenient to choose components that satisfy 
  Eq.~(\ref{relations_16sims}).}  \label{T:16sims}
\end{table}

In Table~\ref{T:16sims}, we suggest a minimal set of 16
simulations needed to simulataneously determine all of the
coefficients in Eqs.~(\ref{uv_2nd_order}).  We have named these
configurations $\{{\rm uneq}0,\ldots,{\rm uneq}15\}$.  As in the
equal-mass case, the $r$th simulation ($r=0,\ldots,15$) has initial
spin components $\{a_{i}^{(r)}, b_{i}^{(r)}\}$, and leads to
final simulated observables $u=u_{}^{(r)}$ and $v=v_{}^{(r)}$.
In each of the proposed simulations, only the initial spin components
listed in Table \ref{T:16sims} are non-zero.  While all 16 of the
simulations $\{{\rm uneq}0,\ldots,{\rm uneq}15\}$ are required to
determine the $u$ expansion coefficients, the 12 simulations $\{{\rm
  uneq}1,\ldots, {\rm uneq}12\}$ suffice to calibrate the $v$
coefficients.  If one is only interested in final quantities with odd
parity (like $s_1$, $s_2$, and $k_3$), one can reduce the computing
time by $\sim25\%$ by only performing the 12 simulations $\{{\rm
  uneq}1,\ldots, {\rm uneq}12\}$.

The simulations in Table \ref{T:16sims} yield a unique solution for
all of the coefficients in Eqs.~(\ref{uv_2nd_order}), as long as one
chooses $a_{3}^{(5)}\neq a_{3}^{(6)}$ and $b_{3}^{(7)}\neq
b_{3}^{(8)}$.  As before, one can choose the values
$\{a_{i}^{(r)}, b_{i}^{(r)}\}$ appearing in this table to make
the inversion of Eqs.~(\ref{uv_2nd_order}) particularly simple.  If
these initial spin components satisfy
\begin{equation}
  \label{relations_16sims}
  \begin{array}{rllllll}
    \alpha_{1}^{}\!\! &
      \!\equiv+a_{1}^{(1)}\!\! &
      \!\!=\!+a_{1}^{(5)}\!\! &
      \!\!=\!+a_{1}^{(\;9\;)}\!\! &
      \!\!=\!+a_{1}^{(11)}\!\! &
      \!\!=\!+a_{1}^{(13)} \\
    \alpha_{2}^{}\!\! &
      \!\equiv+a_{2}^{(2)}\!\! &
      \!\!=\!+a_{2}^{(6)}\!\! &
      \!\!=\!+a_{2}^{(10)}\!\! & 
      \!\!=\!+a_{2}^{(12)}\!\! & 
      \!\!=\!+a_{2}^{(13)} \\
    \alpha_{3}^{}\!\! & 
      \!\equiv+a_{3}^{(5)}\!\! &
      \!\!=\!-a_{3}^{(6)}\!\! &
      \!\!=\!+a_{3}^{(\;9\;)}\!\! &
      \!\!=\!-a_{3}^{(10)}\!\! &
      \!\!=\!+a_{3}^{(15)} \\
    \beta_{1}^{}\!\! &
      \!\equiv+b_{1}^{\;\!(3)}\!\! &
      \!\!=\!+b_{1}^{\;\!(7)}\!\! &
      \!\!=\!+b_{1}^{(\;9\;)}\!\! &
      \!\!=\!+b_{1}^{\;\!(12)}\!\! &
      \!\!=\!+b_{1}^{\;\!(14)} \\
    \beta_{2}^{}\!\! &
      \!\equiv+b_{2}^{\;\!(4)}\!\! &
      \!\!=\!+b_{2}^{\;\!(8)}\!\! &
      \!\!=\!+b_{2}^{\;\!(10)}\!\! &
      \!\!=\!+b_{2}^{\;\!(11)}\!\! &
      \!\!=\!+b_{2}^{\;\!(14)} \\
    \beta_{3}^{}\!\! &
      \!\equiv+b_{3}^{\;\!(7)}\!\! &
      \!\!=\!-b_{3}^{\;\!(8)}\!\! &
      \!\!=\!-b_{3}^{\;\!(11)}\!\! &
      \!\!=\!+b_{3}^{\;\!(12)}\!\! &
      \!\!=\!+b_{3}^{\;\!(15)} \, ,
  \end{array}
\end{equation}
the inverted equations for the spin-expansion coefficients take the
comparatively simple form
\begin{subequations}
  \label{uv_inversion} 
  \begin{equation} \label{u_inversion}
  \begin{array}{rcl}
  u_{}^{000|000}\!&\!\!=\!\!&\!u_{}^{(0)} \\
  u_{}^{200|000}\!&\!\!=\!\!&\!\bar{u}_{}^{(1)}\!/\alpha_{1}^{2} \\
  u_{}^{020|000}\!&\!\!=\!\!&\!\bar{u}_{}^{(2)}\!/\alpha_{2}^{2} \\
  u_{}^{000|200}\!&\!\!=\!\!&\!\bar{u}_{}^{(3)}\!/\beta_{1}^{2} \\
  u_{}^{000|020}\!&\!\!=\!\!&\!\bar{u}_{}^{(4)}\!/\beta_{2}^{2} \\
  u_{}^{100|100}\!&\!\!=\!\!&\![\bar{u}_{}^{(\;9\;)}\!\!-\!
  \bar{u}_{}^{(5)}\!\!-\!\bar{u}_{}^{(3)}]/\alpha_{1}^{}\beta_{1}^{} \\
  u_{}^{010|010}\!&\!\!=\!\!&\![\bar{u}_{}^{(10)}\!\!-\!
  \bar{u}_{}^{(6)}\!\!-\!\bar{u}_{}^{(4)}]/\alpha_{2}^{}\beta_{2}^{} \\
  u_{}^{100|010}\!&\!\!=\!\!&\![\bar{u}_{}^{(11)}\!\!-\!
  \bar{u}_{}^{(8)}\!\!-\!\bar{u}_{}^{(1)}]/\alpha_{1}^{}\beta_{2}^{} \\
  u_{}^{010|100}\!&\!\!=\!\!&\![\bar{u}_{}^{(12)}\!\!-\!
  \bar{u}_{}^{(7)}\!\!-\!\bar{u}_{}^{(2)}]/\alpha_{2}^{}\beta_{1}^{} \\
  u_{}^{110|000}\!&\!\!=\!\!&\![\bar{u}_{}^{(13)}\!\!-\!
  \bar{u}_{}^{(2)}\!\!-\!\bar{u}_{}^{(1)}]/\alpha_{1}^{}\alpha_{2}^{} \\
  u_{}^{000|110}\!&\!\!=\!\!&\![\bar{u}_{}^{(14)}\!\!-\!
  \bar{u}_{}^{(4)}\!\!-\!\bar{u}_{}^{(3)}]/\beta_{1}^{}\beta_{2}^{} \\
  u_{}^{001|000}\!&\!\!=\!\!&\![(\bar{u}_{}^{(5)}\!\!-\!
  \bar{u}_{}^{(1)})\!-\!(\bar{u}_{}^{(6)}\!\!-\!\bar{u}_{}^{(2)})]
  /2\alpha_{3}^{} \\ u_{}^{002|000}\!&\!\!=\!\!&\![(\bar{u}_{}^{(5)}\!\!-\!
  \bar{u}_{}^{(1)})\!+\!(\bar{u}_{}^{(6)}\!\!-\!\bar{u}_{}^{(2)})]
  /2\alpha_{3}^{2} \\
  u_{}^{000|001}\!&\!\!=\!\!&\![(\bar{u}_{}^{(7)}\!\!-\!
  \bar{u}_{}^{(3)})\!-\!(\bar{u}_{}^{(8)}\!\!-\!\bar{u}_{}^{(4)})]
  /2\beta_{3}^{} \\ u_{}^{000|002}\!&\!\!=\!\!&\![(\bar{u}_{}^{(7)}\!\!-\!
  \bar{u}_{}^{(3)})\!+\!(\bar{u}_{}^{(8)}\!\!-\!\bar{u}_{}^{(4)})]
  /2\beta_{3}^{2} \\
  u_{}^{001|001}\!&\!\!=\!\!&\![\bar{u}_{}^{(15)}\!\!-\!
  \bar{u}_{}^{(7)}\!\!-\!\bar{u}_{}^{(5)}\!\!+\!\bar{u}_{}^{(3)}
  \!\!+\!\bar{u}_{}^{(1)}]/\alpha_{3}^{}\beta_{3}^{} 
  \end{array}
  \end{equation}
  \begin{equation}
  \label{v_inversion} 
  \begin{array}{rcl}
  v_{}^{100|000}\!&\!\!=\!\!&\!v_{}^{(1)}\!/\alpha_{1}^{} \\
  v_{}^{010|000}\!&\!\!=\!\!&\!v_{}^{(2)}\!/\alpha_{2}^{} \\
  v_{}^{000|100}\!&\!\!=\!\!&\!v_{}^{(3)}\!/\beta_{1}^{} \\
  v_{}^{000|010}\!&\!\!=\!\!&\!v_{}^{(4)}\!/\beta_{2}^{} \\
  v_{}^{101|000}\!&\!\!=\!\!&\!+[v_{}^{(5)}\!\!-\!v_{}^{(1)}]
  /\alpha_{1}^{}\alpha_{3}^{} \\
  v_{}^{011|000}\!&\!\!=\!\!&\!-[v_{}^{(6)}\!\!-\!v_{}^{(2)}]
  /\alpha_{2}^{}\alpha_{3}^{} \\
  v_{}^{000|101}\!&\!\!=\!\!&\!+[v_{}^{(7)}\!\!-\!v_{}^{(3)}]
  /\beta_{1}^{}\beta_{3}^{} \\
  v_{}^{000|011}\!&\!\!=\!\!&\!-[v_{}^{(8)}\!\!-\!v_{}^{(4)}]
  /\beta_{2}^{}\beta_{3}^{} \\
  v_{}^{001|100}\!&\!\!=\!\!&\!+[v_{}^{(\;9\;)}\!\!-\!
  v_{}^{(5)}\!\!-\!v_{}^{(3)}]/\beta_{1}^{}\alpha_{3}^{} \\
  v_{}^{001|010}\!&\!\!=\!\!&\!-[v_{}^{(10)}\!\!-\!
  v_{}^{(6)}\!\!-\!v_{}^{(4)}]/\beta_{2}^{}\alpha_{3}^{} \\
  v_{}^{100|001}\!&\!\!=\!\!&\!-[v_{}^{(11)}\!\!-\!
  v_{}^{(8)}\!\!-\!v_{}^{(1)}]/\alpha_{1}^{}\beta_{3}^{} \\
  v_{}^{010|001}\!&\!\!=\!\!&\!+[v_{}^{(12)}\!\!-\!
  v_{}^{(7)}\!\!-\!v_{}^{(2)}]/\alpha_{2}^{}\beta_{3}^{} 
  \end{array}
  \end{equation}
\end{subequations}
where we have defined $\bar{u}_{}^{(r)}\equiv u_{}^{(r)}
-u_{}^{000|000}$.  

Once this calibration has been achieved, Eqs.~(\ref{uv_2nd_order})
will predict the simulated observables to second order accuracy,
for {\it any} initial spin configuration.

\subsection{Technical points}
\label{S:comments}

In this final subsection, we collect a few additional remarks that
will be of interest to readers who wish to pursue or extend the
program suggested in this section.

So far, we have discussed the initial spin {\it orientations}
necessary for calibrating the expansion coefficients.  What
about their absolute {\it magnitudes} $|{\bf a}|$ and $|{\bf b}|$?  It
is best to choose the initial spins to be rather {\it small} for two
reasons.  Firstly, the errors introduced by neglecting terms beyond
second order are fractionally smaller for small initial spins.  The
values of the coefficients given by Eqs.~(\ref{wxyz_inversion}) and
(\ref{uv_inversion}) will therefore be closer to their true, unbiased
values.  The second point relates to the way in which initial
conditions for simulations are presently specified.  Most groups
currently take the 3-metric $\gamma_{ij}$ on the initial 3-dimensional
spatial hypersurface to be conformally flat.  While an isolated,
non-spinning (Schwarzschild) black hole has conformally flat spatial
hypersurfaces, a spinning (Kerr) black hole does not
\cite{Garat:2000pn} and neither do BBHs (spinning or
non-spinning).  As the initial spins increase, choosing the initial
$\gamma_{ij}$ to be conformally flat is expected to become an
increasingly poor description of realistic BBH initial data.  This
choice will therefore lead to correspondingly larger systematic errors
in determining the coefficients in the spin expansion.  The initial
spins should be chosen small enough to minimize these problems, yet
large enough that the second-order effects we are seeking are not
swamped by other systematic errors in the numerical simulations.

As these systematic errors are inevitable, it may be fruitful to
reinterpret the coefficients on the left-hand sides of
Eqs.~(\ref{wxyz_inversion}) and (\ref{uv_inversion}).  Instead of
regarding these coefficients as the algebraic solutions to
Eqs.~(\ref{wxyz_2nd_order}) and (\ref{uv_2nd_order}), we consider them
to be {\it estimators} ($\hat{w}_{}^{m_{1}m_{2}m_{3}|
  n_{1}n_{2}n_{3}}$, $\hat{x}_{}^{m_{1}m_{2}m_{3}|n_{1}n_{2}n_{3}}$,
\ldots) constructed from the simulated observables $\{ w^{(r)},
x^{(r)}, \ldots \}$ and the estimated initial spin components
$\{a_{i}^{(r)}, b_{i}^{(r)}\}$.  Once errors are taken into account,
it may be desirable to use larger sets of simulations to beat down the
noise associated with our estimators.  We pursue this approach in
Appendix~\ref{S:MVE}, where we derive minimum-variance estimators
constructed from $N$ simulations, and provide the covariance matrix
for these estimators in terms of the covariance matrices associated
with the estimated observables $\{ w^{(r)}, x^{(r)}, \ldots \}$ and
initial spin components $\{a_{i}^{(r)}, b_{i}^{(r)}\}$.

In the future, it may be interesting to extend this section's
second-order calibration up to third order.  Although all of the
third-order contributions to the mass $m$ and spin ${\bf s}$
identified in Section~\ref{approx} were small corrections, the final
kicks ${\bf k}$ in Case \#4 seemed to exhibit considerable third-order
effects.  Recoils for generic initial spin orientations have not yet
been adequately simulated to determine whether these third-order
effects reflect genuine physical behavior or are merely artifacts of
systematic errors within the numerical codes.  If spin expansions
calibrated to second order according to the program outlined in this
section fail to describe BBH mergers with generic initial spin
orientations, we may want to test whether a third-order expansion can
remedy observed discrepancies.  Calibrating to third order will
require 12 additional simulations (for a total of 10+12=22) in the
equal-mass ($q=1$) case, and 28 additional simulations (for a total of
16+28=44) in the unequal-mass ($q\neq1$) case.  In
Appendix~\ref{3rdOrderCal} we have provided explicit third-order
expansions of the 4 variables $\{ w, x, y, z \}$ in the equal-mass
case.  These expansions can be inverted to obtain formulae for the
third-order coefficients similar to the second-order inversions of
Eqs.~(\ref{wxyz_2nd_order}).  These formulae can then be used to
identify an optimal choice of 22 simulations from which {\it all}
coefficients up to third order can be calibrated.

\section{Discussion}
\label{S:disc}

In this paper, we have developed and tested the ``spin expansion''
formalism.  This is the following simple idea.  Long after merger, we
regard any final (dimensionless) quantity $f$ --- such as the kick
velocity ${\bf k}$ or spin vector ${\bf s}$ of the final Kerr black
hole --- as a function $f(\psi,q,a_{i}^{},b_{i}^{})$ of the 8
``initial'' (dimensionless) quantities $\{\psi,q,a_{i}^{},b_{i}^{}\}$
necessary to specify the initial configuration of a BBH in circular
orbit.  Then we Taylor expand this function around $a_{i}^{} =b_{i}^{}
=0$, and use three symmetries (rotation $R$, parity $P$, and exchange
$X$) to significantly restrict the terms that can appear in the
expansion.  Finally, we interpret the leading-order terms in the
Taylor expansion as leading-order predictions for $f$, while the
next-to-leading terms in the expansion are the next-to-leading
predictions, and so on.  

To us, it seems genuinely surprising that the final state of the
complicated non-linear process of binary black hole merger can be
usefully described by such a simple-minded approach.  This simplicity
should be regarded as another discovery which has come from the recent
breakthroughs in numerical relativity.

In the Introduction to this paper, we listed some of the advantages
--- both practical and conceptual --- of the spin expansion formalism.
It may be helpful to look back at this list, now that we have had a
chance to introduce and explore the formalism in detail.  Here we
would just like to highlight three {\it new} discoveries which came
from applying the spin expansion to simulations in Sec.~\ref{approx},
and which illustrate the potential of this approach.

\subsection{Three highlights}

First, we have discovered a new third-order spin dependence of the
kick velocities in the ``superkick'' configuration considered in
Section~\ref{Case_2}.  These third-order modulations, clearly revealed
in Fig.~\ref{Residfig}, are present in the simulations of
\cite{Campanelli:2007cg,Brugmann:2007zj} but went unnoticed because
with amplitudes less than 100 km/s they are dwarfed by the primary
linear superkicks.  We were able to find them because the spin
expansion made a specific prediction for the next-to-leading
contribution: it told us to look for a contribution to $k_{3}^{}$
proportional to $a^3$ with triple the fundamental (linear) superkick
frequency.  Empirical fitting formulae --- linear in spins, and
inspired by post-Newtonian results --- provide acceptable fits for
these superkick simulations, but our discovery shows that there is
more to learn if one is willing to go beyond these fitting formulae.

Second, we have discovered a new second-order spin dependence of the
radiation energy $E_{\rm rad}$, the radiated angular momentum $J_{\rm
  rad}$ and the final spin $s_{3}^{}$ in the superkick configuration.
The spin expansion predicts that, since these three quantities
$\{E_{\rm rad}, J_{\rm rad}, s_{3}^{}\}$ are all characterized by the
same transformation properties ($P=+1$, $X=+1$), they should all
exhibit the same next-to-leading-order behavior: $A+B{\rm
  cos}(2\phi+{\rm phase})$.  Again this behavior is present in the
simulations of \cite{Campanelli:2007cg, Brugmann:2007zj}, and is
clearly displayed in Fig.~\ref{Case2fig}; but without the guidance of
the spin expansion, it went unnoticed in \cite{Campanelli:2007cg}, and
was dismissed as a possible numerical artifact in
\cite{Brugmann:2007zj}.

Third, we wish to highlight the remarkable agreement between the
predictions of the spin expansion and the simulations of generically
oriented spin configurations \cite{Tichy:2007hk} considered in
Sec.~\ref{Case_5} (Case \#5).  This agreement is illustrated in
Fig.~\ref{Case5res}, where the black points are the simulations
results and the red points are the predictions.  We emphasize that the
red points are genuine predictions --- {\it i.e.}\ there were {\it no
  free parameters} in these fits, since all of the relevant
coefficients had already been calibrated by the simulations in Case
\#1.  The red and black points only disagree for one of the 8
simulations in Case \#5 and, as explained in Sec.~\ref{Case_5}, this
disagreement is easily understood.  So far, mumerical relativists have
focused mostly on highly symmetric configurations like the aligned
case in Section~\ref{Case_1} and the superkick configuration of
Section~\ref{Case_2}.  This is probably because of the expected
complications from non-linear spin precession in the generic case.
Herrmann {\it et al.}  \cite{Herrmann:2007ex} observe these
precessions in their ``S-Series,'' and note that they make it
impossible to use the post-Newtonian-inspired fitting formula for the
final kicks in this case.  Fig.~\ref{Case5res} seems to provide
evidence that our spin expansions continue to apply, even in the
presence of these precession effects.

\subsection{Future directions}

Let us end by briefly mentioning a few directions for further study.

First, it would be extremely fruitful to calibrate the spin expansion
coefficients, up to second or third order.  As explained in
Sec.~\ref{S:prog}, currently available simulations leave many
degeneracies among spin expansion coefficients, even at first and
second order.  To rectify this problem, in Sec.~\ref{S:prog} we
suggest a small set of simulations --- 10 equal-mass simulations and
16 unequal-mass simulations --- and explicitly show how these would
uniquely determine all of the spin-expansion coefficients up to second
order.  Once these coefficients are calibrated in this way, the spin
expansion becomes fully predictive: given {\it any} initial spin
configuration, it predicts the final results $\{m,k_{i}^{},s_{i}^{}\}$
with second-order accuracy.  In addition to facilitating tests of the
spin expansion, it is clear that this result --- a set of simple
formulae which predict the final state of BBH merger given the initial
state --- would be of enormous interest from the standpoint of
astrophysical and cosmological applications.  For example, our spin
expansion precisely encapsulates the relevant information for
incorporating the recent discoveries of numerical relativity into
cosmological simulations of BBH merger in the context of structure
formation.  It would also be interesting from a purely theoretical
standpoint.  The initial spin configurations we have identified in
Section~\ref{S:prog} provide a systematic approach for seeking
qualitatively {\it new} behavior in the unexplored regions of BBH
parameter space.  Any unexpected constraints, patterns, or
relationships among the calibrated coefficients (beyond the ones we
have used thus far in our construction) could indicate interesting new
dynamical effects or symmetries of the system.

Second, we have mentioned that a final quantity $f$ may be regarded as
function on an 8-dimensional space $\{\psi,q,a_{i},b_{i}\}$.  The spin
expansion elucidates the structure of the 6-dimensional subspace
parameterized by $\{a_{i}^{},b_{i}^{}\}$, but we would also like to
know the behavior along the $\psi$ and $q$ directions.  In a follow-up
paper, we consider how post-Newtonian techniques may be used to to
explore the $\psi$-dependence of the spin expansion coefficients.
Some insights may be gained by an analogy with effective field theory,
where the renormalized coupling constants depend on the momentum scale
at which they are defined, although the physical predictions of the
theory do not.  Determining the $q$-dependence of the spin-expansion
coefficients \cite{Rezzolla:2007rd} seems less straightforward, and is
an interesting topic for future research.

\acknowledgments

We thank Antony Lewis and Jonathan Sievers for access to a modified
version of CosmoMC \cite{Lewis:2002ah}, a Fortran 90 Markov-Chain
Monte-Carlo (MCMC) engine that was used to double-check the fits in
this paper.  We would also like to thank Emanuele Berti, Alessandra
Buonanno, Neal Dalal, Peter Diener, Nils Dorband, Larry Kidder, Luis
Lehner, Carlos Lousto, Pedro Marronetti, Samaya Nissanke, Harald
Pfeiffer, Eric Poisson, Denis Pollney, Christian Reisswig, Luciano
Rezzolla, Eric Schnetter, Bela Szilagyi, Wolfgang Tichy, Bill Unruh,
Daniel Wesley, and Yosef Zlochower for useful discussions.

\appendix

\section{Tables of Simulated Final Kicks, Spins, and Masses}
\label{datatables}

\subsection{Case \#1: $q=1$, ${\bf a}_{\perp}={\bf b}_{\perp}=0$}
\label{Case1TabApp}

We use the 28 simulations of \cite{Rezzolla:2007xa} with non-zero
recoils to test our expansions for the final kicks in the case of
equal-mass ($q=1$) binary black holes with spins aligned (or
anti-aligned) with the orbital angular momentum.  This is the largest
and most recent series of simulations for this configuration published
at the time this manuscript was prepared.  The numerical estimates of
$|{\bf k}_{\perp}|$ are avilable in Table 1 of \cite{Rezzolla:2007xa}.
We adopt their $1\sigma$ errors of $8$ km/s for the kick magnitude.

We constrain our expansion for the final spin $s_3$ by performing a
joint fit to the full set of 38 simulations in \cite{Rezzolla:2007xa}
and the 10 simulations of \cite{Marronetti:2007wz} that they consider
to be relatively free from the numerical dissipation of angular
momentum.  We use the proposed $1\sigma$ errors of $0.01$ and $0.02$
for the simulations of \cite{Rezzolla:2007xa} and
\cite{Marronetti:2007wz} respectively.

\begin{table}
  \begin{small}
    \begin{center}
      \begin{tabular}{|c||c|c|c|}
        \hline
        & $a_{3}^{}$ & $b_{3}^{}$ & $m^{{\rm num }}$ \\ \hline
        $A_1$ & $0.2$ & $-0.2$ & $0.9526\pm0.0023$
        \\ \hline
        $A_2$ & $0.4$ & $-0.4$ & $0.9521\pm0.0017$
        \\ \hline
        $A_3$ & $0.6$ & $-0.6$ & $0.9519\pm0.0014$
        \\ \hline
        $A_4$ & $0.8$ & $-0.8$ & $0.9521\pm0.0028$
        \\ \hline \hline
        $B_1$ & $0.584$ & $-0.584$ & $0.9536\pm0.0049$
        \\ \hline
        $B_2$ & $0.584$ & $-0.438$ & $0.9507\pm0.0049$
        \\ \hline
        $B_3$ & $0.584$ & $-0.292$ & $0.9482\pm0.0049$
        \\ \hline
        $B_4$ & $0.584$ & $-0.146$ & $0.9461\pm0.0049$
        \\ \hline
        $B_5$ & $0.584$ & $0$ & $0.9439\pm0.0049$
        \\ \hline
        $B_6$ & $0.584$ & $0.146$ & $0.9412\pm0.0049$
        \\ \hline
        $B_7$ & $0.584$ & $0.292$ & $0.9376\pm0.0049$
        \\ \hline
        $B_8$ & $0.584$ & $0.438$ & $0.9344\pm0.0049$
        \\ \hline
        $B_9$ & $0.584$ & $0.584$ & $0.9315\pm0.0049$
        \\ \hline \hline
        $C_1$ & $-0.90$ & $-0.90$ & $0.970\pm0.004$
        \\ \hline
        $C_2$ & $-0.75$ & $-0.75$ & $0.968\pm0.004$
        \\ \hline
        $C_3$ & $-0.50$ & $-0.50$ & $0.963\pm0.004$
        \\ \hline
        $C_4$ & $-0.25$ & $-0.25$ & $0.958\pm0.004$
        \\ \hline
        $C_5$ & $0.0$ & $0.0$ & $0.951\pm0.004$
        \\ \hline
        $C_6$ & $0.25$ & $0.25$ & $0.944\pm0.004$
        \\ \hline
        $C_7$ & $0.50$ & $0.50$ & $0.933\pm0.004$
        \\ \hline
        $C_8$ & $0.62$ & $0.62$ & $0.926\pm0.004$
        \\ \hline
        $C_9$ & $0.75$ & $0.75$ & $0.916\pm0.004$
        \\ \hline
        $C_{10}$ & $0.82$ & $0.82$ & $0.909\pm0.004$
        \\ \hline
        $C_{11}$ & $0.90$ & $0.90$ & $0.906\pm0.004$
        \\ \hline
      \end{tabular}
    \end{center}
  \end{small}
  \caption{Final mass data for Case \#1: equal-mass ($q=1$) binary black holes
    with spins aligned (or anti-aligned) with the orbital angular
    momentum.  Points $A_{1}$ through $A_{4}$ are from
    \cite{Herrmann:2007ac}, points $B_{1}$ through $B_{9}$ are from
    \cite{Pollney:2007ss}, and points $C_{1}$ through $C_{11}$ are from
    \cite{Marronetti:2007wz}.}
  \label{case1_masstable}
\end{table}

To constrain our expansion for the final mass $m$, we perform a joint
fit to 3 series of simulations \cite{Herrmann:2007ac, Pollney:2007ss,
  Marronetti:2007wz}, as each series consists of only a small number
of individual simulations.  These simulations are summarized in Table
\ref{case1_masstable}.  For points $A_1$ through $A_4$ we assume
fractional errors on the radiated energy identical to those provided
in \cite{Herrmann:2007ac} for the final kicks.  For points $B_1$
through $B_9$ we assume errors on the final masses of 0.5\% of the
initial energy $M_{\rm ADM}$, as \cite{Pollney:2007ss} only claims to
conserve energy to this accuracy.  Finally, we use the
highest-resolution simulations of all 11 initial data sets listed in
Table I of \cite{Marronetti:2007wz}.  While they claim that numerical
dissipation of angular momentum makes estimates of $s_3$ unreliable
for $a_3, b_3 > 0.75$, the final masses are largely unaffected as
shown in their Fig.~4.  We therefore use both simulations with $a_3,
b_3 > 0.75$, and assume for all simulations errors on $m$ of $0.004$
consistent with their claimed resolution limits.

\subsection{Case \#2: $q=1$, ${\bf a}_{\perp}=-{\bf b}_{\perp}$,
  $a_{3}=b_{3}=0$} \label{Case2TabApp}

\begin{table}
  \begin{small}
    \begin{center}
      \begin{tabular}{|c||c|c|c|c|c|}
        \hline
        & $a$ & $\phi$ & $k_{3}^{{\rm num}}$ & $J_{\rm rad}/M^2$ & $\% E_{\rm rad}$
        \\ \hline
        $A_1$ & $0.515$ & $1.571$ & $1833\pm30$ & $0.248\pm0.003$ & $3.63\pm0.01$
        \\ \hline
        $A_2$ & $0.515$ & $0.785$ & $1093\pm10$ & $0.244\pm0.003$ & $3.53\pm0.01$
        \\ \hline
        $A_3$ & $0.515$ & $3.142$ & $352\pm10$ & $0.246\pm0.004$ & $3.57\pm0.01$
        \\ \hline
        $A_4$ & $0.515$ & $4.712$ & $-1834\pm30$ & $0.249\pm0.003$ & $3.63\pm0.01$
        \\ \hline
        $A_5$ & $0.515$ & $3.304$ & $47\pm10$ & $0.245\pm0.005$ & $3.55\pm0.02$
        \\ \hline
        $A_6$ & $0.515$ & $0.0$ & $-351\pm10$ & $0.246\pm0.003$ & $3.57\pm0.02$
        \\ \hline \hline
        & $a$ & $\phi$ & $k_{3}^{{\rm num}}$ & $s_3$ & ---
        \\ \hline
        $B_1$ & $0.723$ & $0.0$ & $2680\!\pm\!94$ & $0.6859\!\pm\!0.0009$ &
        \\ \hline
        $B_2$ & $0.723$ & $0.524$ & $2310\!\pm\!94$ & $0.6856\!\pm\!0.0009$ &
        \\ \hline
        $B_3$ & $0.723$ & $1.047$ & $1150\!\pm\!94$ & $0.6897\!\pm\!0.0009$ &
        \\ \hline
        $B_4$ & $0.723$ & $1.571$ & $-340\!\pm\!94$ & $0.6931\!\pm\!0.0009$ &
        \\ \hline
        $B_5$ & $0.723$ & $2.094$ & $-1590\!\pm\!94$ & $0.6934\!\pm\!0.0009$ &
        \\ \hline
        $B_6$ & $0.723$ & $2.618$ & $-2400\!\pm\!94$ & $0.6892\!\pm\!0.0009$ &
        \\ \hline
        $B_7$ & $0.723$ & $3.142$ & $-2690\!\pm\!94$ & $0.6859\!\pm\!0.0009$ &
        \\ \hline
        $B_8$ & $0.723$ & $3.665$ & $-2320\!\pm\!94$ & $0.6856\!\pm\!0.0009$ &
        \\ \hline
        $B_9$ & $0.723$ & $4.189$ & $-1160\!\pm\!94$ & $0.6897\!\pm\!0.0009$ &
        \\ \hline
        $B_{10}$ & $0.723$ & $4.712$ & $320\!\pm\!94$ & $0.6932\!\pm\!0.0009$ &
        \\ \hline
        $B_{11}$ & $0.723$ & $5.236$ & $1580\!\pm\!94$ & $0.6935\!\pm\!0.0009$ &
        \\ \hline
        $B_{12}$ & $0.723$ & $5.760$ & $2400\!\pm\!94$ & $0.6892\!\pm\!0.0009$ &
        \\ \hline
      \end{tabular}
    \end{center}
  \end{small}
  \caption{Data for Case \#2: equal-mass $(q=1)$ binary black holes with equal
    and opposite spins in the orbital plane.  Points $A_1 - A_6$ are from
    \cite{Campanelli:2007cg}, while points $B_1 - B_{12}$ are from
    \cite{Brugmann:2007zj}.  These papers present final quantities
    different from those discussed in this paper, but whose spin
    dependence can be readily analyzed in our formalism.  The scalar
    $J_{\rm rad}$ is the total angular momentum radiated in gravitational
    waves, listed in units of $M^2$ where $M \equiv M_a + M_b$ is the
    sum of the horizon masses of the initial binary black holes.  $\%
    E_{\rm rad}$ is the percentage of the initial energy radiated in
    gravitational waves.}
  \label{super_table}
\end{table}

The data points for this case are summarized in Table
\ref{super_table}.  Campanelli {\it et al.} \cite{Campanelli:2007cg}
provides error estimates for their final observables, which we assume
represent true $1\sigma$ statistical error bars.  Br\"{u}gmann {\it et
  al.} does not provide error estimates for the individual simulated
data points $B_i$, however they do claim $95\%$ confidence limits of
$\pm2\%$ on their maximum kick amplitude of $2,725$ km/s and $\pm5
\times 10^{-4}$ on their mean spin $a_0 = 0.6891$ as determined from
the black-hole ringdown.  We assume these values correspond to
$2\sigma$ error bars on each parameters, and that they were derived
from $12$ independent data points.  This leads to crude $1\sigma$
errors of $1/2 \times \sqrt{12} \times 0.02 \times 2,725 {\rm km/s} =
94 {\rm km/s}$ on each kick and $1/2 \times \sqrt{12} \times 0.0005 =
0.0009$ on each final spin.

\subsection{Case \#3: Herrmann {\it et al} ``B-series''}
\label{Case3TabApp}

The data for this case are summarized in Table \ref{Case3_table}.

\begin{table}
  \begin{small}
    \begin{center}
      \begin{tabular}{|c||c|c|c|c|c|}
        \hline
        & $a$ & $\phi$ & $|{\bf k}|^{{\rm num}}$ & $J_{\rm rad}/L_{0}^{z}$
        & $E_{\rm rad}/M$
        \\ \hline
        $A$ & $0.6$ & $0.0$ & $285\!\pm\!12$ & --- & ---
        \\ \hline \hline
        $B_1$ & $0.6$ & $0.349$ & $427\pm64$ & $0.24\!\pm\!0.036$ & 
        $0.033\!\pm\!0.005$
        \\ \hline
        $B_2$ & $0.6$ & $0.524$ & $544\pm82$ & $0.24\!\pm\!0.036$ & 
        $0.033\!\pm\!0.005$
        \\ \hline
        $B_3$ & $0.6$ & $0.873$ & $761\pm114$ & $0.25\!\pm\!0.038$ & 
        $0.034\!\pm\!0.005$
        \\ \hline
        $B_4$ & $0.6$ & $1.222$ & $908\pm136$ & $0.25\!\pm\!0.038$ & 
        $0.034\!\pm\!0.005$
        \\ \hline
        $B_5$ & $0.6$ & $1.396$ & $945\pm142$ & $0.25\!\pm\!0.038$ & 
        $0.034\!\pm\!0.005$
        \\ \hline
        $B_6$ & $0.6$ & $1.571$ & $963\pm144$ & $0.25\!\pm\!0.038$ & 
        $0.034\!\pm\!0.005$
        \\ \hline
      \end{tabular}
    \end{center} 
  \end{small}
  \caption{Data for Case \#3: equal-mass $(q=1)$ binary black holes with equal
    and opposite spins in the ${\bf e}^{(1)}-{\bf e}^{(3)}$ plane.  Points
    $A$ is from \cite{Herrmann:2007ac} while points $B_1 - B_6$ are from
    \cite{Herrmann:2007ex}.  $J_{\rm rad}/L_{0}^{z}$ is the ratio of the
    total radiated angular momentum to the initial orbital angular
    momentum, while $E_{\rm rad}/M$ is the ratio of the total radiated
    energy to the sum of the initial horizon masses.  We do not include
    estimates of the radiated energy and angular momentum for point $A$
    because this simulation began at a different initial separation and
    orbital angular momentum from points $B_1 - B_6$.  Herrmann {\it et
      al.}  \cite{Herrmann:2007ex} claimed $15\%$ errors on their
    reported numbers which we treat here as true $1\sigma$ bars.}
  \label{Case3_table}
\end{table}
  
\subsection{Case \#4: Herrmann {\it et al} ``S-series''}
\label{Case4TabApp}

The data for this case are summarized in Table \ref{Case4_table}.

\begin{table}
  \begin{small}
    \begin{center}
      \begin{tabular}{|c||c|c|c|c|c|}
        \hline
        & $\phi$ & $|{\bf k}|^{{\rm num}}$ & $J_{\rm final}^z/M^2$
        & $E_{\rm rad}/M$ & $T_{\rm max}$
        \\ \hline
        $A_1$ & $0^{\circ}$ & $854\!\pm\!128$ & $0.68\!\pm\!0.102$ &
        $0.046\!\pm\!0.0069$ & $192.3$
        \\ \hline
        $A_2$ & $15^{\circ}$ & $1401\!\pm\!210$ & $0.68\!\pm\!0.102$ &
        $0.044\!\pm\!0.0066$ & $189.5$
        \\ \hline
        $A_3$ & $30^{\circ}$ & $2000\!\pm\!300$ & $0.67\!\pm\!0.101$ &
        $0.044\!\pm\!0.0066$ & $184.1$
        \\ \hline
        $A_4$ & $45^{\circ}$ & $2030\!\pm\!305$ & $0.66\!\pm\!0.099$ &
        $0.043\!\pm\!0.0065$ & $177.3$
        \\ \hline
        $A_5$ & $60^{\circ}$ & $1218\!\pm\!183$ & $0.65\!\pm\!0.098$ &
        $0.040\!\pm\!0.0060$ & $168.6$
        \\ \hline
        $A_6$ & $75^{\circ}$ & $230\!\pm\!35$ & $0.64\!\pm\!0.096$ &
        $0.037\!\pm\!0.0056$ & $159.1$
        \\ \hline
        $A_7$ & $90^{\circ}$ & $1462\!\pm\!219$ & $0.62\!\pm\!0.093$ &
        $0.034\!\pm\!0.0051$ & $148.6$
        \\ \hline
        $A_8$ & $105^{\circ}$ & $1979\!\pm\!297$ & $0.60\!\pm\!0.090$ &
        $0.033\!\pm\!0.0050$ & $138.6$
        \\ \hline
        $A_9$ & $120^{\circ}$ & $1787\!\pm\!268$ & $0.58\!\pm\!0.087$ &
        $0.032\!\pm\!0.0048$ & $130.5$
        \\ \hline
        $A_{10}$ & $135^{\circ}$ & $1234\!\pm\!185$ & $0.56\!\pm\!0.084$ &
        $0.030\!\pm\!0.0045$ & $124.1$
        \\ \hline
        $A_{11}$ & $150^{\circ}$ & $689\!\pm\!103$ & $0.55\!\pm\!0.083$ &
        $0.029\!\pm\!0.0044$ & $119.5$
        \\ \hline
        $A_{12}$ & $165^{\circ}$ & $335\!\pm\!50$ & $0.55\!\pm\!0.083$ &
        $0.028\!\pm\!0.0042$ & $117.7$
        \\ \hline
        $A_{13}$ & $180^{\circ}$ & $188\!\pm\!28$ & $0.55\!\pm\!0.083$ &
        $0.028\!\pm\!0.0042$ & $117.7$
        \\ \hline
        $A_{14}$ & $195^{\circ}$ & $157\!\pm\!24$ & $0.55\!\pm\!0.083$ &
        $0.028\!\pm\!0.0042$ & $120.5$
        \\ \hline
        $A_{15}$ & $210^{\circ}$ & $173\!\pm\!26$ & $0.56\!\pm\!0.084$ &
        $0.030\!\pm\!0.0045$ & $125.5$
        \\ \hline
        $A_{16}$ & $225^{\circ}$ & $223\!\pm\!33$ & $0.57\!\pm\!0.086$ &
        $0.032\!\pm\!0.0048$ & $132.7$
        \\ \hline
        $A_{17}$ & $240^{\circ}$ & $268\!\pm\!40$ & $0.59\!\pm\!0.089$ &
        $0.034\!\pm\!0.0051$ & $141.4$
        \\ \hline
        $A_{18}$ & $285^{\circ}$ & $253\!\pm\!38$ & $0.65\!\pm\!0.098$ &
        $0.039\!\pm\!0.0059$ & $174.1$
        \\ \hline
        $A_{19}$ & $300^{\circ}$ & $406\!\pm\!61$ & $0.66\!\pm\!0.099$ &
        $0.042\!\pm\!0.0063$ & $181.8$
        \\ \hline
        $A_{20}$ & $315^{\circ}$ & $399\!\pm\!60$ & $0.67\!\pm\!0.101$ &
        $0.045\!\pm\!0.0068$ & $187.7$
        \\ \hline
        $A_{21}$ & $330^{\circ}$ & $354\!\pm\!53$ & $0.68\!\pm\!0.102$ &
        $0.046\!\pm\!0.0069$ & $191.8$
        \\ \hline
        $A_{22}$ & $345^{\circ}$ & $459\!\pm\!69$ & $0.68\!\pm\!0.102$ &
        $0.046\!\pm\!0.0069$ & $193.2$
        \\ \hline
      \end{tabular}
    \end{center} 
  \end{small}
  \caption{Data for Case \#4: equal-mass $(q=1)$ BBHs belonging to the
    ``S-Series'' of \cite{Herrmann:2007ex}.  The spins have magnitudes
    $a = 0.6$ and orientations ${\bf a} = (-a, 0, 0)$ and ${\bf b} = (a
    \sin \phi, 0, a \cos \phi)$.  $J_{\rm final}^z/M^2$ is the
    $z-$component of the final black hole's spin in units of the sum $M$
    of the initial horizon masses, and $E_{\rm rad}/M$ is total radiated
    energy in units of $M$.  Herrmann {\it et al.}
    \cite{Herrmann:2007ex} claimed $15\%$ errors on their reported
    numbers which we treat here as true $1\sigma$ bars.  $T_{\rm max}$,
    measured in units of $M$, is an estimate of the merger time defined
    as the coordinate time between the beginning of the simulation and
    when the Newman-Penrose quantity $\Psi_4$ is maximized.}
  \label{Case4_table}
\end{table}

\subsection{Case \#5: The generic case (Tichy-Marronetti)}
\label{Case5TabApp}

The data for this case are summarized in Table \ref{Case5_table}.
Even at second order in the initial spin magnitude $a$, there are too
many independent non-degenerate coefficients to fit with only 8
simulations.  We therefore only attempt to fit the final masses
$M_f/M$ and spin magnitudes $J_f/M^2$ as these can be fit with linear
order terms in our formalism.  We assume errors on these quantities
that are 20\% of the radiated energy $(M_{\infty}^{\rm ADM}- M_f)/M$
and radiated angular momentum $(J_{\infty}^{\rm ADM} - J_f)/M^2$.

\begin{table}
  \begin{small}
    \begin{center}
      \begin{tabular}{|c||c|c|c|c|c|c|}
        \hline
        & $\theta_a$ & $\phi_a$ & $\theta_b$ & $\phi_b$ & $M_f/M$ & $J_f/M^2$
        \\ \hline
        $A_1$ & $90^{\circ}$ & $180^{\circ}$ & $90^{\circ}$ & $0^{\circ}$ &
        $0.95\!\pm\!0.0070$ & $0.67\!\pm\!0.053$ 
        \\ \hline
        $A_2$ & $90^{\circ}$ & $225^{\circ}$ & $90^{\circ}$ & $315^{\circ}$ &
        $0.95\!\pm\!0.0070$ & $0.72\!\pm\!0.051$ 
        \\ \hline
        $A_3$ & $45^{\circ}$ & $90^{\circ}$ & $135^{\circ}$ & $270^{\circ}$ &
        $0.95\!\pm\!0.0070$ & $0.68\!\pm\!0.050$ 
        \\ \hline
        $A_4$ & $45^{\circ}$ & $270^{\circ}$ & $135^{\circ}$ & $270^{\circ}$ &
        $0.952\!\pm\!0.0066$ & $0.73\!\pm\!0.048$
        \\ \hline
        $A_5$ & $60^{\circ}$ & $90^{\circ}$ & $60^{\circ}$ & $90^{\circ}$ &
        $0.96\!\pm\!0.0050$ & $0.64\!\pm\!0.036$ 
        \\ \hline
        $A_6$ & $90^{\circ}$ & $270^{\circ}$ & $0^{\circ}$ & $0^{\circ}$ &
        $0.94\!\pm\!0.0088$ & $0.81\!\pm\!0.068$ 
        \\ \hline
        $A_7$ & $90^{\circ}$ & $240^{\circ}$ & $0^{\circ}$ & $0^{\circ}$ &
        $0.94\!\pm\!0.0088$ & $0.80\!\pm\!0.070$ 
        \\ \hline
        $A_8$ & $90^{\circ}$ & $210^{\circ}$ & $0^{\circ}$ & $0^{\circ}$ &
        $0.94\!\pm\!0.0088$ & $0.80\!\pm\!0.070$
        \\ \hline
      \end{tabular}
    \end{center} 
  \end{small}
  \caption{Data for Case \#5: equal-mass $(q=1)$ BBHs with generic spin
    orientations taken from \cite{Tichy:2007hk}.  The initial spins have
    magnitudes $a = 0.8$ and orientations given by traditional spherical
    coordinates, ${\bf a} = (a\sin \theta_a \cos \phi_a, a\sin \theta_a
    \sin \phi_a, a\cos \theta_a)$ and ${\bf b} = (a\sin \theta_b \cos
    \phi_b, a\sin \theta_b \sin \phi_b, a\cos \theta_b)$.}
  \label{Case5_table}
\end{table}

\section{Supplementary Equations}
\label{grossness}

\subsection{Relations between coefficients for Case \#1}
\label{Case1App}

Here are the relations between the ``new'' coefficients $A$, $B$, and
$C$, and the original expansion coefficients ${\bf
  k}_{\!\perp}^{m_{1}m_{2}m_{3}|n_{1}n_{2}n_{3}}$:
\begin{subequations}
  \begin{eqnarray}
    A\!&\!\!=\!\!&\!\frac{|{\bf k}_{\!\perp}^{002|000}|}
    {|{\bf k}_{\!\perp}^{001|000}|} \cos \Theta \\
    B\!&\!\!=\!\!&\!\frac{{\bf k}_{\!\perp}^{001|000}\!\cdot\!
      {\bf k}_{\!\perp}^{003|000}}{|{\bf k}_{\!\perp}^{001|000}|^{2}}
    \!+\!\frac{1}{2}\!\left( \frac{|{\bf k}_{\!\perp}^{002|000}|}
    {|{\bf k}_{\!\perp}^{001|000}|} \right)^{\!2}\!\sin^2 \Theta \\
    C\!&\!\!=\!\!&\!\frac{{\bf k}_{\!\perp}^{001|000}\!\cdot\!
      ({\bf k}_{\!\perp}^{002|001}\!\!-\!{\bf k}_{\!\perp}^{003|000})}
    {|{\bf k}_{\!\perp}^{001|000}|^{2}}\!+\!2B
  \end{eqnarray}
\end{subequations}
where $\Theta$ is the angle between ${\bf k}_{\!\perp}^{001|000}$ and
${\bf k}_{\!\perp}^{002|000}$.

\subsection{Relations between coefficients for Case \#2} \label{Case2App}

Here are the relations between the ``old'' coefficients $f_{}^{m_1 m_2
  m_3 | n_1 n_2 n_3}$ and the ``new'' coefficients $f_{}^{(i,j)}$ of
subsection \ref{Case_2}:
\begin{subequations}
  \label{Case_2_relations}
  \begin{eqnarray}
    \label{Case_2_even_relations}
    \!\!&\!\!\!\!&\!\!\begin{array}{rcl}  
      ^{c}f_{}^{(0,0)}\!&\!\!=\!\!&\!f_{}^{000|000} \\
      ^{c}f_{}^{(2,0)}\!&\!\!=\!\!&\!f_{}^{200|000}
      \!+\!f_{}^{020|000}\!-\!\frac{1}{2}f_{}^{100|100}\!-\!
      \frac{1}{2}f_{}^{010|010}\qquad \\
      ^{c}f_{}^{(2,2)}\!&\!\!=\!\!&\!
      f_{}^{200|000}\!-\!f_{}^{020|000}\!-\!\frac{1}{2}f_{}^{100|100}
      \!+\!\frac{1}{2}f_{}^{010|010} \\
      ^{s}f_{}^{(2,2)}\!&\!\!=\!\!&\!f_{}^{110|000}
      \!-\!f_{}^{100|010}
    \end{array} \\
    \label{Case_2_odd_relations}
    \!\!&\!\!\!\!&\!\!\begin{array}{rcl}
      ^{c}f_{}^{(1,1)}\!&\!\!=\!\!&\! 2f_{}^{100|000} \\
      ^{s}f_{}^{(1,1)}\!&\!\!=\!\!&\! 2f_{}^{010|000} \\
      ^{c}f_{}^{(3,1)}\!&\!\!=\!\!&\! 
      \frac{3}{2}f_{}^{300|000}\!-\!\frac{3}{2}f_{}^{200|100}\!+\!
      \frac{1}{2}f_{}^{120|000} \\
      \!&\!\!-\!\!&\!
      \frac{1}{2}f_{}^{020|100}\!-\!\frac{1}{2}f_{}^{110|010} \\
      ^{s}f_{}^{(3,1)}\!&\!\!=\!\!&\!
      \frac{1}{2}f_{}^{210|000}\!-\!\frac{1}{2}f_{}^{110|100}\!-\!
      \frac{1}{2}f_{}^{200|010} \\
      \!&\!\!+\!\!&\!
      \frac{3}{2}f_{}^{030|000}\!-\!\frac{3}{2}f_{}^{020|010} \\
      ^{c}f_{}^{(3,3)}\!&\!\!=\!\!&\!
      \frac{1}{2}f_{}^{300|000}\!-\!\frac{1}{2}f_{}^{200|100}\!-\!
      \frac{1}{2}f_{}^{120|000} \\
      \!&\!\!+\!\!&\!
      \frac{1}{2}f_{}^{020|100}\!+\!\frac{1}{2}f_{}^{110|010} \\
      ^{s}f_{}^{(3,3)}\!&\!\!=\!\!&\! 
      \frac{1}{2}f_{}^{210|000}\!-\!\frac{1}{2}f_{}^{110|100}\!-\!
      \frac{1}{2}f_{}^{200|010} \\
      \!&\!\!-\!\!&\!
      \frac{1}{2}f_{}^{030|000}\!+\!\frac{1}{2}f_{}^{020|010}
    \end{array}
  \end{eqnarray}
\end{subequations}

\subsection{Relations between coefficients for Case \#3} \label{Case3App}

Here are the relations between the ``old'' coefficients $f_{}^{m_1 m_2
  m_3 | n_1 n_2 n_3}$ and the ``new'' coefficients $f_{}^{(i,j)}$ of
subsection~\ref{Case_3}.  Cosine terms with {\it even} $i,j$ appear in
the expansion of scalars {\it even} under $PX$, the observables $m$
and $s_3$.
\begin{equation}
  \label{Case_3_m_relations}
  \begin{array}{rcl}
    ^{c}f_{}^{(0,0)}\!&\!\!=\!\!&\!f_{}^{000|000} \\
    ^{c}f_{}^{(2,0)}\!&\!\!=\!\!&\!
    f_{}^{002|000}\!+\!f_{}^{200|000}\!-\!\frac{1}{2}f_{}^{001|001}
    \!-\!\frac{1}{2}f_{}^{100|100} \\
    ^{c}f_{}^{(2,2)}\!&\!\!=\!\!&\!f_{}^{002|000}
    \!-\!f_{}^{200|000}\!-\!\frac{1}{2}f_{}^{001|001}\!+\!
    \frac{1}{2}f_{}^{100|100}
  \end{array}
\end{equation}
Cosine terms with {\it odd} $i,j$ appear in the expansion of ${\bf
  k}_{\perp}$ because it is a scalar {\it odd} under $PX$.
\begin{equation}
  \label{Case_3_k_perp_relations}
  \begin{array}{rcl}
    ^{c}f_{}^{(1,1)}\!&\!\!=\!\!&\!2f_{}^{001|000} \\
    ^{c}f_{}^{(3,1)}\!&\!\!=\!\!&\!
    \frac{3}{2}f_{}^{003|000}\!-\!\frac{3}{2}f_{}^{002|001}
    \!+\!\frac{1}{2}f_{}^{201|000} \\
    \!&\!\!-\!\!&\!
    \frac{1}{2}f_{}^{101|100}\!-\!\frac{1}{2}f_{}^{200|001} \\
    ^{c}f_{}^{(3,3)}\!&\!\!=\!\!&\!
    \frac{1}{2}f_{}^{003|000}\!-\!\frac{1}{2}f_{}^{002|001}
    \!-\!\frac{1}{2}f_{}^{201|000} \\
    \!&\!\!+\!\!&\!
    \frac{1}{2}f_{}^{101|100}\!+\!\frac{1}{2}f_{}^{200|001})
  \end{array}
\end{equation}
Sine terms with {\it odd} $i,j$ appear in the expansion of $k_3$
because it is a pseudoscalar {\it odd} under $PX$.
\begin{equation}
  \label{Case_3_k_3_relations}
  \begin{array}{rcl}
    ^{s}f_{}^{(1,1)}\!&\!\!=\!\!&\!2f_{}^{100|000} \\
    ^{s}f_{}^{(3,1)}\!&\!\!=\!\!&\!\frac{1}{2}f_{}^{102|000} 
    \!-\!\frac{1}{2}f_{}^{101|001}\!+\!\frac{1}{2}f_{}^{100|002} \\
    \!&\!\!+\!\!&\! 
    \frac{3}{2}f_{}^{300|000}\!-\!\frac{3}{2}f_{}^{200|100} \\
    ^{s}f_{}^{(3,3)}\!&\!\!=\!\!&\!\frac{1}{2}f_{}^{102|000} 
    \!-\!\frac{1}{2}f_{}^{101|001}\!+\!\frac{1}{2}f_{}^{100|002} \\
    \!&\!\!-\!\!&\!
    \frac{1}{2}f_{}^{300|000}\!+\!\frac{1}{2}f_{}^{200|100})
  \end{array}
\end{equation}
Sine terms with {\it even} $i,j$ appear in the expansion of ${\bf
  s}_{\perp}$ because it is a pseudoscalar {\it even} under $PX$.
\begin{equation}
  \label{Case_3_s_perp_relations}
  \begin{array}{rcl}
    ^{s}f_{}^{(2,2)}\!&\!\!=\!\!&\!f_{}^{101|000}\!-\!f_{}^{100|001}
  \end{array}
\end{equation}

\subsection{Relations between coefficients for Case \#4} \label{Case4App}

Here are the relations between the ``old'' coefficients $f_{}^{m_1 m_2
  m_3 | n_1 n_2 n_3}$ and the ``new'' coefficients $f_{}^{(i,j)}$ of
subsection~\ref{Case_4}.  Coefficients in the expansions of $J_{\rm
  final}^z/M^2$ and $E_{\rm rad}/M$ behave like those for $m$, which
are provided here.
\begin{equation}
  \label{Case_4_m_relations}
  \begin{array}{rcl}
    ^{c}m^{(0,0)}\!&\!\!=\!\!&\!m^{000|000} \\
    ^{c}m^{(1,0)}\!&\!\!=\!\!&\!0 \\
    ^{c}m^{(1,1)}\!&\!\!=\!\!&\!m^{001|000} \\
    ^{s}m^{(1,1)}\!&\!\!=\!\!&\!0 \\
    ^{c}m^{(2,0)}\!&\!\!=\!\!&\!\frac{1}{2}(m^{002|000}\!+\!3m^{200|000}) \\
    ^{c}m^{(2,1)}\!&\!\!=\!\!&\!0 \\
    ^{s}m^{(2,1)}\!&\!\!=\!\!&\!-m^{100|100} \\
    ^{c}m^{(2,2)}\!&\!\!=\!\!&\!\frac{1}{2}(m^{002|000}\!-\!m^{200|000}) \\
    ^{s}m^{(2,2)}\!&\!\!=\!\!&\!0 \\
    ^{c}m^{(3,0)}\!&\!\!=\!\!&\!0 \\
    ^{c}m^{(3,1)}\!&\!\!=\!\!&\!m^{200|001}\!+\!
    \frac{1}{4}(3m^{003|000}\!+\!m^{201|000}) \\
    ^{s}m^{(3,1)}\!&\!\!=\!\!&\!0 \\
    ^{c}m^{(3,2)}\!&\!\!=\!\!&\!0 \\
    ^{s}m^{(3,2)}\!&\!\!=\!\!&\!-\frac{1}{2} m^{101|100} \\
    ^{c}m^{(3,3)}\!&\!\!=\!\!&\!\frac{1}{4}(m^{003|000}\!-\!m^{201|000}) \\
    ^{s}m^{(3,3)}\!&\!\!=\!\!&\!0
  \end{array}
\end{equation}
We next provide expressions for the coefficients in the expansions of
${\bf k}_{\perp}$, $k_3$, and $|{\bf k}|^2$ in
Eq.~(\ref{Case4kickfit}) in terms of the original coefficients of our
general expansion (\ref{expansions}).
\begin{subequations}
  \begin{eqnarray}
    ^{c}{\bf k}_{\perp}^{(1)}\!&\!\!=\!\!&\!
    -{\bf k}_{\perp}^{001|000}a\!+\!({\bf k}_{\perp}^{200|001}
    \!-\!{\bf k}_{\perp}^{201|000})a_{}^{3} \\
    ^{s}{\bf k}_{\perp}^{(1)}\!&\!\!=\!\!&\!
    {\bf k}_{\perp}^{101|100}a_{}^{3} \\
    ^{c}{\bf k}_{\perp}^{(2)}\!&\!\!=\!\!&\!
    ({\bf k}_{\perp}^{200|000}\!-\!{\bf k}_{\perp}^{002|000})a_{}^{2} \\
    ^{c}{\bf k}_{\perp}^{(3)}\!&\!\!=\!\!&\!
    ({\bf k}_{\perp}^{201|000}\!-\!{\bf k}_{\perp}^{003|000})a_{}^{3}
  \end{eqnarray}
\end{subequations}
\begin{subequations}
  \begin{eqnarray}
    ^{c\!}k_{3}^{(0)}\!&\!\!=\!\!&\!-k_{3}^{100|000}a
    \!+\!(k_{3}^{200|100}\!-\!k_{3}^{300|000})a_{}^{3} \\
    ^{s\!}k_{3}^{(0)}\!&\!\!=\!\!&\!-k_{3}^{100|000}a
    \!+\!(k_{3}^{200|100}\!-\!k_{3}^{300|000})a_{}^{3} \\
    ^{c\!}k_{3}^{(1)}\!&\!\!=\!\!&\!-k_{3}^{100|001}a_{}^{2} \\
    ^{s\!}k_{3}^{(1)}\!&\!\!=\!\!&\!-k_{3}^{101|000}a_{}^{2} \\
    ^{c\!}k_{3}^{(2)}\!&\!\!=\!\!&\!
    -(k_{3}^{200|100}\!+\!k_{3}^{100|002})a_{}^{3} \\
    ^{s\!}k_{3}^{(2)}\!&\!\!=\!\!&\!
    +(k_{3}^{300|000}\!-\!k_{3}^{102|000})a_{}^{3}
  \end{eqnarray}
\end{subequations}
\begin{subequations}
  \begin{eqnarray}
    ^{c\!}K_{}^{(0)}\!&\!\!=\!\!&\!{}^{c}k_{3}^{(0)2}
    \!+\!{}^{s\!\!\;}k_{3}^{(0)2} \\
    ^{s\!}K_{}^{(0)}\!&\!\!=\!\!&\!2\,{}^{c}k_{3}^{(0)}
    {}^{s\!\!\;}k_{3}^{(0)} \\
    ^{c\!}K_{}^{(1)}\!&\!\!=\!\!&\!2[{}^{c}k_{3}^{(0)}
    {}^{c}k_{3}^{(1)}\!\!+\!{}^{s\!\!\;}k_{3}^{(0)}
    {}^{s\!\!\;}k_{3}^{(1)}] \\
    ^{s\!}K_{}^{(1)}\!&\!\!=\!\!&\!2[{}^{c}k_{3}^{(0)}
    {}^{s\!\!\;}k_{3}^{(1)}\!\!+\!{}^{s\!\!\;}k_{3}^{(0)}
    {}^{s\!\!\;}k_{3}^{(1)}] \\
    ^{c\!}K_{}^{(2)}\!&\!\!=\!\!&\!\big|^{c}{\bf k}_{\perp}^{(1)}\!\big|^2
    \!+\!\big|^{s}{\bf k}_{\perp}^{(1)}\!\big|^2\!+\!{}^{c}k_{3}^{(1)2}
    \!\!+\!{}^{s\!\!\;}k_{3}^{(1)2}\!\!-\!{}^{s\!\!\;}k_{3}^{(0)2}\quad 
    \nonumber\\ 
    \!&\!\!+\!\!&\!2[{}^{c}k_{3}^{(0)}{}^{c}k_{3}^{(2)}
      \!\!+\!{}^{s\!\!\;}k_{3}^{(0)}{}^{s\!\!\;}k_{3}^{(2)}] \\
    ^{s\!}K_{}^{(2)}\!&\!\!=\!\!&\!2[{}^{s}{\bf k}_{\perp}^{(1)}
    \!\!\cdot\!{}^{c}{\bf k}_{\perp}^{(1)}\!\!+\!{}^{c}k_{3}^{(0)}
    {}^{s\!\!\;}k_{3}^{(2)}\!\!+\!{}^{s\!\!\;}k_{3}^{(0)}{}^{c}k_{3}^{(2)}
    \nonumber\\
    \!&\!\!+\!\!&\!{}^{c}k_{3}^{(1)}{}^{s\!\!\;}k_{3}^{(1)}] \\
    ^{c\!}K_{}^{(3)}\!&\!\!=\!\!&\!2[{}^{c}{\bf k}_{\perp}^{(1)}
    \!\!\cdot\!{}^{c}{\bf k}_{\perp}^{(2)}\!\!+\!{}^{c}k_{3}^{(1)}
    {}^{c}k_{3}^{(2)}\!\!+\!{}^{s\!\!\;}k_{3}^{(1)}{}^{s\!\!\;}k_{3}^{(2)} 
    \nonumber\\ 
    \!&\!\!-\!\!&\!{}^{s\!\!\;}k_{3}^{(0)}{}^{s\!\!\;}k_{3}^{(1)}] \\
    ^{s\!}K_{}^{(3)}\!&\!\!=\!\!&\!2[{}^{s}{\bf k}_{\perp}^{(1)}\!\!
    \cdot\!{}^{c}{\bf k}_{\perp}^{(2)}\!\!+\!{}^{c}k_{3}^{(1)}
    {}^{s\!\!\;}k_{3}^{(2)}\!\!+\!{}^{s\!\!\;}k_{3}^{(1)}{}^{c}k_{3}^{(2)}] \\
    ^{c\!}K_{}^{(4)}\!&\!\!=\!\!&\!\big|^{c}{\bf k}_{\perp}^{(2)}\!\big|^2
    \!\!-\!\big|{}^{s}{\bf k}_{\perp}^{(1)}\!\big|^2\!\!+\!2\,
    {}^{c}{\bf k}_{\perp}^{(1)}\!\!\cdot\!{}^{c}{\bf k}_{\perp}^{(3)}
    \!\!+\!{}^{c}k_{3}^{(2)2}
    \nonumber\\
    \!&\!\!+\!\!&\!{}^{s\!\!\;}k_{3}^{(2)2}\!\!-\!{}^{s\!\!\;}k_{3}^{(1)2}
    \!\!-\!2\,{}^{s\!\!\;}k_{3}^{(0)}{}^{s\!\!\;}k_{3}^{(2)} \\
    ^{s\!}K_{}^{(4)}\!&\!\!=\!\!&\!2[{}^{s}{\bf k}_{\perp}^{(1)}
    \!\!\cdot\!{}^{c}{\bf k}_{\perp}^{(3)}\!\!+\!{}^{c}k_{3}^{(2)}
    {}^{s\!\!\;}k_{3}^{(2)}] \\
    ^{c\!}K_{}^{(5)}\!&\!\!=\!\!&\!2[{}^{c}{\bf k}_{\perp}^{(2)}
    \!\!\cdot\!{}^{c}{\bf k}_{\perp}^{(3)}\!\!-\!{}^{s\!\!\;}k_{3}^{(1)}
    {}^{s\!\!\;}k_{3}^{(2)}] \\
    ^{c\!}K_{}^{(6)}\!&\!\!=\!\!&\!\big|^{c}{\bf k}_{\perp}^{(3)}\!\big|^2
    \!\!-\!{}^{s\!\!\;}k_{3}^{(2)2}
  \end{eqnarray}
\end{subequations}

\section{Third-order Spin Expansions}
\label{3rdOrderCal}

In Section~\ref{S:10sims}, we identified 10 equal-mass initial spin
configurations which when simulated could be used to calibrate all the
coefficients appearing in spin expansions of the 4 variables $\{ w, x,
y, z \}$ up to second order.  Here we provide the corresponding
third-order terms appearing in those same spin expansions.  If
desired, these formulae can be used to identify 12 additional
equal-mass spin configurations with which these third-order terms may
be calibrated.  The third-order terms in the expansion for $w$ $(P
=+1, X =+1)$ are
\begin{subequations}
  \label{wxyz_3rd_order} 
  \begin{eqnarray}
    w\!&\!\!=\!\!&\!\ldots
    \nonumber \\
    \!&\!\!+\!\!&\!w_{}^{201|000}(a_{1}^{2}a_{3}^{}\!+\!b_{1}^{2}b_{3}^{})
    \!+\!w_{}^{021|000}(a_{2}^{2}a_{3}^{}\!+\!b_{2}^{2}b_{3}^{})
    \nonumber \\ 
    \!&\!\!+\!\!&\!w_{}^{200|001}(a_{1}^{2}b_{3}^{}\!+\!b_{1}^{2}a_{3}^{})
    \!+\!w_{}^{020|001}(a_{2}^{2}b_{3}^{}\!+\!b_{2}^{2}a_{3}^{}) 
    \nonumber \\
    \!&\!\!+\!\!&\!w_{}^{111|000}(a_{1}^{}a_{2}^{}a_{3}^{}
    \!+\!b_{1}^{}b_{2}^{}b_{3}^{})
    \!+\!w_{}^{110|001}(a_{1}^{}a_{2}^{}b_{3}^{}
    \!+\!b_{1}^{}b_{2}^{}a_{3}^{}) 
    \nonumber\\
    \!&\!\!+\!\!&\!w_{}^{101|010}(a_{1}^{}b_{2}^{}a_{3}^{}
    \!+\!b_{1}^{}a_{2}^{}b_{3}^{})
    \!+\!w_{}^{100|011}(a_{1}^{}b_{2}^{}b_{3}^{}
    \!+\!b_{1}^{}a_{2}^{}a_{3}^{}) 
    \nonumber\\
    \!&\!\!+\!\!&\!w_{}^{101|100}a_{1}^{}b_{1}^{}(a_{3}^{}\!+\!b_{3}^{})
    \!+\!w_{}^{011|010}a_{2}^{}b_{2}^{}(a_{3}^{}\!+\!b_{3}^{}) 
    \nonumber \\
    \label{w_3rd_order}
    \!&\!\!+\!\!&\!
    w_{}^{002|001}a_{3}^{}b_{3}^{}(a_{3}^{}\!+\!b_{3}^{})\!+\! 
    w_{}^{003|000}(a_{3}^{3}\!+\!b_{3}^{3}) \, .
  \end{eqnarray}
  The corresponding third-order terms in the expansion for $x$
  ($P=+1$, $X=-1$) may be obtained from the above equation for $w$ by
  making the substitution $w_{}^{m_{1}m_{2}m_{3}|n_{1}n_{2}n_{3}}\to
  x_{}^{m_{1}m_{2}m_{3}| n_{1}n_{2}n_{3}}$, and changing ``$+$'' to
  ``$-$'' when it appears in parentheses:
  $(\ldots\!+\!\ldots)\to(\ldots\!-\!\ldots)$.
  
  The third-order terms in the expansion for $y$ ($P=-1$, $X=+1$) are
  \begin{eqnarray}
    y\!&\!\!=\!\!&\!\ldots 
    \nonumber\\
    \!&\!\!+\!\!&\!
    y_{}^{200|100}a_{1}^{}b_{1}^{}(a_{1}^{}\!+\!b_{1}^{})\!+\!
    y_{}^{020|010}a_{1}^{}b_{1}^{}(a_{2}^{}\!+\!b_{2}^{}) 
    \nonumber\\
    \!&\!\!+\!\!&\!
    y_{}^{110|100}a_{2}^{}b_{2}^{}(a_{1}^{}\!+\!b_{1}^{})\!+\!
    y_{}^{110|010}a_{2}^{}b_{2}^{}(a_{2}^{}\!+\!b_{2}^{}) 
    \nonumber \\  
    \!&\!\!+\!\!&\!
    y_{}^{101|001}a_{3}^{}b_{3}^{}(a_{1}^{}\!+\!b_{1}^{})\!+\!
    y_{}^{011|001}a_{3}^{}b_{3}^{}(a_{2}^{}\!+\!b_{2}^{}) 
    \nonumber \\
    \!&\!\!+\!\!&\!
    y_{}^{120|000}(a_{1}^{}a_{2}^{2}\!+\!b_{1}^{}b_{2}^{2})\!+\!  
    y_{}^{210|000}(a_{2}^{}a_{1}^{2}\!+\!b_{2}^{}b_{1}^{2})
    \nonumber\\
    \!&\!\!+\!\!&\!  
    y_{}^{020|100}(b_{1}^{}a_{2}^{2}\!+\!a_{1}^{}b_{2}^{2})\!+\!
    y_{}^{200|010}(b_{2}^{}a_{1}^{2}\!+\!a_{2}^{}b_{1}^{2}) 
    \nonumber \\
    \!&\!\!+\!\!&\!
    y_{}^{102|000}(a_{1}^{}a_{3}^{2}\!+\!b_{1}^{}b_{3}^{2})\!+\!
    y_{}^{012|000}(a_{2}^{}a_{3}^{2}\!+\!b_{2}^{}b_{3}^{2}) 
    \nonumber\\
    \!&\!\!+\!\!&\!
    y_{}^{100|002}(a_{1}^{}b_{3}^{2}\!+\!b_{1}^{}a_{3}^{2})\!+\!
    y_{}^{010|002}(a_{2}^{}b_{3}^{2}\!+\!b_{2}^{}a_{3}^{2}) 
    \nonumber\\
    \label{y_3rd_order}
    \!&\!\!+\!\!&\!y_{}^{300|000}(a_{1}^{3}\!+\!b_{1}^{3})
    \!+\!y_{}^{030|000}(a_{2}^{3}\!+\!b_{2}^{3}) \, .
  \end{eqnarray}
\end{subequations}
The third order terms in the expansion for $z$ ($P=-1$, $X=-1$)
again may be obtained from the above equation for $y$ by making
the substitution $y_{}^{m_{1}m_{2}m_{3}|n_{1}n_{2}n_{3}}\to
z_{}^{m_{1}m_{2}m_{3}|n_{1}n_{2}n_{3}}$ and changing ``$+$''
to ``$-$'' when it appears inside parentheses: $(\ldots+\ldots)
\to(\ldots-\ldots)$.

\section{Minimum-Variance Estimators for the Spin Coefficients}
\label{S:MVE}

In Sec.~\ref{S:prog} we showed how a small number of simulations (10
in the equal-mass case, 16 for unequal masses) can be used to uniquely
determine the 10 or 16 coefficients appearing to second order in the
spin expansion.  In the absence of systematic errors these are all the
simulations that would be required, but further simulations may be
useful once these errors are taken into account.  In this Appendix, we
will explicitly construct minimum-variance unbiased estimators for the
coefficients in the spin expansion from $N$ noisy simulations of known
covariance.

We proceed in two steps.  In step one, we solve the problem under the
assumption that the uncertainties in the initial spins are negligible
compared to those in the final quantities.  In step two, we explore
how our approach might be modified to include the effects of these
initial spin uncertainties.

\subsection{Neglecting initial spin uncertainties}

Imagine a generic final quantity $f \in \{ w, x, y, z, u, v \}$ such
as those described in Section~\ref{S:prog}. In the $i$th simulation,
with initial spin configuration $\{a_{1}^{(i)},
a_{2}^{(i)},a_{3}^{(i)},b_{1}^{(i)},b_{2}^{(i)},b_{3}^{(i)}\}$, this
quantity will have an estimated value $\hat{f}_i$ that when averaged
over different assumptions for the systematic errors is equal to its
true value $f_i$:
\begin{equation} \label{E:meanf}
  \langle \hat{f}_i \rangle = f_i \, .
\end{equation}
The systematic errors introduce uncertainty in the values of $f$
estimated for each simulation, and this uncertainty can be described
by the covariance matrix for $f$,
\begin{equation} \label{E:covf}
  F_{ij} \equiv \langle \hat{f}_i \hat{f}_j \rangle -
  \langle \hat{f}_i \rangle \langle \hat{f}_j \rangle \, .
\end{equation}
Consider $N$ initial spin configurations, which correspond to the $N$
true final values $f_{i}$ ($i=1,\ldots,N$).  If we truncate the spin
expansion at finite order then, as seen in Eqs.~(\ref{wxyz_2nd_order})
and (\ref{uv_2nd_order}), these $N$ values $f_{i}$ are a linear
combination of $D$ spin expansion coefficients $c_{j}$.  Thus we can
write the relationship in matrix form as
\begin{equation}
  \label{f_eq}
  {\bf f}={\bf A}{\bf c}
\end{equation}
where ${\bf f}$ is a column vector with $N$ elements $f_{i}$, ${\bf
  c}$ is a column vector with $D$ elements $c_{j}$, and ${\bf A}$ is
an $N\times D$ matrix whose $N$ rows consist of the $D$
  combinations of initial spin components in each simulation
  multiplying the coefficients $c_j$.  For example, when we rewrite
Eq.~(\ref{x_2nd_order}) in the form of Eq.~(\ref{f_eq}), we have ${\bf
  f}=\{x_{1},\ldots,x_{N}\}$ and ${\bf
  c}=\{x_{}^{001|000},x_{}^{002|000},\ldots,x_{}^{100|010}\}$, with
$D=6$.  The elements of the matrix $A$ are then easily read off from
Eq.~(\ref{x_2nd_order}), {\it
  e.g.}\ $A_{41}=(a_{3}^{(4)}-b_{3}^{(4)})$.

At least $D$ simulations are required to determine the $D$
coefficients $c_j$.  If this minimum number of simulations $N = D$ are
performed and the spin configurations of these simulations are chosen
such that the $D \times D$ matrix ${\bf A}$ is invertible, then there
is a unique estimator 
\begin{equation} \label{E:D2est}
  \hat{{\bf c}} = {\bf A}^{-1} \hat{{\bf f}}
\end{equation}
such that
\begin{equation} \label{E:D2mean}
  \langle \hat{{\bf c}} \rangle = {\bf A}^{-1} \langle \hat{{\bf f}}
  \rangle = {\bf A}^{-1}{\bf A}{\bf c} = {\bf c} \, .
\end{equation}
Here we have assumed that the initial spin components are known
exactly and all the uncertainty lies in the estimated final quantities
$\hat{{\bf f}}$.  We will relax this assumption later in this
Appendix.  The covariance matrix $C_{ij}$ for this estimator is
\begin{eqnarray} \label{E:D2cov}
  C_{ij} &=& \langle \hat{c}_i \hat{c}_j \rangle - \langle \hat{c}_i
  \rangle \langle \hat{c}_j \rangle \nonumber \\
  &=& A^{-1}_{ik} A^{-1}_{jl} \langle \hat{f}_k \hat{f}_l \rangle -
  A^{-1}_{ik}A^{-1}_{jl}\langle\hat{f}_k\rangle\langle\hat{f}_l\rangle 
  \nonumber \\
  &=& A^{-1}_{ik} A^{-1}_{jl} F_{kl} \, ,
\end{eqnarray}
where here and throughout this Appendix we adopt the Einstein
convention of summing over repeated indices.  In the absence of
systematic errors $(F_{ij} = 0)$, $D$ simulations would suffice to
determine the spin coefficients with perfect accuracy $(C_{ij} = 0)$.

With systematic errors present $(F_{ij}\neq0)$, a larger set of
simulations $(N > D)$ can be used to construct estimators $\hat{{\bf
    c}}$ with lower variance provided these additional simulations are
at least partially {\it uncorrelated}.  As the estimated final
quantities $\hat{{\bf f}}$ are linear in the spin coefficients, our
estimator generalizes to
\begin{equation} \label{E:Cest}
  \hat{{\bf c}} = {\bf W} \hat{{\bf f}} \, ,
\end{equation}
where ${\bf W}$ is now a $D \times N$ matrix of linear weights.  For
this estimator to be unbiased
\begin{equation} \label{E:Cbias}
  \langle \hat{{\bf c}} \rangle = {\bf W} \langle \hat{{\bf f}}
  \rangle = {\bf W}{\bf A}{\bf c} = {\bf c}
\end{equation}
implying that ${\bf W}$ must satisfy the constraint
\begin{equation} \label{E:cons}
  {\bf W}{\bf A} = {\bf I} \, ,
\end{equation}
where ${\bf I}$ is the $D\times D$ identity matrix.  The set of $N$
spin configurations to be simulated must be chosen such that ${\bf A}$
has a left inverse.  Eq.~(\ref{E:cons}) consists of $D^2$ constraints
on the $DN$ elements of ${\bf W}$, leaving additional freedom for $N >
D$ to choose ${\bf W}$ to minimize the covariance
\begin{eqnarray} \label{E:Ccov}
  C_{ij} &=& \langle \hat{c}_i \hat{c}_j \rangle - \langle \hat{c}_i
  \rangle \langle \hat{c}_j \rangle \nonumber \\
  &=& W_{ik} W_{jl} \langle\hat{f}_k\hat{f}_l\rangle -
  W_{ik} W_{jl}\langle\hat{f}_k\rangle\langle\hat{f}_l\rangle\nonumber \\
  &=& W_{ik} W_{jl} F_{kl} \, ,
\end{eqnarray}
or in matrix notation
\begin{equation} \label{E:Ccovmat}
  {\bf C} = {\bf W} {\bf F} {\bf W}^T \, .
\end{equation}
As ${\bf C}$ is a real, symmetric matrix, it can be decomposed into a
real, diagonal eigenvalue matrix $\boldsymbol{\sigma}$ and an
orthogonal eigenvector matrix ${\bf O}$
\begin{equation} \label{E:Ceigen}
  {\bf C} = {\bf O}\boldsymbol{\sigma}{\bf O}^T \, .
\end{equation}
The columns of ${\bf O}$ give the uncorrelated linear combinations of
estimators $\hat{c}_i$, while the elements of $\boldsymbol{\sigma}$
give the variances of these combinations.  We specifically seek the
weight matrix ${\bf W}$ that minimizes the sum of these eigenvalues
\begin{equation} \label{E:trace}
  {\rm Tr} \, \boldsymbol{\sigma} = {\rm Tr} \, {\bf C} \, .
\end{equation}

We can determine this ${\bf W}$ by the method of Lagrange multipliers,
with a Lagrangian given by
\begin{equation} \label{E:Lagrange}
  L = {\rm Tr}[{\mathbf C} + \boldsymbol{\lambda}^{T} ({\bf W}{\bf
      A}-{\bf I})]\, ,
\end{equation}
where $\boldsymbol{\lambda}$ is a $D \times D$ matrix of Lagrange
multipliers.  Setting the partial derivatives $\partial L/\partial
\lambda_{ij}$ to zero yields the $D \times D$ constraint equation
(\ref{E:cons}), while $\partial L/\partial W_{ij} = 0$ provides the
additional $D \times N$ matrix equation
\begin{equation} \label{E:MV}
  2{\bf W}{\bf F} + \boldsymbol{\lambda}{\bf A}^T = 0 \, .
\end{equation}
Eqs.~(\ref{E:cons}) and (\ref{E:MV}) thus provide $D(N+D)$ linear
equations for the $D^2$ elements of $\boldsymbol{\lambda}$ and
$DN$ elements of ${\bf W}$.  Solving these equations, we find
\begin{equation} \label{E:opt}
  {\bf W} = ({\bf A}^T{\bf F}^{-1}{\bf A})^{-1}{\bf A}^T{\bf F}^{-1} \, .
\end{equation}

To summarize the analysis so far: If we can neglect the errors in the
initial spin components $\{a_{i}^{(r)},b_{i}^{(r)}\}$ that go into the
construction of ${\bf A}$, then the minimum variance unbiased
estimators $\hat{c}_{i}$ for the spin-expansion coefficients $c_{i}$
are given by Eq.~(\ref{E:Cest}), with weight matrix ${\bf W}$ given by
Eq.~(\ref{E:opt}).  These optimal estimators $\hat{c}_{i}$ will have
covariance matrix given by
\begin{equation} \label{E:var1}
  {\bf C} = ({\bf A}^T{\bf F}^{-1}{\bf A})^{-1} \, .
\end{equation}

\subsection{Including initial spin uncertainties}

Now let us consider the effect of errors in the initial spin
components $\{a_{i}^{(r)}, b_{i}^{(r)}\}$.  One source of these errors
is that numerical relativists do not know how to specify the proper
initial data corresponding to physical binaries of arbitrary spin.
This problem is usually addressed by waiting for the non-physical
``junk radiation'' to exit the system, after which the binary is
presumed to settle down into a {\it physical} spin configuration.
This physical spin configuration will generally be slightly different
than the one relativists had intended to specify, introducing error
into the initial spin components $\{a_{i}^{(r)}, b_{i}^{(r)}\}$.
Techniques exist to measure BBH spins in simulations
\cite{Brown:1992br, Ashtekar:2003hk, Campanelli:2006fy, Cook:2007wr,
  Owen:2008}, so one could measure the initial spins {\it after} the
junk radiation had exited the system, and use these components rather
than those supposedly specified by the numerical initial data.
However, while at large binary separations the initial spin components
of the two black holes are well defined, at finite separations the
components are gauge-dependent and different techniques for measuring
them yield different results.

To formally address the systematic errors in ${\bf A}$ we promote it
to an estimator $\hat{{\bf A}}$ that on average will provide the
correct values
\begin{equation} \label{E:Amean}
  \langle \hat{A}_{ij} \rangle = A_{ij}
\end{equation}
but will now have a non-zero covariance
\begin{equation} \label{E:Acov}
  S_{ijkl} \equiv  \langle \hat{A}_{ij} \hat{A}_{kl} \rangle - \langle
  \hat{A}_{ij} \rangle \langle \hat{A}_{kl} \rangle \, .
\end{equation}

Deriving the truly optimal weight matrix ${\bf W}$ that simultaneously
minimizes contributions to the covariance from errors in both the
final quantities $\hat{{\bf f}}$ and initial spins $\hat{{\bf A}}$
will be challenging.  Since ${\bf W}$ is constructed from ${\bf A}$ as
seen in Eq.~(\ref{E:opt}), it is itself now an estimator $\hat{{\bf
    W}}$ with its own covariance matrix
\begin{equation} \label{E:Wcov}
  T_{ijkl} \equiv  \langle \hat{W}_{ij} \hat{W}_{kl} \rangle - \langle
  \hat{W}_{ij} \rangle \langle \hat{W}_{kl} \rangle \, .
\end{equation}
The covariance matrix of our estimator $\hat{{\bf c}}$ will now be
given by
\begin{eqnarray} \label{E:Ccov2}
  C_{ij} &=& \langle \hat{c}_i \hat{c}_j \rangle - \langle \hat{c}_i
  \rangle \langle \hat{c}_j \rangle \nonumber \\
  &=& \langle \hat{W}_{ik} \hat{W}_{jl} \hat{f}_k \hat{f}_l \rangle -
  c_i c_j \, .
\end{eqnarray}
To make further progress, we make the possibly invalid assumption that
errors in our estimators $\hat{{\bf W}}$ and $\hat{{\bf f}}$ are
{\it uncorrelated}
\begin{equation} \label{E:uncorr}
  \langle \hat{W}_{ik} \hat{W}_{jl} \hat{f}_k \hat{f}_l \rangle =
  \langle \hat{W}_{ik} \hat{W}_{jl} \rangle \langle \hat{f}_k \hat{f}_l
  \rangle \, .
\end{equation}
This allows us to reduce Eq.~(\ref{E:Ccov2}) to
\begin{equation} \label{E:CcovF}
  C_{ij} = W_{ik} W_{jl} F_{kl} + T_{ikjl} f_k f_l + T_{ikjl} F_{kl} \, .
\end{equation}
The first term in Eq.~(\ref{E:CcovF}) is the familiar error of
Eq.~(\ref{E:Ccov}), while the second and third terms proportional to
$T_{ijkl}$ reflect the increased covariance due to errors in the
initial spins.  One might next hope to insert this ${\bf C}$ into the
Lagrangian $L$ of Eq.~(\ref{E:Lagrange}) and obtain a new $N \times D$
matrix equation $\partial L/\partial A_{ij} = 0$ to constrain ${\bf
  A}$.  Two problems immediately come to mind with this approach.
Firstly, once ${\bf A}$ has been promoted to an estimator
Eqs.~(\ref{E:cons}) and (\ref{E:MV}) become non-linear in ${\bf W}$,
${\bf A}$, and $\boldsymbol{\lambda}$ and hence much more difficult to
solve.  Secondly, only $6N$ of the $DN$ elements of ${\bf A}$ may be
chosen independently as there are only 6 initial spin components in
each of the $N$ simulations.  Possibly this could be addressed by
adding new terms to the Lagrangian with new Lagrange multipliers,
although this might be difficult to implement.

Leaving the determination of a truly optimal estimator for future
work, we choose to stick with the estimator $\hat{{\bf c}}$ defined by
the weight matrix ${\bf W}$ of Eq.~(\ref{E:opt}).  This estimator
should remain nearly optimal provided the initial spin errors are
subdominant to the other errors coming from the simulations themselves
and from truncating the spin expansion at finite order.  Now we can
account for the errors in the initial spins $\{a_{i}^{(r)},
b_{i}^{(r)}\}$ as follows.  If these errors have a known probability
distribution ({\it e.g.}\ if they are assumed to be Gaussian, with
known covariance matrix), then we can Monte Carlo many realizations of
$\{a_{i}^{(r)},b_{i}^{(r)}\}$, and use these to compute many
realizations of ${\bf A}$.  Next, by inserting these ${\bf
  A}$-realizations into Eqs.~(\ref{E:opt}) and (\ref{E:Cest}), we
obtain many realizations of $\hat{{\bf c}}$.  The mean of these ${\bf
  c}$-realizations is our best guess for ${\bf c}$, while the
covariance of these realizations gives an estimate of the ``extra''
uncertainty in the estimator $\hat{{\bf c}}$ due to the initial spin
uncertainties.  We can add this ``extra'' covariance to the right hand
side of Eq.~(\ref{E:var1}) to estimate the {\it total} covariance
$C_{ij}$.

As an alternative to Monte Carlo, we can make further analytic
progress by assuming that the errors $\boldsymbol{\delta}{\bf A}$ in
$\hat{{\bf A}}$ are small, in which case the errors
$\boldsymbol{\delta}{\bf W}$ in $\hat{{\bf W}}$ can be linearized in
$\boldsymbol{\delta}{\bf A}$.  Defining
\begin{equation} \label{E:Ndef}
  {\bf N} \equiv {\bf A}^T{\bf F}^{-1}{\bf A} \, ,
\end{equation}
we linearize Eq.~(\ref{E:opt}) to obtain
\begin{eqnarray} \label{E:delW}
  \boldsymbol{\delta}{\bf W} &=& {\bf N}^{-1}
  (\boldsymbol{\delta}{\bf A}^T - \boldsymbol{\delta}{\bf A}^T
             {\bf F}^{-1}{\bf A}{\bf N}^{-1}{\bf A}^T \nonumber \\
             && - {\bf A}^T{\bf F}^{-1}\boldsymbol{\delta}{\bf A}
             {\bf N}^{-1}{\bf A}^T){\bf F}^{-1} \, .
\end{eqnarray}
This equation allows us to propagate errors and express the covariance
$T_{ijkl}$ of $\hat{{\bf W}}$ in terms of the covariance $S_{ijkl}$
of $\hat{{\bf A}}$.  Defining 
\begin{equation}
  {\bf M}\equiv {\bf F}^{-1}{\bf A}{\bf N}^{-1}{\bf A}^{T}
\end{equation}
we obtain
\begin{eqnarray} \label{E:SpinProp}
  T_{ijkl}\!&\!=\!&\!\langle \delta W_{ij} \delta W_{kl} \rangle 
  \nonumber \\
  \!&\!=\!&\! N_{ia}^{-1} N_{kb}^{-1} F_{jc}^{-1} F_{ld}^{-1} \Big[
  S_{cadb}+S_{eafb} {\bf M}_{ec}{\bf M}_{fd}\nonumber \\
  \!&\!+\!&\!S_{efgh} ({\bf F}^{-1}{\bf A})_{ea}({\bf A}{\bf N}^{-1})_{cf}
  ({\bf F}^{-1}{\bf A})_{gb}({\bf A}{\bf N}^{-1})_{dh}\nonumber\\
  \!&\!-\!&\!2S_{caef}({\bf F}^{-1}{\bf A})_{eb}
  ({\bf A}{\bf N}^{-1})_{df}-2S_{caeb}{\bf M}_{ed}\nonumber \\
  \!&\!+\!&\!2S_{eafg}{\bf M}_{ec}({\bf F}^{-1}{\bf A})_{fb}
  ({\bf A}{\bf N}^{-1})_{dg}\Big]\,.
\end{eqnarray}
Inserting Eq.~(\ref{E:SpinProp}) in (\ref{E:CcovF}) provides an
approximate analytic expression for the ``extra'' covariance of the
estimator $\hat{{\bf c}}$ caused by uncertainties in the initial
spins.

\section{Generalization to non-circular (eccentric) orbits}
\label{S:ecc}

We focused on circular orbits in the body of this paper as
gravitational radiation is expected to circularize orbits of most
astrophysical systems long before the final stage of the merger
\cite{Peters:1963ux}.  However, our approach readily generalizes to
initially non-circular orbits so we felt that a few brief remarks on
this subject might be appropriate here.  Recall from
Sec.~\ref{S:formalism} that in the circular case, the initial
conditions are specified by 8 dimensionless parameters: the mass ratio
$q$ and the dimensionless spins $\{ {\bf a}, {\bf b} \}$ at some
initial instant labelled by $\psi$.  We can extend our spin expansion
to non-circular orbits by specifying the difference in linear momentum
${\bf p} \equiv {\bf p}_{A}^{} - {\bf p}_{B}^{}$ between the two BBHs.
As ${\bf p}$ lies in the orbital plane, our initial conditions are now
specified by 8+2=10 numbers.

Apart from this modification, the analysis proceeds just as in
Sec.~\ref{S:formalism}.  We define the same orthonormal triad $\{ {\bf
  e}^{(1)}, {\bf e}^{(2)},{\bf e}^{(3)} \}$, and consider the maps
from the initial quantities to the final quantities
\begin{equation}
  f=f(\psi,p_{i}^{},q,a_{i}^{},b_{i}^{}).
\end{equation}
As in Sec.~\ref{S:formalism}, we can constrain these maps through
symmetry considerations.  Under parity $P$ or exchange $X$, we have
${\bf p}\to-{\bf p}$ and $\{{\bf e}_{}^{(1)}, {\bf
  e}_{}^{(2)}\}\to-\{{\bf e}_{}^{(1)}, {\bf e}_{}^{(2)}\}$.  Thus,
the components $p_{1}^{}$ and $p_{2}^{}$ are invariant under both 
$P$ and $X$.  The maps therefore satisfy
\begin{subequations}
  \begin{equation}
    f(\psi,p_{i}^{},q,a_{i}^{},b_{i}^{})=(\pm)_{P}^{}
    f(\psi,p_{i}^{},q,\tilde{a}_{i}^{},\tilde{b}_{i}^{})
  \end{equation}
  and 
  \begin{equation}
    f(\psi,p_{i}^{},q,a_{i}^{},b_{i}^{})=(\pm)_{PX}^{}
    f(\psi,p_{i}^{},1/q,b_{i}^{},a_{i}^{}).
  \end{equation}
\end{subequations}
Since the components $p_{1}^{}$ and $p_{2}^{}$ have eigenvalues of
$+1$ under $P$ and $X$, the series expansions introduced in
Sec.~\ref{S:series} still hold, but now the coefficients
$f{}^{m_{1}m_{2}m_{3}|n_{1}n_{2}n_{3}}$ are functions of $\psi$, $q$,
{\it and} $p_{i}^{}$.  It is probably useful to Taylor expand these
coefficinets around the point $p_{i}^{}=p_{i,{\rm circ}}^{}$, where
$p_{i,{\rm circ}}^{}$ is the linear momentum for a circular {\it
  non}-spinning orbit at orbital ``separation'' $\psi$.  In the
Newtonian limit, the three parameters $\{\psi,p_{1}^{},p_{2}^{}\}$
specify the semi-major axis, eccentricity, and longitude of pericenter
associated with elliptical orbits.  As the BBHs inspiral, they will
trace a trajectory through this 3-dimensional parameter space.  Using
this to relate coefficients defined at different points in the
parameter space will be pursued in future work.

\end{document}